\documentclass[fleqn,usenatbib]{mnras}

\usepackage{newtxtext,newtxmath}
\newcommand{\Teff}{\mbox{$T_{\mathrm{eff}}$}}
\newcommand{\logg}{\mbox{$\log{g}$}}
\newcommand{\htohe}{\mbox{$\log(\mathrm{H/He})$}}

\newcommand{\Ion}[2]{#1{\,\sc#2}}

\newcommand{\Msun}{\mbox{$\mathrm{M}_{\odot}$}}

\newcommand{\specnum}{3673}
\newcommand{\totalnum}{4064}
\newcommand{\totminusbit}{627}
\newcommand{\bitnum}{3437}
\newcommand{\numexposures}{31\,842}

\newcommand{\numnoexp}{4852}
\newcommand{\numwd}{2706}
\newcommand{\numbinary}{66}
\newcommand{\numcontaminant}{817}
\newcommand{\numunclass}{84}
\newcommand{\numsdss}{2149}
\newcommand{\numdas}{1958}

\usepackage{stmaryrd}

\usepackage[T1]{fontenc}

\DeclareRobustCommand{\VAN}[3]{#2}
\let\VANthebibliography\thebibliography
\def\thebibliography{\DeclareRobustCommand{\VAN}[3]{##3}\VANthebibliography}

\usepackage{graphicx}	
\usepackage{amsmath}	
\usepackage{caption}
\usepackage{xcolor}
\usepackage{float}

\title[DESI EDR White Dwarf Catalogue]{The DESI Early Data Release White Dwarf Catalogue}

\author[C.\,J.\,Manser et al.]{\noindent
Christopher J. Manser,$^{1,2}$\thanks{E-mail: c.j.manser92@googlemail.com}
Paula Izquierdo,$^{2}$
Boris T. G\"ansicke,$^{2}$
Andrew Swan,${^{2}}$
Detlev Koester,${^{3}}$
\newauthor
Akshay Robert,${^{4}}$
Siyi Xu,${^{5}}$
Keith Inight,${^{2}}$
Ben Amroota,${^{1}}$
N.~P.~Gentile Fusillo,${^{6}}$
Sergey E. Koposov,${^{7,8,9}}$
\newauthor
Bokyoung Kim,${^{7}}$
Arjun Dey,${^{10}}$
Carlos Allende Prieto,${^{11,12}}$
J.~Aguilar,${^{13}}$
S.~Ahlen,${^{14}}$
R.~Blum,${^{10}}$
\newauthor
D.~Brooks,${^{4}}$
T.~Claybaugh,${^{13}}$
A.~P.~Cooper,${^{15}}$
K.~Dawson,${^{16}}$
A.~de la Macorra,${^{17}}$
P.~Doel,${^{4}}$
\newauthor
J.~E.~Forero-Romero,${^{18,19}}$
E.~Gaztañaga,${^{20,21,22}}$
S.~Gontcho A Gontcho,${^{13}}$
K.~Honscheid,${^{23,24,25}}$
\newauthor
T.~Kisner,${^{13}}$
A.~Kremin,${^{13}}$
A.~Lambert,${^{13}}$
M.~Landriau,${^{13}}$
L.~Le~Guillou,${^{26}}$
Michael~E.~Levi,${^{13}}$
T.~S.~Li,${^{27}}$
\newauthor
A.~Meisner,${^{10}}$
R.~Miquel,${^{28,29}}$
J.~Moustakas,${^{30}}$
J.~Nie,${^{31}}$
N.~Palanque-Delabrouille,${^{13,32}}$
\newauthor
W.~J.~Percival,${^{33,34,35}}$
C.~Poppett,${^{13,36,37}}$
F.~Prada,${^{38}}$
M.~Rezaie,${^{39}}$
G.~Rossi,${^{40}}$
E.~Sanchez,${^{41}}$
\newauthor
E.~F.~Schlafly,${^{42}}$
D.~Schlegel,${^{13}}$
M.~Schubnell,${^{43,44}}$
H.~Seo,${^{45}}$
J.~Silber,${^{13}}$
G.~Tarl\'{e},${^{44}}$
B.~A.~Weaver,${^{10}}$
\newauthor
Z.~Zhou,${^{31}}$
H.~Zou${^{31}}$
\\
$^{1}$ Astrophysics Group, Department of Physics, Imperial College London, Prince Consort Rd, London, SW7 2AZ, UK \\
$^{2}$ Department of Physics, University of Warwick, Coventry CV4 7AL, UK \\
$^{3}$ Institut f\"{u}r Theoretische Physik und Astrophysik, University of Kiel, 24098 Kiel, Germany \\
$^{4}$ Department of Physics and Astronomy, University College London, London WC1E 6BT, UK \\
$^{5}$ Gemini Observatory/NSF's NOIRLab, 670 N. A'ohoku Place, Hilo, Hawaii, 96720, USA \\
$^{6}$ Department of Physics, Universita' degli Studi di Trieste, Via A. Valerio 2,  34127, Trieste, Italy\\  
$^{7}$ Institute for Astronomy, University of Edinburgh, Royal Observatory, Blackford Hill, Edinburgh EH9 3HJ, UK \\
$^{8}$ Institute of Astronomy, University of Cambridge, Madingley Road, Cambridge
CB3 0HA, UK\\
$^{9}$ Kavli Institute for Cosmology, University of Cambridge, Madingley Road,
Cambridge CB3 0HA, UK\\
$^{10}$ NSF’s NOIRLab, 950 N. Cherry Avenue, Tucson, AZ 85719, USA \\
$^{11}$ Instituto de Astrof\'isica de Canarias, V\'ia L\'actea, 38205 La Laguna, Tenerife, Spain \\
$^{12}$ Universidad de La Laguna, Departamento de Astrof\'isica, 38206 La Laguna, Tenerife, Spain \\
$^{13}$ Lawrence Berkeley National Laboratory, 1 Cyclotron Road, Berkeley, CA 94720, USA \\
$^{14}$ Physics Dept., Boston University, 590 Commonwealth Avenue, Boston, MA 02215, USA \\
$^{15}$ Institute of Astronomy and Department of Physics, National Tsing Hua University, 101 Kuang-Fu Rd. Sec. 2, Hsinchu 30013, Taiwan \\
$^{16}$ Department of Physics and Astronomy, The University of Utah, 115 South 1400 East, Salt Lake City, UT 84112, USA\\
$^{17}$ Instituto de F\'{\i}sica, Universidad Nacional Aut\'{o}noma de M\'{e}xico,  Cd. de M\'{e}xico  C.P. 04510,  M\'{e}xico \\
$^{18}$ Departamento de F\'isica, Universidad de los Andes, Cra. 1 No. 18A-10, Edificio Ip, CP 111711, Bogot\'a, Colombia \\
$^{19}$ Observatorio Astron\'omico, Universidad de los Andes, Cra. 1 No. 18A-10, Edificio H, CP 111711 Bogot\'a, Colombia \\
$^{20}$ Institut d'Estudis Espacials de Catalunya (IEEC), 08034 Barcelona, Spain \\
$^{21}$ Institute of Cosmology \& Gravitation, University of Portsmouth, Dennis Sciama Building, Portsmouth, PO1 3FX, UK \\
$^{22}$ Institute of Space Sciences, ICE-CSIC, Campus UAB, Carrer de Can Magrans s/n, 08913 Bellaterra, Barcelona, Spain \\
$^{23}$ Center for Cosmology and AstroParticle Physics, The Ohio State University, 191 West Woodruff Avenue, Columbus, OH 43210, USA \\
$^{24}$ Department of Physics, The Ohio State University, 191 West Woodruff Avenue, Columbus, OH 43210, USA \\
$^{25}$ The Ohio State University, Columbus, 43210 OH, USA \\
$^{26}$ Sorbonne Universit\'{e}, CNRS/IN2P3, Laboratoire de Physique Nucl\'{e}aire et de Hautes Energies (LPNHE), FR-75005 Paris, France \\
$^{27}$ Department of Astronomy \& Astrophysics, University of Toronto, Toronto, ON M5S 3H4, Canada \\
$^{28}$ Instituci\'{o} Catalana de Recerca i Estudis Avan\c{c}ats, Passeig de Llu\'{\i}s Companys, 23, 08010 Barcelona, Spain \\
$^{29}$ Institut de F\'{i}sica d'Altes Energies (IFAE), The Barcelona Institute of Science and Technology, Campus UAB, 08193 Bellaterra Barcelona, Spain \\
$^{30}$ Department of Physics and Astronomy, Siena College, 515 Loudon Road, Loudonville, NY 12211, USA \\
$^{31}$ National Astronomical Observatories, Chinese Academy of Sciences, A20 Datun Rd., Chaoyang District, Beijing, 100012, P.R. China \\
$^{32}$ IRFU, CEA, Universit\'{e} Paris-Saclay, F-91191 Gif-sur-Yvette, France \\
$^{33}$ Department of Physics and Astronomy, University of Waterloo, 200 University Ave W, Waterloo, ON N2L 3G1, Canada \\
$^{34}$ Perimeter Institute for Theoretical Physics, 31 Caroline St. North, Waterloo, ON N2L 2Y5, Canada \\
$^{35}$ Waterloo Centre for Astrophysics, University of Waterloo, 200 University Ave W, Waterloo, ON N2L 3G1, Canada \\
$^{36}$ Space Sciences Laboratory, University of California, Berkeley, 7 Gauss Way, Berkeley, CA  94720, USA\\
$^{37}$ University of California, Berkeley, 110 Sproul Hall \#5800 Berkeley, CA 94720, USA\\
$^{38}$ Instituto de Astrof\'{i}sica de Andaluc\'{i}a (CSIC), Glorieta de la Astronom\'{i}a, s/n, E-18008 Granada, Spain \\
$^{39}$ Department of Physics, Kansas State University, 116 Cardwell Hall, Manhattan, KS 66506, USA \\
$^{40}$ Department of Physics and Astronomy, Sejong University, Seoul, 143-747, Korea \\
$^{41}$ CIEMAT, Avenida Complutense 40, E-28040 Madrid, Spain \\
$^{42}$ Space Telescope Science Institute, 3700 San Martin Drive, Baltimore, MD 21218, USA \\
$^{43}$ Department of Physics, University of Michigan, Ann Arbor, MI 48109, USA \\
$^{44}$ University of Michigan, Ann Arbor, MI 48109, USA \\
$^{45}$ Department of Physics \& Astronomy, Ohio University, Athens, OH 45701, USA \\
}
\date{Accepted XXX. Received YYY; in original form ZZZ}

\pubyear{2023}

\begin{document}
\label{firstpage}
\pagerange{\pageref{firstpage}--\pageref{lastpage}}
\maketitle

\begin{abstract}
The Early Data Release (EDR) of the Dark Energy Spectroscopic Instrument (DESI) comprises spectroscopy obtained from 2020 December 14 to 2021 June 10. White dwarfs were targeted by DESI both as calibration sources and as science targets and were selected based on \textit{Gaia} photometry and astrometry. Here we present the DESI EDR white dwarf catalogue, which includes 2706 spectroscopically confirmed white dwarfs of which approximately 1630 ($\simeq 60$\,per\,cent) have been spectroscopically observed for the first time, as well as 66 white dwarf binary systems. We provide spectral classifications for all white dwarfs, and discuss their distribution within the \textit{Gaia} Hertzsprung–Russell diagram. We provide atmospheric parameters derived from spectroscopic and photometric fits for white dwarfs with pure hydrogen or helium photospheres, a mixture of those two, and white dwarfs displaying carbon features in their spectra. We also discuss the less abundant systems in the sample, such as those with magnetic fields, and cataclysmic variables. The DESI EDR white dwarf sample is significantly less biased than the sample observed by the Sloan Digital Sky Survey, which is skewed to bluer and therefore hotter white dwarfs, making DESI more complete and suitable for performing statistical studies of white dwarfs.  
\end{abstract}
\begin{keywords}
white dwarfs -- surveys -- catalogues -- techniques: spectroscopic
\end{keywords}



\section{Introduction}
White dwarfs are the stellar remnants left over from the evolution of main sequence stars born with masses $\leq 8$\,\Msun\ \citep{ibenetal97-1, dobbieetal06-1}. Sustained against gravity by the pressure of degenerate electron gas, white dwarfs are Earth-sized and intrinsically faint objects. White dwarfs emerge from the giant branch evolution of their progenitors extremely hot, $\simeq10^5$\,K, but in the absence of nuclear fusion, they gradually cool. Throughout their evolution, white dwarfs have similar optical colours as much more luminous and distant quasars and main sequence stars. Consequently, the selection of foreground white dwarfs has been difficult against the much more numerous background contaminants. 

Early samples were based on proper motion searches \citep[e.g.][]{giclasetal65-1}, and while they were fairly representative of the intrinsic white dwarf population, only yielded a few hundred confirmed white dwarfs. Photometric surveys for blue-excess objects and individual follow-up spectroscopy identified $\sim1000$ white dwarfs \citep{greenetal86-1, homeieretal98-1, christliebetal01-1}~--~however, these samples were heavily biased towards hot ($\Teff\gtrsim10\,000$\,K) and young ($\lesssim1$\,Gyr) systems. Major progress was made possible by wide-area multi-object spectroscopic surveys, including the Two-degree-Field Galaxy Redshift Survey (2dF, \citealt{vennesetal02-1}), the Large Sky Area Multi-Object Fiber Spectroscopic Telescope Survey (LAMOST, \citealt{guoetal15-2}), and the Sloan Digital Sky Survey (SDSS, \citealt{kleinmanetal04-1,kleinmanetal13-1,eisensteinetal06-1,kepleretal21-1}), with the latter producing the largest and most homogeneous white dwarf sample to date. Yet, even the SDSS white dwarf sample was still subject to complex selection effects \citep{richardsetal02-1,kleinmanetal04-1}, as white dwarfs were targeted serendipitously.

An unbiased selection of white dwarfs became only possible with photometry and astrometry from \textit{Gaia}, leading to a deep $(G\simeq20)$ and homogeneous all-sky sample of $\simeq359\,000$ high-confidence white dwarf candidates \citep{fusilloetal19-1,fusilloetal21-1}. Follow-up spectroscopy is essential to both confirm the white dwarf nature of these candidates, and determine their physical properties.

Research areas making use of large samples of white dwarfs with accurately determined characteristics include, but are not limited to:

\begin{itemize}
\item The spectral evolution of white dwarfs as they age \citep{koester76-1, pelletieretal86-1, bergeronetal90-1, althausetal05-1, cunninghametal20-1, rollandetal20-1, bedardetal22-1, blouinetal23-1, camisassaetal23-1}.
\item The composition and evolution of planetary systems \citep{zuckerman+becklin87-1, jura03-1, zuckermanetal07-1, gaensickeetal12-1, zuckerman15-1, vanderburgetal15-1,swanetal19-1, schreiberetal19-1, hollandsetal21-1, izquierdoetal21-1, kaiseretal21-1, kleinetal21-1, trierweileretal23-1,rogersetal24-1, swanetal24-1}. 
\item The effects of magnetic fields in the range of $\simeq 10^2$-$10^6$\,kG on the atom and stellar atmospheres \citep{kempetal70-1, angeletal74-2, reidetal01-1, schimeczek+wunner14-1, schimeczek+wunner14-2, hardyetal23-1, hardyetal23-2, manseretal23-1, redingetal23-1}.
\item The physics of accretion discs in isolated and binary systems \citep{kraft62-1, tapia77-1, horne+marsh86-1, marsh+horne88-1, gaensickeetal09-1, manseretal16-1, cauleyetal18-1, manseretal21-1, steeleetal21-1, cunninghametal22-1, inightetal23-1, okuyaetal23-1}. 
\item The processes of energy transfer in, and modelling of white dwarf atmospheres \citep{koester10-1, tremblayetal11-1, tremblayetal19-1, baueretal18-1, blouinetal18-1,cukanovaiteetal18-1, cunninghametal19-1, bedardetal20-1}. 
\end{itemize}

Spectroscopic follow-up of the \textit{Gaia} white dwarf candidate sample will be obtained as part of four of the next generation multi-object spectroscopic surveys.  While the William Herschel Telescope Enhanced Area Velocity Explorer (WEAVE, \citealt{daltonetal12-1, daltonetal16-1} and the 4-metre Multi-Object Spectroscopic Telescope (4MOST, \citealt{dejongetal16-1}) are expected to come online in the near future, SDSS-V \citep{kollmeieretal17-1,sdssv23-1} and the Dark Energy Spectroscopic Instrument (DESI, \citealt{DESI16-1, DESI16-2,desi23-1}) survey are already collecting white dwarf spectra. The DESI survey obtained commissioning and survey validation observations from 2020 December 14 to 2021 June 10 \citep{desi24-1}, which have been released as the DESI Early Data Release (EDR, \citealt{desi23-1}). The data observed in this release include more than 4000 white dwarf candidates.

In this paper we present the DESI EDR sample of white dwarfs and associated systems. We describe the observations used in this work in Section\,\ref{sec:obs}. We then describe how we constructed the DESI EDR white dwarf sample in Section\,\ref{sec:WD_cat}, and broadly describe the subsystems included along with a comparison to the Sloan Digital Sky Survey (SDSS) in Section\,\ref{WD_HRD_discussion}. In Section\,\ref{sec:pop_studs} we give a more detailed discussion and analysis of white dwarf sub-types of interest, and conclude with our findings in Section\,\ref{sec:conc}.

\section{Observations}\label{sec:obs}

\subsection{DESI}

DESI on the Mayall 4-m telescope at Kitt Peak National Observatory (KPNO) is a multi-object spectroscopic instrument capable of collecting fibre spectroscopy on up to $\simeq5000$ targets per pointing \citep{desi22-1}. The fibres are positioned by robot actuators and are grouped into ten petals which feed ten identical three-arm spectrographs, each spanning 3600\,--\,9824\,\AA\ at a full width at half maximum resolution of $\simeq1.8$\,\AA. The inter-exposure sequence which includes telescope slewing, spectrograph readout and focal plane reconfiguration can be completed in as little as $\simeq2$\,min \citep{desi22-1}. The acquired data is then wavelength- and flux-calibrated with the DESI processing and reduction pipeline \citep[for a full description see][]{guyetal23-1}. 

The DESI survey started main-survey operations on 2021 May 14 and will obtain spectroscopy of more than 40\,million galaxies and quasars over five years to explore the nature of dark matter. During sub-optimal observing conditions (e.g., poor seeing or high lunar illumination), observations switch focus to nearby bright galaxies \citep{ruiz-maciasetal20-1,hahnetal22-1} and stars \citep{allendeprietoetal20-1, cooperetal23-1}. As part of this `BRIGHT' program, the DESI survey is targeting white dwarfs.

\begin{figure}
	\includegraphics[width=1\columnwidth]{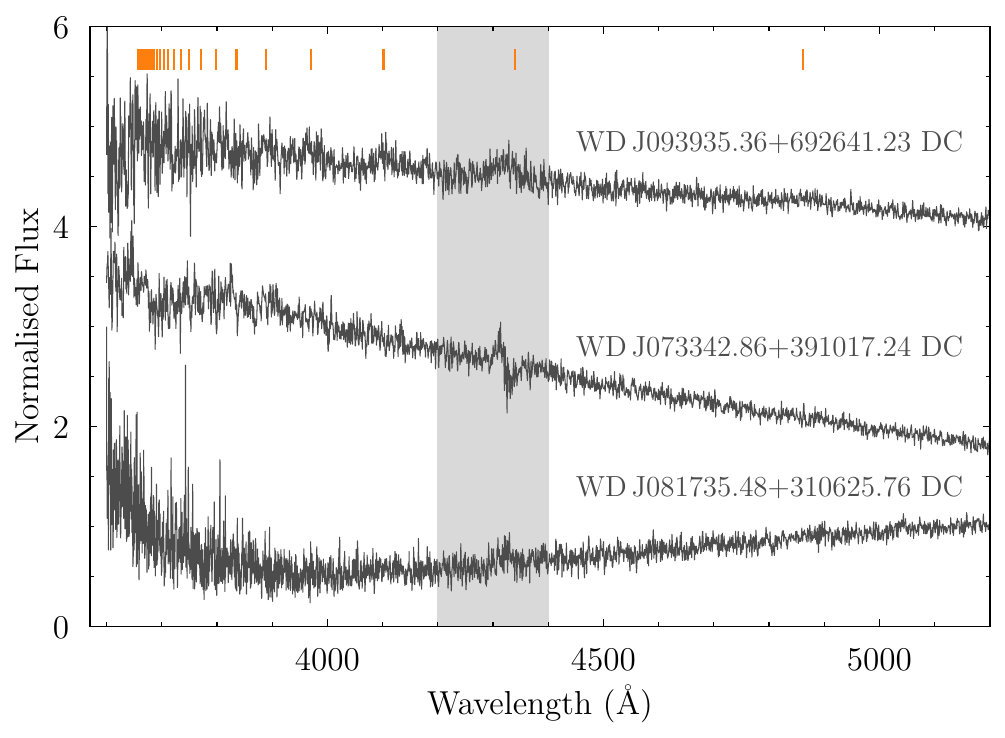}
    \caption{Three featureless `DC' (see Table\,\ref{t-wdclass}) white dwarfs identified in the DESI EDR sample showcasing some of the spectral issues affecting a small number of systems. Orange tabs show the Balmer series up to the Balmer jump, and the vertical gray region covering the wavelength range $4200 < \lambda < 4400$\,\AA\ highlights the region where the spectrograph collimator coating has reduced reflectivity and calibration residuals.}
    \label{fig:calibration}
\end{figure}

\subsubsection{DESI targeting of white dwarfs}

The target selection of white dwarfs for the DESI survey is described in \cite{cooperetal23-1} \S\,4.4.1. For the main survey, these selection criteria use \textit{Gaia} DR2 photometry and \textit{Gaia} EDR3 astrometry, and are based on equations $1 - 7$ defined in GF19. DESI EDR observations also follow these same criteria but use \textit{Gaia} DR2 astrometry. Applying the white dwarf target selection within the DESI footprint results in 66\,811 white dwarf candidates, where over 99\,per\,cent of these targets are expected to be allocated a fibre during the main survey (see fig.\,1 and table 2 of \citealt{cooperetal23-1}). However, as fibre assignment is done on the fly immediately before an observation is made, the details of what individual white dwarfs are observed by DESI is not predictable in advance \citep{schlaflyetal23-1}.

Targets selected as white dwarf candidates can be easily identified by their target bitmask (as detailed in \citealt{myersetal23-1} and \citealt{desi23-1}), and \bitnum\ of them were assigned fibres as part of the DESI EDR sample. We additionally cross-matched the entire DESI EDR sample with the GF19 \textit{Gaia} DR2 white dwarf catalogue using the unique \textit{Gaia} DR2 source ID to identify additional white dwarf candidates that may have been observed by other programs in the main survey. This resulted in an additional \totminusbit\ targets, for a total number of \totalnum\ DESI EDR white dwarf candidates.

The majority of white dwarfs targeted in the DESI EDR were observed more than once, and for the construction of the white dwarf catalogue we produced uncertainty-weighted co-adds using all the individual exposures for each white dwarf candidate and combined all spectra obtained in each of the three spectral arms. Across the \totalnum\ white dwarf candidates targeted, a total of \numexposures\ single-object exposures were obtained. However, during commissioning and the early parts of survey validation some observed spectra were of poor quality, leading to \numnoexp\ exposures with a median signal-to-noise ratio $<0.5$ in all three spectral arms. As a result, 391 white dwarf candidates observed during DESI EDR had unusable spectroscopy, resulting in a set of \specnum\ white dwarf candidates with usable spectra.

\subsubsection{Known issues with DESI EDR spectral calibration}

There are two known instrumental and processing issues that affect a small subset of DESI spectroscopy (Fig.\,\ref{fig:calibration}):

\textbf{Sensitivity at $\simeq4300$\,\AA:} The reflectivity of the collimators in the blue arm of the DESI spectrographs is reduced in the wavelength range $\simeq4200-4400$\,\AA, resulting in a drop in sensitivity in that region (see fig.\,24 of \citealt{guyetal23-1}). The exact shape of the reflectivity profile as a function of wavelength varies with the location in the mirror and with the incidence angle of the light. This sensitivity issue affects both the flux calibration and the sky subtraction and can result in the creation of non-physical continuum shapes in this region, such as in the DESI spectrum of WD\,J073342.86+391017.24 (Fig.\,\ref{fig:calibration}).

\textbf{Flux calibration:} The flux calibration of DESI is performed using F-stars with a preference for low-metallicity halo stars. \cite{guyetal23-1} used hydrogen-atmosphere DA white dwarfs as an independent test of the flux calibration. For the majority of the DESI spectral range the average residuals are relatively flat at a level of $\simeq2$\,per\,cent (see fig.\,41 of \citealt{guyetal23-1}). However, around the Balmer and Paschen series emission and absorption features are present in the average residuals, which are suggestive of an imperfect fit to the spectral features of the standard stars. This is most notable at wavelengths $\simeq3750$\,\AA, where a $\simeq6$\,per\,cent jump coincides with the Balmer jump. For some very cool white dwarfs (i.e., the featureless DC white dwarfs) this increase in flux can appear quite dramatic and starts affecting the spectrum at redder wavelengths. While some features in the average residual from fitting DAs are likely due to imperfections in the white dwarf models, such as the very broad ($\sim$\,100\,\AA) Paschen absorption features, many of the other features are reproduced in the continuum fitting of moderate redshift quasars observed by DESI, highlighting these as calibration issues (see fig.\,5 of \citealt{perezetal23-1}).

\begin{table}
\caption{Description of the identifiers for the varied spectral features detected in DESI EDR white dwarf spectra. A white dwarf spectral class is given the degenerate ``D" prefix, with any combination of features that are present, for example ``DA", ``DABZ", or ``DBH". The top six identifiers relate to the chemical composition of the atmosphere, the next four are used to label additional characteristics of the spectra. Systems we could identify as white dwarfs but could not further classify were given the classification `WD'. \label{t-wdclass}}
\begin{center}
\begin{tabular}{ll}
\hline
Identifier & Definition                                     \\
\hline
A          & H\,\textsc{i} lines present                    \\
B          & He\,\textsc{i} lines present                   \\
C          & Featureless spectrum                           \\
O          & He\,\textsc{ii} lines present                  \\
Q          & C features present in atomic or molecular form \\
Z          & Metal lines present\,$^{a}$                    \\\hline
P          & Peculiar or unidentified features              \\
H          & Zeeman splitting present                       \\
e          & Emission lines present                         \\
:          & Tentative or uncertain classification          \\
\hline
\multicolumn{2}{|l|}{$^{\textrm{a}}$ Some of these detections may include interstellar medium absorption} \\
\multicolumn{2}{|l|}{~~features, which is more common for hotter white dwarfs at larger}\\
\multicolumn{2}{|l|}{~~distances.}
\end{tabular}
\end{center}
\end{table}

\subsection{SDSS}

The SDSS has been taking multi-band photometry and multi-fibre spectroscopy since 2000, using a 2.5-m telescope located at the Apache Point Observatory in New Mexico \citep{gunnetal06-1}. We retrieved archival SDSS spectroscopy (DR17, \citealt{abdurroufetal22-1}) and crossmatched these with the DESI EDR sample of white dwarf candidates, which resulted in 1427 unique sources common to both surveys and a total of 2149 SDSS spectra. We use these SDSS spectra to validate our fitting routines (Section\,\ref{sec:pop_studs}) using different spectroscopic sources for the same target and comparing with previous literature results. 

\begin{table*}
\centering
\caption{Classifications\,$^{\textrm{a}}$ of the white dwarf candidates visually inspected from the DESI EDR.}
\label{t-cat}
\begin{tabular}{lrlrlrlrlrlr}
\hline 
Class                 & Number & Class  & Number & Class   & Number & Class & Number & Class             & Number  & Class                 & Number \\
\hline
\textbf{White dwarfs} & 2706   & DAH:   & 12     & DBAZ   & 4      & DAe:   & 5      & DQZ	             & 1       & \textbf{Contaminants} & 817    \\ 
DA	                  & 1958   & DQ	    & 50     & DBAZ:  & 3      & DAHe   & 2      & DZAP	             & 1       & EXGAL	               & 445    \\
DA:	                  & 18     & DQ:    & 5      & DH:	  & 6      & DAHe:  & 2      & DZBA	             & 1       & STAR	               & 329    \\
DC	                  & 197    & DZ	    & 49     & DZA	  & 6      & DAZe:  & 1      & DZBA:	         & 2       & sdX	               & 42     \\
DC:	                  & 23     & DBZ	& 19     & DZAB   & 4      & DAOH:  & 1      & DZQ               & 1       & sdX:	               & 1      \\
DB	                  & 141    & DBA	& 15     & DZB	  & 4      & DAQ:   & 1      & WD	             & 20      & \textbf{UNCLASS}	   & 84     \\
DB:	                  & 6      & DBA:	& 1      & DCH:   & 3      & DAZH:  & 1      & \textbf{Binaries} & 66      &                       &        \\
DAZ	                  & 54     & DAO	& 10     & DAP	  & 2      & DO	    & 1      & WD+MS	         & 53      &                       &        \\
DAZ:	              & 11     & DAO:	& 1      & DAB	  & 1      & DQA    & 1      & CV	             & 12      &                       &        \\
DAH	                  & 53     & DBZA	& 7      & DABZ   & 1      & DQH	& 1      & DA+DQ	         & 1       &                       &        \\                   
\hline
\multicolumn{12}{|l|}{$^{\textrm{a}}$ In the text we present classifications such as D(AB), where bracketed sub-classifications collate any permutation of the spectral classifications, } \\
\multicolumn{12}{|l|}{~~e.g., D(AB) includes both DAB and DBA white dwarfs.} \\
\end{tabular}
\end{table*}

\subsection{Photometry}

\subsubsection{Multi-colour photometry}

We retrieved photometric data from four catalogues, namely \textit{Gaia}~DR2, \textit{Gaia}~EDR3, SDSS~DR18 and Pan-STARRS~DR2 \citep{Chambers2019}. Zero-point corrections were applied to the SDSS \textit{u}, \textit{i} and \textit{z} bands as suggested by \cite{eisensteinetal06-1} to convert to actual AB magnitudes. \textit{Gaia}~DR2 photometry and astrometry was used for the target selection in the DESI EDR and for our cross-match  to identify potential white dwarf candidates observed through other target selections (see Section\,\ref{sec:WD_cat}). \textit{Gaia} EDR3 (or DR2 if the former was not available) photometry was employed to compare our internal photometric analysis with an independent test. 

\subsubsection{Zwicky Transient Facility (ZTF)}

ZTF is a robotic time-domain survey using the Palomar 48-inch Schmidt Telescope \citep{bellmetal19-1, mascietal19-1}. Utilising a 47\,$\mathrm{deg}^2$ field of view, ZTF can scan the entire sky in $\simeq$\,two days, making it a powerful survey for identifying photometrically variable sources at optical wavelengths. We obtained archival ZTF data from DR18 \citep{mascietal19-1}, which includes data up to 2023 May 7. We inspected the ZTF lightcurves of Cataclysmic Variables identified in the DESI EDR sample to assist in their classifications (Section\,\ref{sec:CV}).

\begin{figure}
	\includegraphics[width=\columnwidth]{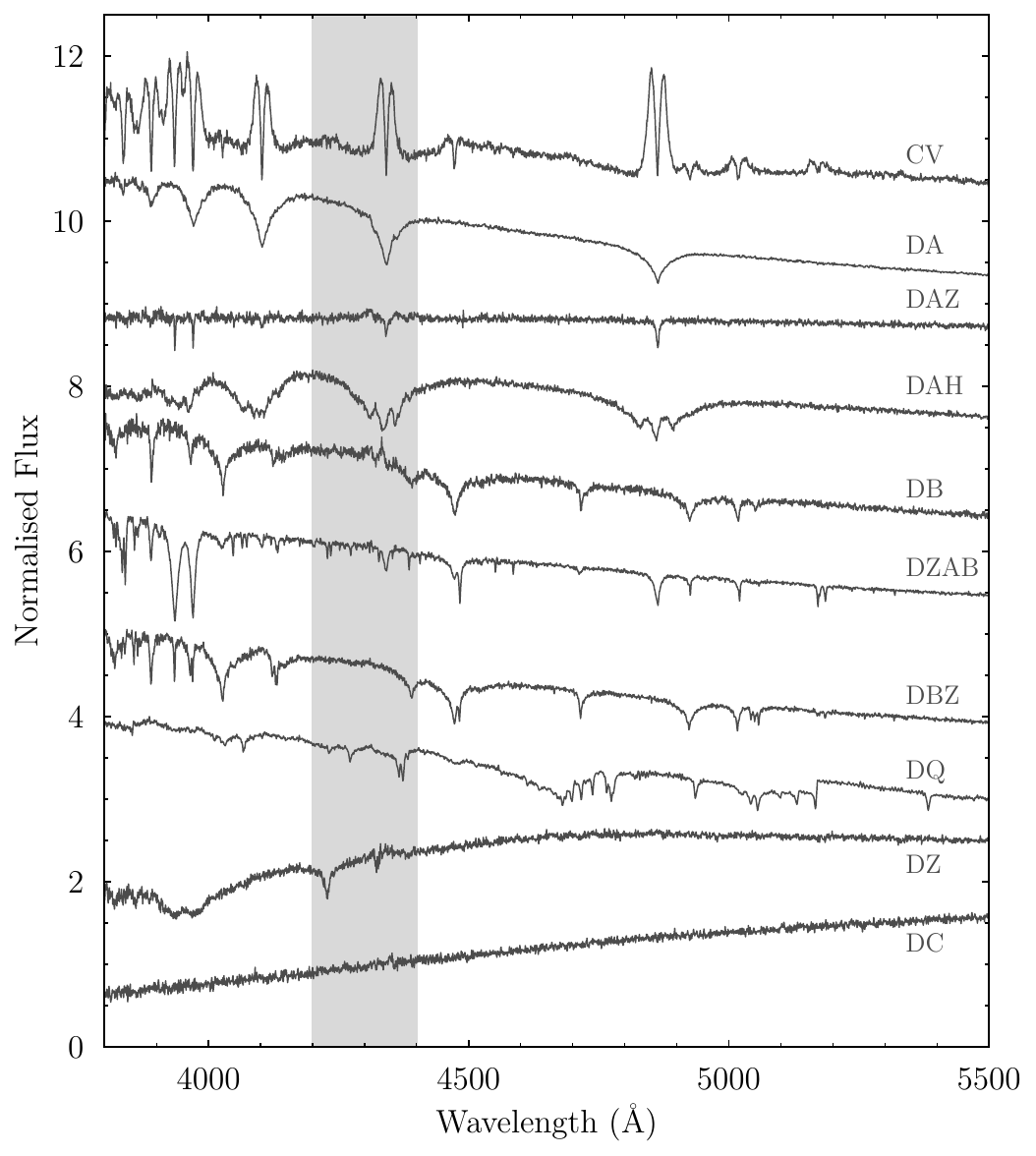}
    \caption{A subset of the white dwarf systems spectra within the DESI EDR showcasing some of the spectral types. The wavelength range $4200 < \lambda < 4400$\,\AA\ is subject to relatively poor calibration due to an issue with the spectrograph collimator coating. Adapted from \protect\cite{cooperetal23-1}.}
    \label{fig:wdspectra}
\end{figure}

\section{White dwarf catalogue}\label{sec:WD_cat}



\begin{figure*}
	\includegraphics[width=2\columnwidth]{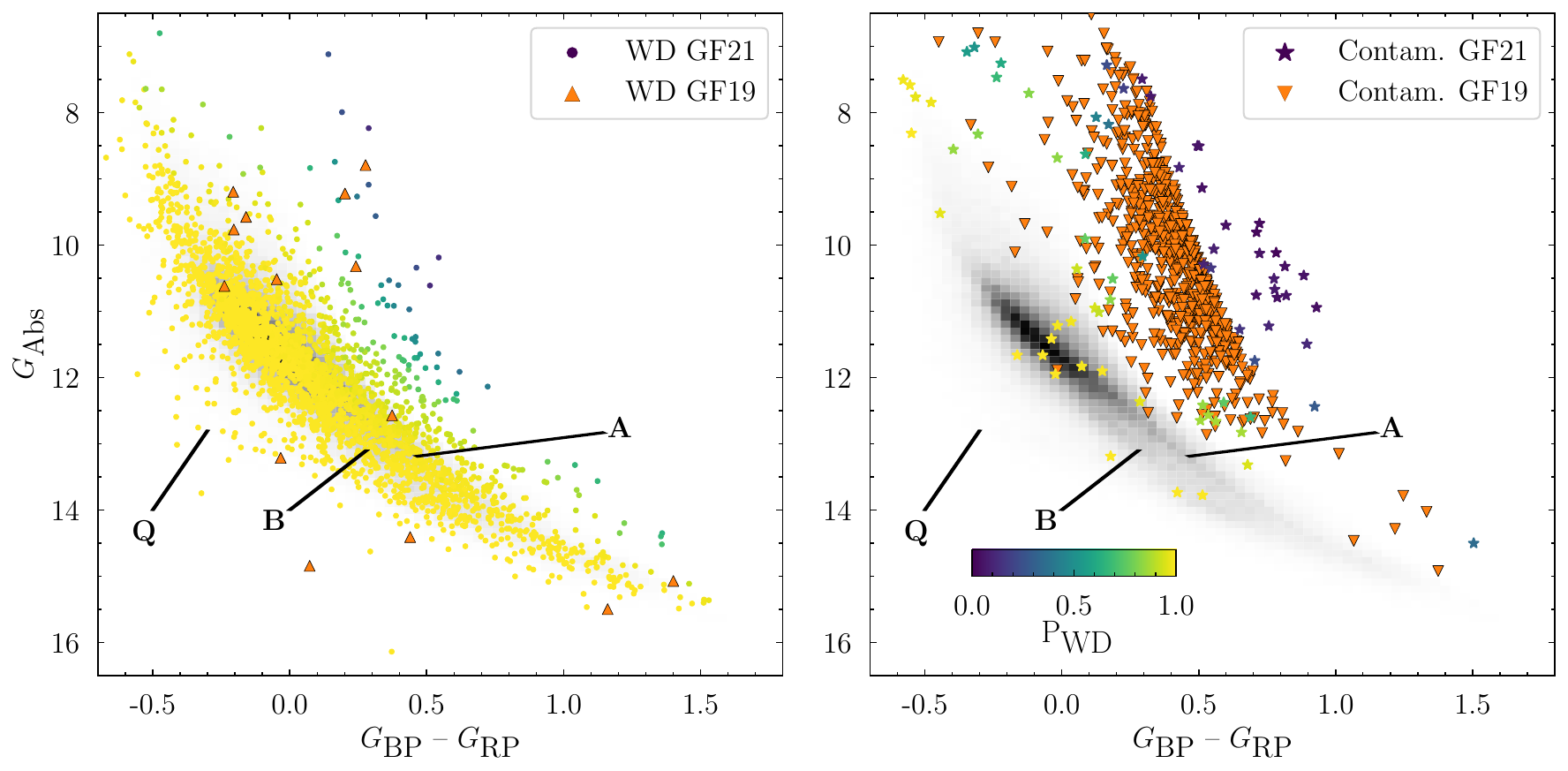}
    \caption{The \textit{Gaia} Hertzsprung-Russell Diagram (HRD) showing spectroscopically confirmed white dwarf systems (left panel) and contaminants (right panel), observed by DESI. For reference, the high-confidence ($P_{\textrm{WD}} > 0.95$) sample of \textit{Gaia} EDR3 white dwarfs brighter than $G=20$ are shown as a 2D gray-scale histogram in both panels \citep[][GF21]{fusilloetal21-1}, although it is completely obscured by the DESI EDR white dwarf sample in the left panel. DESI-identified white dwarf systems and contaminants with entries in the GF21 \textit{Gaia} EDR3 white dwarf catalogue are denoted as circles and stars in the left and right panels, respectively. 14 white dwarf systems and 714 contaminants do not have entries in the GF21 \textit{Gaia} EDR3 catalogue and only appear in the GF19 \textit{Gaia} DR2 white dwarf catalogue and we plot them as up-facing (left panel) and down-facing (right panel) orange triangles with a black edge colour. White dwarf systems and contaminants are coloured based on their probability of being a white dwarf, $P_{\textrm{WD}}$, from GF21. The A, B, and Q branches defined in \protect\cite{gaia18-hrd} are indicated by black lines (see text). The sharp upper-edge in the distribution of the contaminants is due to the selection criteria defined by GF19 (see their fig.\,1).}
    \label{fig:HRD_WDvsContam}
\end{figure*}

\subsection{Spectral Classification}

The spectral classification system for white dwarfs has been defined by \citet{sionetal83-1}, and is purely based on the morphology of the spectroscopic data. We summarise in Table\,\ref{t-wdclass} the identifiers associated with specific features in the spectrum of a white dwarf, which relate both to the chemical composition of the atmosphere, as well as to other characteristic features that may be present in the spectra. The majority of white dwarfs in any magnitude or volume-limited sample are ``DA'' white dwarfs, i.e. their spectra only show Balmer absorption lines. Spectral identifiers can be combined if multiple features are identified, resulting in composite spectral types such as ``DBAZ''. The original definition of \citet{sionetal83-1} defined that ``weaker or secondary spectroscopic features'' are used for additional spectral identifiers, leaving a fair amount of ambiguity with respect to deciding the sequencing of the spectral type. We adopted an overall weighted visual assessment of the different spectral features, rather than focusing on the strength of specific individual lines. It is important to note that the \textit{equivalent width} of features of a specific element are not directly linked to its \textit{abundance} within the atmosphere, hence care has to be taken not to mistake spectral types as a reflection of the dominant species within the atmosphere of a given white dwarf.

The \specnum\ DESI EDR spectra of white dwarf candidates were visually classified by three of the authors, who carefully reviewed all systems where their initial classifications were discrepant, followed by a final inspection of all spectra within any individual spectral class. In total, we spectroscopically confirm \numwd\ white dwarfs. We also provide classifications for \numbinary\ binary white dwarf systems, including white dwarf-main sequence binaries (WD$+$MS), cataclysmic variables (CVs), and double degenerate binaries such as the DA$+$DQ system WD\,J092053.05+692645.36 (hereafter WD\,J0920+6926) are discussed in Section~\ref{sec:hrd-exotic}. We also identified \numcontaminant\ contaminant extragalactic sources, main-sequence stars and subdwarfs, with an additional \numunclass\ spectra that could not be confidently classified. These classifications are listed in Table\,\ref{t-cat}, and some example DESI spectra of white dwarf systems are shown in Fig.\,\ref{fig:wdspectra}. 

\begin{figure*}
	\includegraphics[width=2\columnwidth]{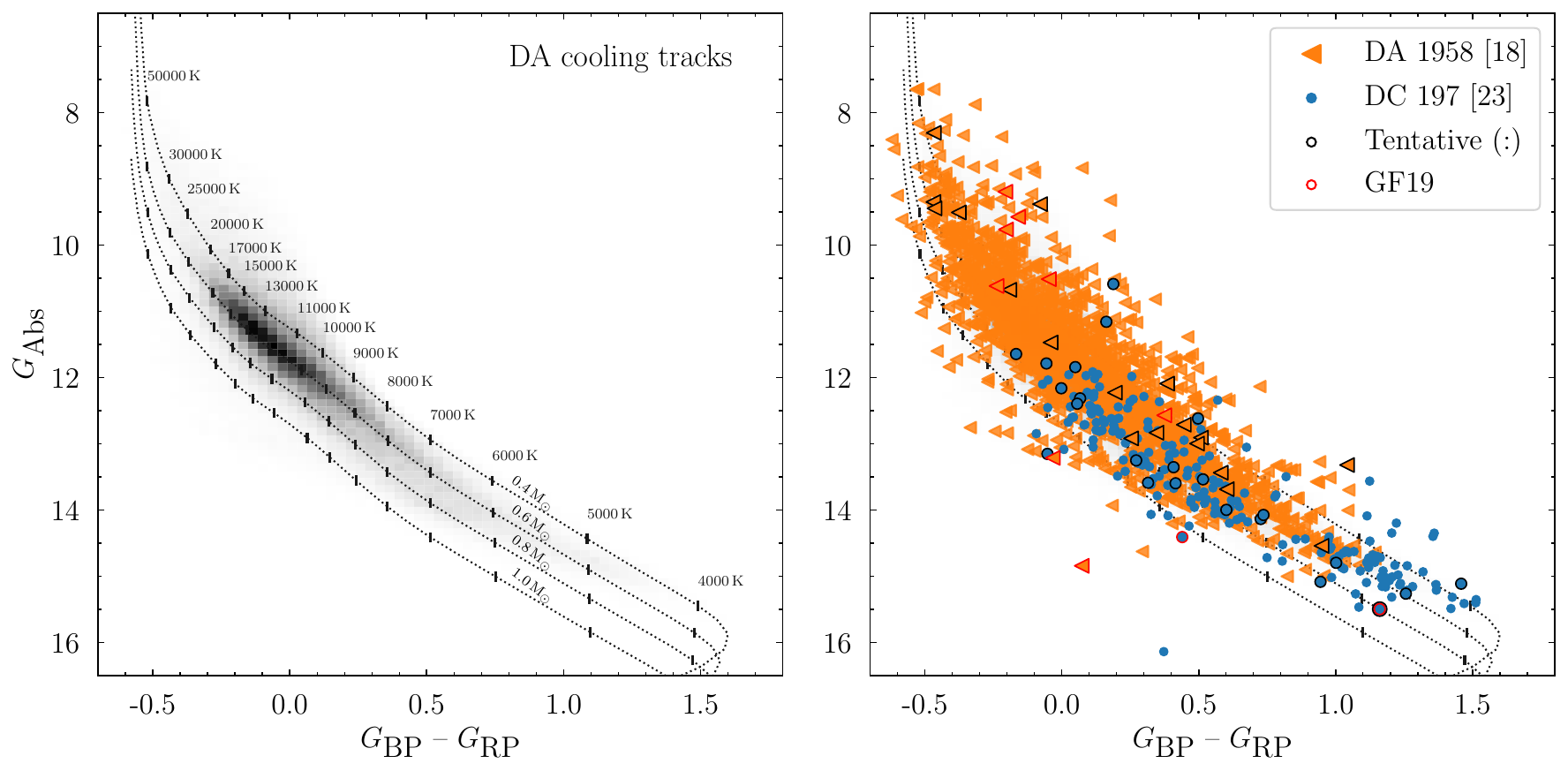}
    \includegraphics[width=2\columnwidth]{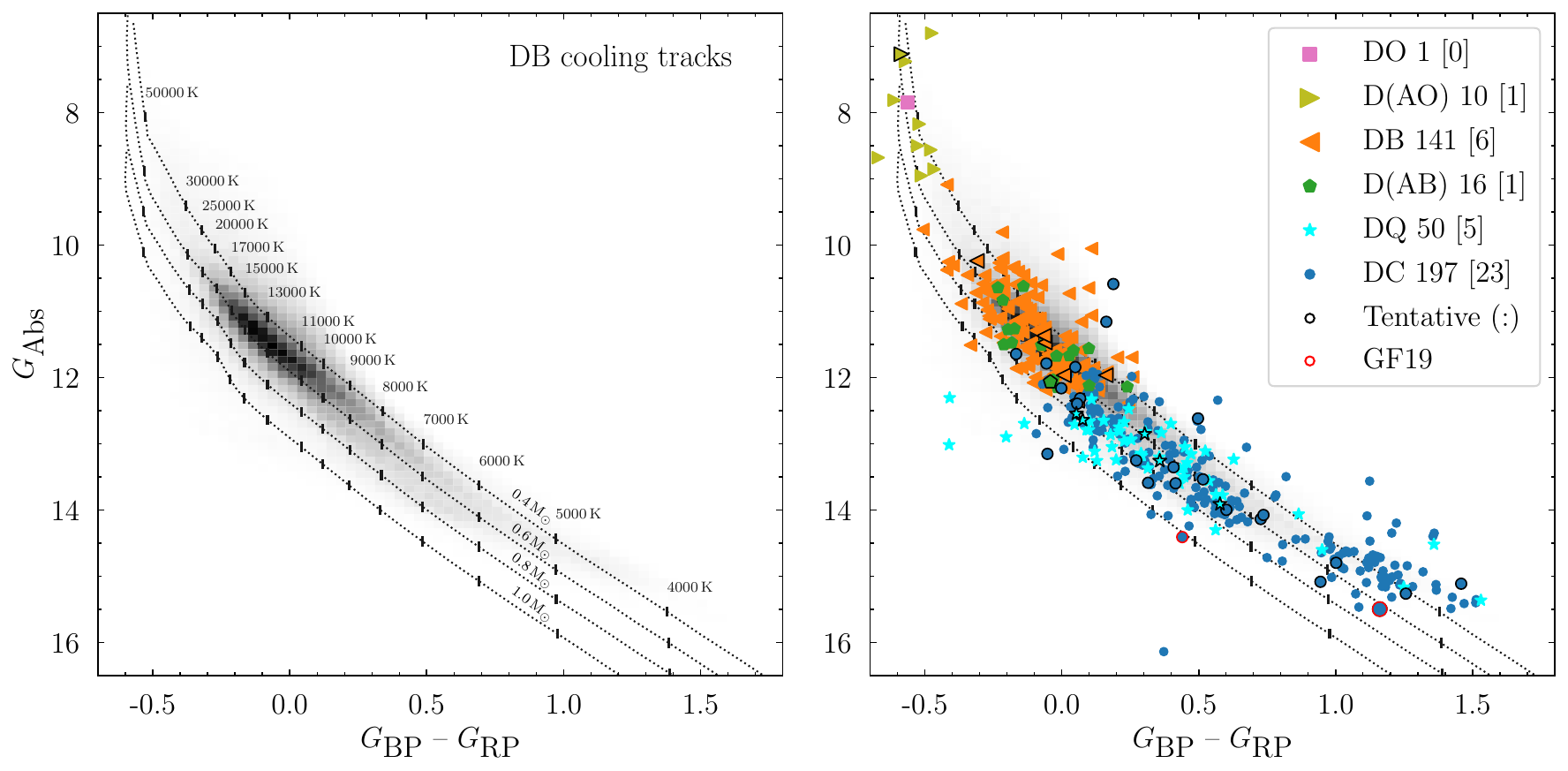}
    \caption{High-confidence ($P_{\textrm{WD}} > 0.95$) \textit{Gaia} white dwarfs brighter than $G=20$ identified by GF21 are shown as a 2D gray-scale histogram in all panels. Left panels: cooling tracks from \protect\cite{bedardetal20-1} for pure DA (top) and DB (bottom) white dwarfs are plotted as dashed lines for masses of 0.4\,\Msun, 0.6\,\Msun, 0.8\,\Msun\ and 1.0\,\Msun\, with vertical tabs highlighting $T_{\textrm{eff}}$ values. Right panels: the same as the left panels but with white dwarfs commonly associated with H- and He-dominated atmospheres identified by DESI (top and bottom, respectively). Bracketed classifications in the legend collate any combination of the spectral classifications, e.g., D(AB) includes both DAB and DBA white dwarfs. The number of confidently classified white dwarfs associated with each classification are given, with the number of tentative (:) classifications provided additionally for each white dwarf type in square-brackets. In both the top right and bottom right panels tentatively identified systems are highlighted by a black border. Confirmed white dwarfs not present in GF21 but are present in GF19 are highlighted by a red border. The spectrum of the very faint ($G_\mathrm{Abs}\simeq16$) but blue ($G_\mathrm{Bp}-G_\mathrm{Rp}\simeq0.38$) DC white dwarf, WD\,J0951+6454, is shown in Fig.\,\ref{fig:wdj0951}.}
    \label{fig:HRD_WDsystems1}
\end{figure*}

The DESI EDR white dwarf catalogue is provided as a FITS file\footnote{This FITS file can be obtained at \url{https://zenodo.org/records/10620344}.}, with the details of the extensions included and the content of the individual columns described in detail in Appendix\,\ref{sec:catalogue}. The first extension contains the full list of \totalnum\ white dwarf candidates targeted by DESI with our spectral classifications (where applicable) and auxiliary data. The second extension includes the results of our cross-match between the DESI EDR sample and SDSS with \numsdss\ entries giving the unique SDSS plate, modified Julian date (MJD), fibre identifiers for each SDSS spectrum, the separation between a given white dwarf candidate and the location associated with the SDSS spectrum, and the number of spectra associated with each object. Subsequent extensions include our best fit parameters to the DA, DB, DBA, and DQ systems.

\subsection{New identifications vs. known systems}

To determine the number of new white dwarfs the DESI EDR sample has identified, we cross-matched our sample with the Montreal White Dwarf Database (MWDD, \citealt{dufouretal17-1}) and the SIMBAD astronomical database \citep{wengeretal00-1}. The results of this cross-match are included in our DESI EDR catalogue (Table\,\ref{t-WD_EDR_catalogue}), and in the discussion of spectral types and sub-classes below we exclude systems within the DESI EDR with uncertain (:) classifications. For the largest sample of systems, i.e., DA, DB, DC white dwarfs, we use the results of our crossmatch to determine which white dwarfs have been identified as such for the first time. 1201 DAs (61\,per\,cent), 151 DCs (77\,per\,cent), 68 DBs (48\,per\,cent), 5 DAB/DBAs (31\,per\,cent), and 3 DAO/DOs (27\,per\,cent) have been spectroscopically confirmed as a white dwarf for the first time with DESI, adding over $\simeq$1400 new systems. This is somewhat expected from the cooler distribution of DESI white dwarfs compared to the SDSS sample (Fig.\,\ref{fig:HRD_compare}), where DCs and DAs are the dominant spectral types. As SDSS preferentially samples hotter white dwarfs, the DB, DAB/DBA and DAO/DO subclasses are better examined and fewer new systems have been identified by DESI.

We do a more in-depth literature search for four sub-groups of interest to us; DQ white dwarfs (Section\,\ref{sec:DQ}, Table\,\ref{t-DQx}), metal-enriched systems (Section\,\ref{sec:DxZ}, Table\,\ref{t-DxZ1}), magnetic white dwarfs (Section\,\ref{sec:DxH}, Table\,\ref{t-DxH}), and CVs (Section\,\ref{sec:CV} , Table\,\ref{t-CVs}). In these searches, we identify whether DESI has newly discovered (or confirmed in the case of candidates in the literature) these white dwarfs as members of their given sub-classes, and discuss them in more depth in the given sections.

\section{The DESI EDR white dwarf sample in the context of their location on the \textit{Gaia} HRD}\label{WD_HRD_discussion}

We illustrate the distribution of the DESI EDR white dwarf sample within the \textit{Gaia} Hertzsprung-Russell Diagram (HRD) in Fig.\,\ref{fig:HRD_WDvsContam} separated into spectroscopically confirmed white dwarf systems (including binaries, left) and contaminants (right), with the 84 unclassified systems (`UNCLASS') being omitted (see Table\,\ref{t-cat}). The DESI EDR white dwarf sample closely follows the distribution of the magnitude limited sample of high confidence (probability of being a white dwarf  $P_{\textrm{WD}} > 0.95$) white dwarfs identified in the \textit{Gaia} EDR3 catalogue of GF21. The DESI white dwarf sample clearly reproduces the A and B branches, populated by standard-mass white dwarfs with H- and He-dominated photospheres, respectively \citep{gaia18-hrd}. The Q-branch is also present in the sample \citep{gaia18-hrd}, and is thought to be due to delayed cooling by crystallisation of high-mass white dwarfs \citep{tremblayetal19-1}, settling of $^{22}$Ne \citep{chengetal19-1}, or white dwarf-subgiant mergers \citep{shenetal23-1}. The majority of white dwarf systems observed by DESI have high $P_{\textrm{WD}}$ values in both GF19 and GF21. The $P_{\textrm{WD}}$ variable was estimated based on the probability of a candidate being an \textit{isolated} white dwarf, and so the binary sample that dominates the systems that sit above the white dwarf cooling track have a lower $P_{\textrm{WD}}$. The contaminants are similarly located above the white dwarf cooling track, with many of them having very small $P_{\textrm{WD}}$ values. Additionally between the construction of the GF19 and GF21 catalogues, 714 (87.3\,per\,cent) of the contaminant population have been excluded from the more reliable \textit{Gaia} EDR3-based selection criteria of GF21 compared with that of DR2 used by GF19, whereas only 14 (0.5\,per\,cent) of the confirmed white dwarf systems have dropped out of the GF21 white dwarf catalogue (Fig.\,\ref{fig:sysindr2}). Overall, this showcases both the high fidelity of the selection methods and the robustness of the $P_{\textrm{WD}}$ value calculated by GF19 and GF21.

In the following subsections we will describe the DESI EDR white dwarf sample within the \textit{Gaia} HRD, split up into the various spectral classes (Table\,\ref{t-cat}). The systems are loosely grouped to illustrate the evolution of hydrogen-dominated atmospheres (DA $\shortrightarrow$ DC) and helium-dominated atmospheres (DO $\shortrightarrow$ DB $\shortrightarrow$ DC/DQ), as well as classes that share similar physical characteristics (e.g., metal enrichement, the presence of magnetic fields or binary companions). These groupings are not fully representative however, as some white dwarf classes may be populated by multiple evolutionary channels, such as DC white dwarfs evolving from both H- and He-dominated atmosphere white dwarfs \citep{bergeron01-1,kowalski+saumon06-1,kilicetal09-1,caronetal23-1}, as well as different physical processes that can alter the spectral type of a white dwarf, such as convective mixing or accretion from various sources \citep{fontaineetal01-1,fusilloetal17-1,rollandetal18-1,cunninghametal20-1,rollandetal20-1}. Cooling tracks\footnote{\url{http://www.astro.umontreal.ca/~bergeron/CoolingModels}} for DA- and DB-type white dwarfs are presented to help give context to the samples, but are not applicable to all objects.

\subsection{Typical white dwarfs}\label{sec:the_usual_suspects}

\begin{figure}
    \includegraphics[width=1\columnwidth]{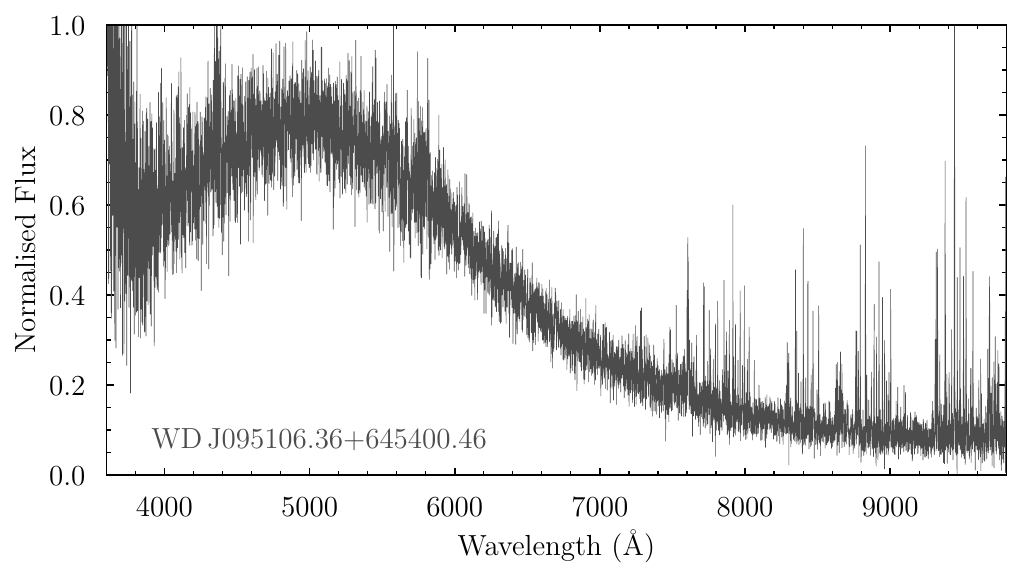}
    \caption{DESI EDR spectrum of the ultra-blue white dwarf WD\,J0951+6454. The steep increase in flux below $\simeq$\,3800\,\AA\ is likely due to the calibration issues shown in Fig.\,\ref{fig:calibration}. Additionally there are residual sky features in the red part of the spectrum past $\simeq$\,7500\,\AA. The suppression of flux in the red part of the spectrum is due to collisionally induced absorption \citep{blouinetal17-1}}
    \label{fig:wdj0951}
\end{figure}

\begin{figure*}
    \includegraphics[width=2\columnwidth]{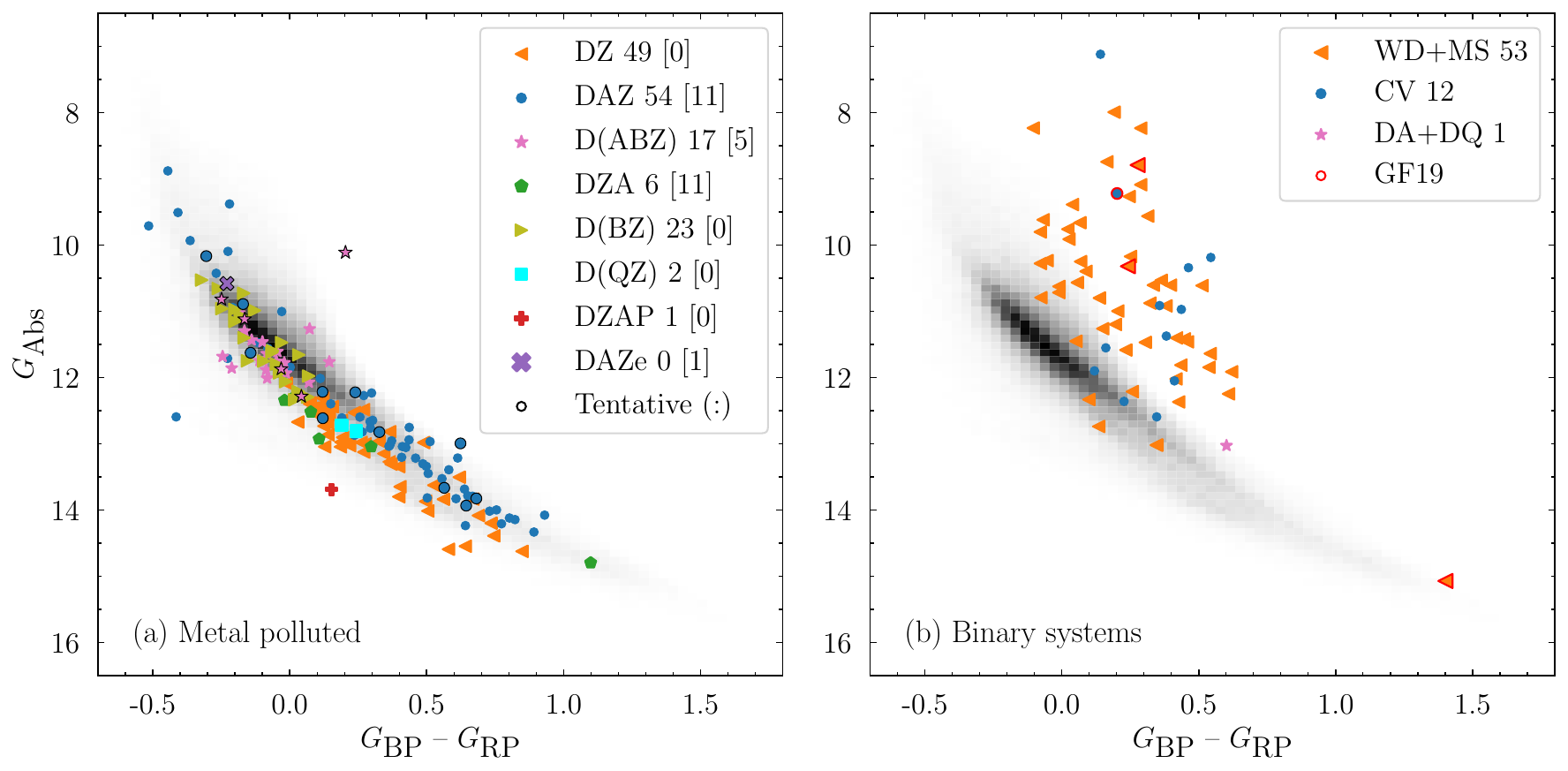}
    \includegraphics[width=2\columnwidth]{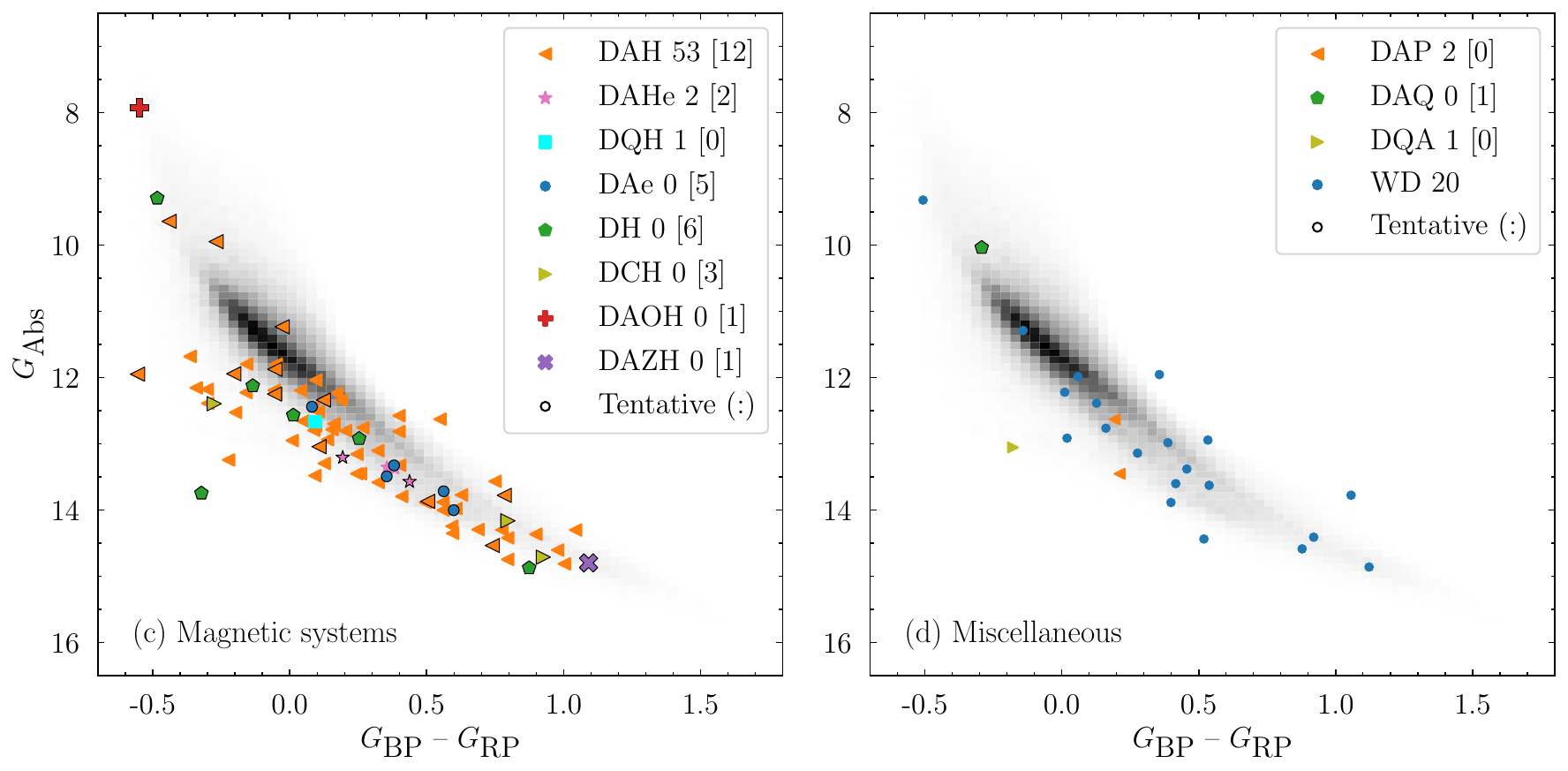}
    \caption{Same as Fig.\,\ref{fig:HRD_WDsystems1} showing additional types of white dwarf systems identified by DESI with no cooling models presented. The number of confidently classified white dwarfs associated with each classification are given, with tentative (:) classifications given additionally in square-brackets. (a) White dwarfs with metals present in the atmosphere commonly associated with the accretion of planetary material. (b) White dwarf binary systems. (c) White dwarfs showing evidence for hosting a magnetic field, including DAe white dwarfs, where the presence of emission is thought to be associated with magnetic fields \citep{elmsetal23-1}. (d) Miscellaneous white dwarfs in addition to systems which were identified as a white dwarf but no further classification could be given.}
    \label{fig:HRD_WDsystems2}
\end{figure*}

Figure\,\ref{fig:HRD_WDsystems1} shows the majority of the white dwarf sample, which is dominated by the DA, DC, and DB spectral types, adding up to a total of 86.3\,per\,cent of the DESI EDR confirmed white dwarfs. The DAs make up 72.8\,per\,cent of the sample, and cover the majority of the cooling sequence. A small subset of high-mass DAs lie on the Q-branch described above. DA white dwarfs cool until they reach $\simeq5000$\,K where the Balmer lines become undetectable and these systems transition to featureless DCs.

While He-dominated atmosphere white dwarfs cool along a similar path in the \textit{Gaia} HRD, their spectral evolution through this cooling sequence is more complex. The hottest pure He-atmosphere white dwarfs in our sample are the DO systems that show He\,\textsc{ii} lines in their spectra and are thought to form after a late shell flash in which the white dwarf progenitor burns all the remaining H in the envelope \citep{herwigetal99-1,althausetal05-1,werner+herwig06-1} or through the convective dilution or convective mixing processes, in which a thin H layer is diluted by the deeper convective He one \citep{fontaine+wesemael87-1,cunninghametal20-1}. As DO white dwarfs cool, they can either transition into a DB as He becomes neutral and produces He\,{\textsc{i}} features ($\sim$~1/3 of DOs follow this path) or into a DA, following the DO-to-DA transition due to the upward diffusion of leftover H \citep{wesemael+fontaine85-1,flemingetal86-1,liebert86-1}. As DBs cool further (first path), He\,\textsc{i} features disappear below $\simeq$\,10\,000\,K, and these white dwarfs then split into two categories; those with featureless DC spectra, and those showing the presence of C, i.e., DQ white dwarfs. The existence of C in the spectra of these white dwarfs is thought to be due to dredge-up, when the deepening He convection-zone reaches the C-rich core \citep{pelletieretal86-1}. Not all DQ white dwarfs appear to share this origin however, as a number sit on the Q-branch, suggesting they are both hotter and have a higher mass. These `warm' or `hot' DQs are thought to form instead from binary mergers \citep{dunlap+clemens15-1,kawkaetal23-1}. Many He-atmosphere white dwarfs also show spectroscopically detectable H in their atmospheres, such as the DAO/DOA and DAB/DBA white dwarfs. These systems by-and-large overlap with those that do not show the presence of H.

The DC white dwarfs appear to group in two regions, one group roughly between $0.00 < (G_{\textrm{BP}}-G_{\textrm{RP}}) < 0.75$, with the other appearing cooler at $(G_{\textrm{BP}}-G_{\textrm{RP}}) > 1.00$. The bluer sample of DCs is likely dominated by white dwarfs with He-rich atmospheres, and overlap with the cool DQ systems. Like their DQ counterparts, these bluer DCs likely also have trace C in their atmospheres\footnote{DC white dwarfs with evidence of trace C have been dubbed as `stealth-DQs' and `DQ-manqu\'{e}'}: while by definition they have featureless optical spectra, a large fraction of DCs experience some form of C dredge-up during their lifetime. This dredge-up is evidenced by non-negligible changes in their optical spectral energy distribution due to the addition of extra electrons from C \citep{camisassaetal23-1, blouinetal23-1}, in addition to a reduction of their flux in the ultraviolet from C absorption \citep{blouinetal23-2}. The redder sample of DCs is thought to be dominated by H-dominated white dwarfs \citep{caronetal23-1}, although these DCs appear to deviate quite significantly from the DA cooling sequence, which is centred roughly on the 0.6\,\Msun\ evolutionary track. This may suggest imperfections in the cooling sequences used at these cooler temperatures \citep{caronetal23-1}, but the origin of this deviation is unknown.

A single DC white dwarf, WD\,J095106.36+645400.46 (hereafter WD\,J0951+6454, Fig.\,\ref{fig:wdj0951}), lies significantly below the main white dwarf cooling sequence and is the least luminous white dwarf in the DESI EDR white dwarf sample  ($G_\mathrm{Abs}\simeq16$, $G_\mathrm{Bp}-G_\mathrm{Rp}\simeq0.38$). The system sits on the \textit{ultra-blue sequence} of infra-red faint targets identified by \cite{kilicetal20-1} and \cite{bergeronetal22-1}. These systems appear to be comprised of relatively cool ($\simeq$\,4000\,K) white dwarfs. The spectra of these systems are unusual due to the presence of collisionally-induced absorption generated from molecular H colliding with He dominating the opacity \citep{borysowetal01-1}.

\subsection{Exotic white dwarf systems}
\label{sec:hrd-exotic}
Figure\,\ref{fig:HRD_WDsystems2} showcases the rarer white dwarf systems and spectral types arising from interactions with external bodies or additional physical characteristics beyond the thermal evolution of white dwarfs depicted in Fig.\,\ref{fig:HRD_WDsystems1}. The top left panel of Fig.\,\ref{fig:HRD_WDsystems2} displays all systems where metals (Z) heavier than helium, and excluding carbon, are spotted. We identify a total of 152 metal-enriched white dwarfs, the vast majority of which are likely the result of the accretion of planetary material that survived the evolution of the post main sequence evolution planet-hosting main-sequence star \citep{zuckerman+becklin87-1,grahametal90-1,debes+sigurdsson02-1,jura03-1,zuckermanetal03-1}. White dwarfs redder than $(G_{\textrm{BP}}-G_{\textrm{RP}}) > 0.1$, corresponding to $T_{\textrm{eff}}$\,$\lesssim$\,12\,000\,K, dominate those with metal enrichment, which is more striking when additionally taking into account the magnitude-limited bias to hotter white dwarfs. The key reason for the apparent increase in the fraction of metal-enriched white dwarfs is the strength of the Ca H/K resonance lines, and the fact that \Ion{Ca}{ii} is gradually becoming more populated in cooler atmospheres~--~resulting in a spectroscopic detection of Ca even at low photospheric abundances \citep{dufouretal07-2,koesteretal11-1,hollandsetal17-1,hollandsetal18-1}. Observations of white dwarf samples at higher resolution and signal-to-noise ratios, and also in the far-ultraviolet (wavelengths below $\simeq$\,2000\,\AA) where the detection of additional elements can be made, show that $\simeq25-50$\,per\,cent of all white dwarfs are enriched with metals consistent with the accretion of planetary bodies \citep{zuckermanetal03-1, zuckermanetal07-1, barstowetal14-1, koesteretal14-1}.

\begin{figure}
    \includegraphics[width=1\columnwidth]{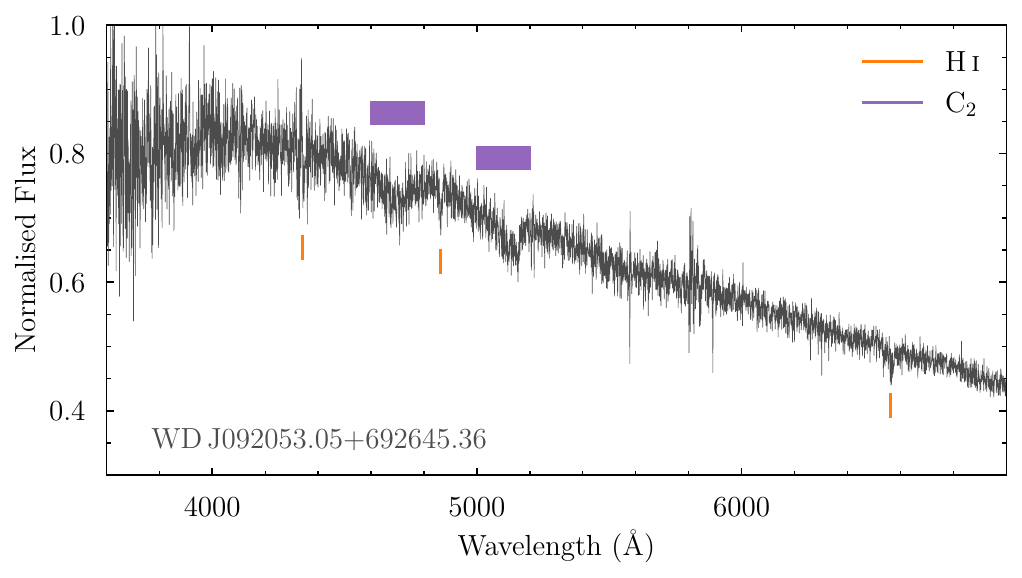}
    \caption{DESI EDR spectrum of the likely DA+DQ binary system WD\,J0920+6926. The system is located slightly above the white dwarf cooling sequence, suggestive of either being of low mass, or a binary, or both. DAQ/DQA white dwarfs are usually hotter and present atomic carbon features in their spectra. H Balmer lines and C$_2$ Swan bands are highlighted by colored tabs.}
    \label{fig:wdj0920}
\end{figure}

We also identify 66 white dwarf binary systems which occupy a region that lies above the white dwarf cooling sequence due to the contribution of flux from two stars rather than one increasing the systems $G_{\textrm{Abs}}$ (Fig.\,\ref{fig:HRD_WDsystems2}, top right panel). The majority of these systems are white dwarf-main sequence binaries, where contributions from largely blue white dwarfs and red M dwarfs can be seen simultaneously in the DESI EDR spectra. Twelve of these systems are of the accreting binary cataclysmic variable type, of which five are newly confirmed (see Section\,\ref{sec:CV}), and are largely dominated by disc emission. The last system in this sample is a double white dwarf binary WD\,J0920+6926 (Fig.\,\ref{fig:wdj0920}). The spectra of white dwarfs with both H and C in their spectrum (DAQ/DQA white dwarfs) usually show atomic C\,\textsc{i} lines which arise in hotter atmospheres and are thought to be, similar to the warm DQs, the product of binary mergers \citep{koester+kepler19-1,hollandsetal20-1}. However, WD\,J0920+6926 displays both weak and narrow Balmer lines, in addition to the molecular C Swan bands which are corroborated by the \Teff\ of $6250\pm100$\,K we obtain from our spectroscopic fitting (see Section\,\ref{sec:DQ}) that is much cooler than for the known DAQ/DQA systems. These arguments, along with its position above the white dwarf cooling sequence have led to our classification of this system as a DA+DQ system. Only one DA+DQ binary has previously been identified and characterised, NLTT~16249 \citep{vennes+kawka12-1, vennesetal12-1}, making WD\,J0920+6926 an interesting target for follow-up observations.

\begin{figure*}
	\includegraphics[width=2\columnwidth]{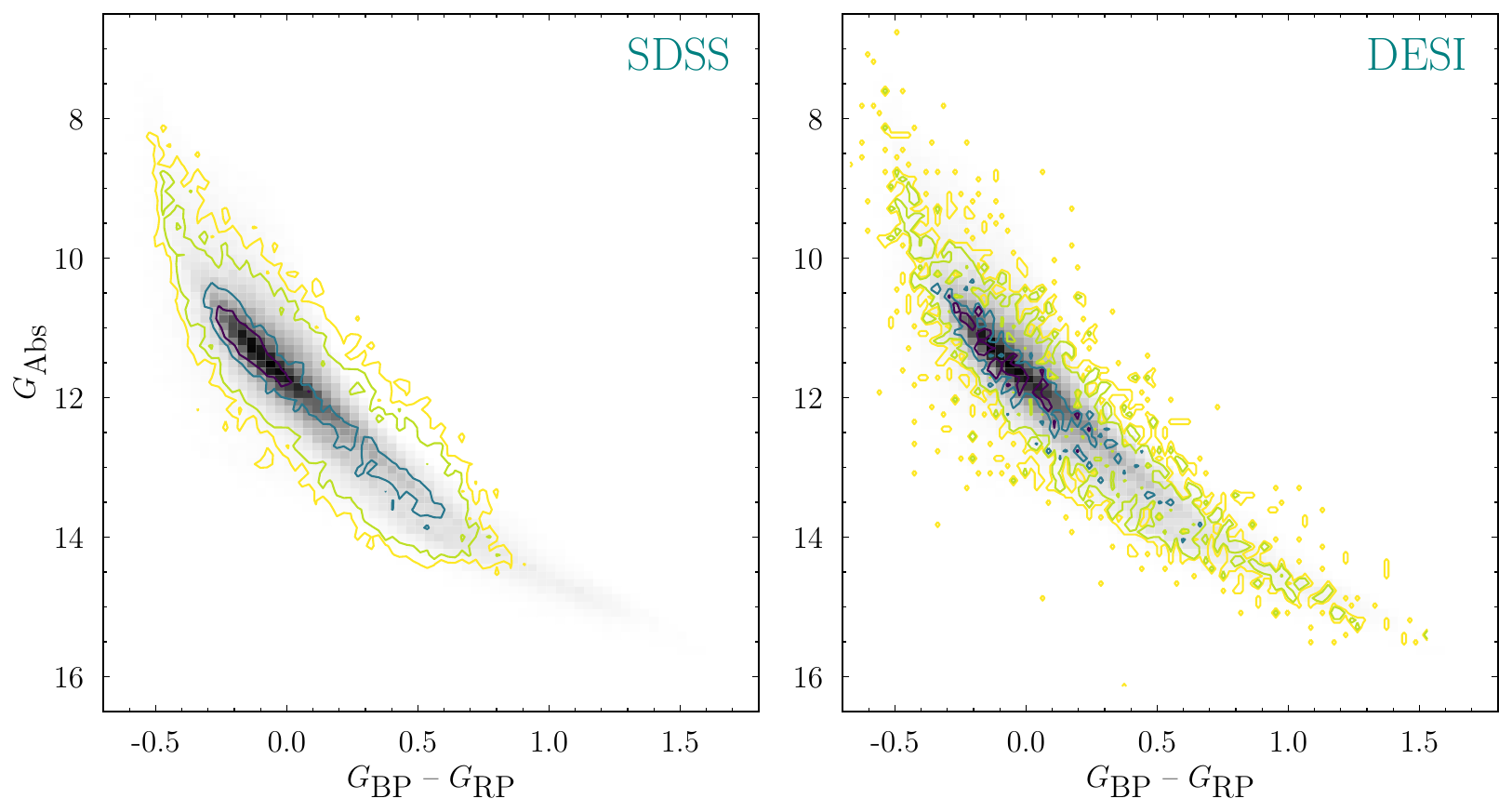}
    \caption{The distributions of the spectroscopically confirmed sample of 
    \textit{Gaia}-SDSS white dwarfs of GF21 (left) and the sample of 
    DESI EDR white dwarfs (right) are shown as normalised contours in the \textit{Gaia} HRD, where the yellow, lime, teal and purple lines correspond to values of 0.05, 0.1, 0.35 and 0.55 respectively (from largest to smallest). Spectroscopically-identified binaries have been excluded from both data sets. The magnitude-limited ($G<20$) sample of  high-confidence ($P_{\textrm{WD}} > 0.95$) white dwarf candidates \textit{Gaia} EDR3 from GF21 is underlied as a 2D gray-scale histogram. While the DESI EDR sample closely follows the magnitude limited distribution of high-probability white dwarf candidates, the SDSS sample reveals significant selection effects (see text). The higher granularity of the DESI EDR contours is due to the order smaller number of stars  compared with the \textit{Gaia}-SDSS sample.}
    \label{fig:HRD_compare}
\end{figure*}

\begin{figure}
	\includegraphics[width=1\columnwidth]{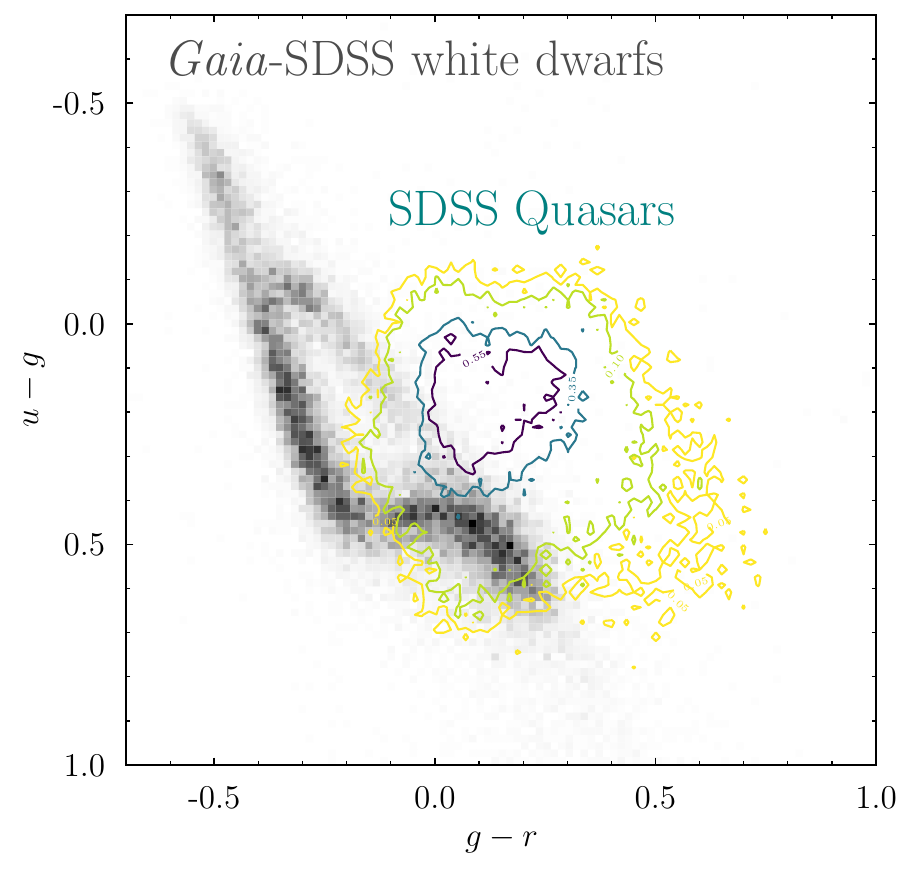}
    \caption{A histogram (black) of the \textit{Gaia}-SDSS spectroscopically confirmed white dwarf sample of GF21, with a sample of SDSS-observed quasars plotted as contours \protect\cite{schneideretal07-1}. The maximum of the Quasar distribution is set to one, and the yellow, lime, teal and purple contours correspond to values of 0.05, 0.1, 0.35 and 0.55 respectively (from largest to smallest). The over-density of white dwarfs observed by SDSS on the \textit{Gaia} HRD seen in Fig.\,\ref{fig:HRD_compare} is likely due to serendipitous observations from SDSS Quasar target selection \protect\citep{richardsetal02-1}.}
    \label{fig:SDSS-QSOs}
\end{figure}

The bottom left panel of Fig.\,\ref{fig:HRD_WDsystems2} contains systems that either show evidence for Zeeman-splitting of spectral features or are associated with magnetic fields. Only 56 white dwarfs in the sample have a confident detection of magnetic fields, with all but three of them being classed as DAH white dwarfs. The population of magnetic white dwarfs is dimmer and redder than what would be expected if they uniformly sample the magnitude limited population of white dwarfs provided by \textit{Gaia}, and is split into two populations. The first group of magnetic white dwarfs appears to follow the main white dwarf cooling track although slightly fainter suggesting that these white dwarfs are either bluer and/or heavier than their non-magnetic counterparts. The second population of magnetic white dwarfs lies close to, or on, the Q-branch. These systems appear to have higher masses than their redder counterparts and are likely the result of binary mergers \citep{reghos+tout95-1, toutetal08-1, nordhausetal11-1, garcia-berroetal12-1, wickramasingheetal14-1}. 

In the bottom right panel of Fig.\,\ref{fig:HRD_WDsystems2} we show the remaining white dwarf systems that did not nicely fall into the loose categories above, in addition to 20 systems we identified as white dwarfs but no further classification could be given.

\subsection{Advantages over the SDSS white dwarf sample}

The DESI survey has already observed over 47\,000 white dwarf candidates \citep{manseretal23-1}, making it one of the largest surveys of white dwarfs to date, surpassing the $\simeq30\,000$ white dwarfs observed by the SDSS \citep{harrisetal03-1,kleinmanetal04-1,kleinmanetal13-1,eisensteinetal06-1,fusilloetal15-1, fusilloetal19-1, fusilloetal21-1, kepleretal21-1}. Most of the SDSS white dwarfs were serendipitously targeted as quasar candidates or blue-excess objects, resulting in strong selection effects with respect to the underlying galactic population of white dwarfs. The left panel of Fig.\,\ref{fig:HRD_compare} compares the distribution of the \textit{Gaia}-SDSS sample of spectroscopically confirmed isolated white dwarfs from GF21 (as contour lines) with the entire magnitude limited sample of high-confidence ($P_\mathrm{WD} > 0.95)$ white dwarf candidates from GF21 (gray 2D histogram). It can be clearly seen that the SDSS sample is heavily biased towards bluer, hotter white dwarfs, with very few systems extending past $(G_{\textrm{BP}} - G_{\textrm{RP}}) > 0.8$. Additionally, there is an over-density of white dwarfs in the \textit{Gaia}-SDSS sample around $ 0.25 < (G_{\textrm{BP}} - G_{\textrm{RP}}) < 0.6$. Conversely, the DESI EDR sample shows a much better agreement with the magnitude limited sample of high-confidence white dwarf candidates. This should not be surprising given the DESI EDR white dwarf target selection was based on the sample selected by GF21, but highlights the selection effects within the SDSS sample. 

The biases present in the SDSS white dwarf sample have been previously reported in other studies (e.g., section 5.2.1 of \citealt{gaensickeetal09-1}), and are mainly due to two broad selection effects.

\textbf{(i) Serendipitous discoveries in the SDSS quasar search:} the bulk of white dwarfs identified by SDSS were selected originally as blue quasar candidates, resulting in an excess of white dwarfs in the range $ 0.25 < (G_{\textrm{BP}} - G_{\textrm{RP}}) < 0.6$, corresponding (roughly) to a range of effective temperatures $7000$\,K$ < T_{\textrm{eff}} < 9000$\,K. However, to reduce the total number of stars observed as part of the quasar search, the SDSS target selection adopted a number of exclusion boxes, which add further complexity to the biases affecting the SDSS sample (see sect.\,3.5.1 in \citealt{richardsetal02-1} and sect.\,5.2.1 in \citealt{gaensickeetal09-2}).

\textbf{(ii) Blue selection of white dwarfs:} Blue and hot ($T_\mathrm{eff}\gtrsim12\,000$) were included as ancillary science targets for spectroscopy in SDSS~III \citep{dawsonetal13-1} and, in smaller numbers as flux standards, in SDSS~IV \citep{dawsonetal16-1}, resulting in an additional selection biases.

In summary, whereas SDSS provided a large number of spectroscopic observations of white dwarfs, it was subject to selection effects that are extremely difficult to quantify. As DESI follows up \textit{Gaia} white dwarf candidates, which are only subject to a magnitude-limit, the forthcoming large sample of DESI white dwarfs will be very well suited for detailed statistical studies of their physical properties. 


\section{White dwarf sub-type studies}\label{sec:pop_studs}



\subsection{DA/DB/D(AB) model atmospheres and fitting procedure}\label{sec:spec_fit}

The properties of the population can be determined if their stellar parameters (\Teff\ and \logg) are known. Nearly 80 per cent of the stars in the catalogue show spectral lines from only H or He (types DA, DAB, DB and DBA), allowing reliable determination of their parameters by fitting white dwarf models to their data using established techniques. The methods we use are described in detail in Appendix\,\ref{sec:fitting_methods}. In summary, grids of synthetic spectra and photometry are generated using the atmosphere codes of \cite{koester10-1}, varying \Teff, \logg, and (where appropriate) the H/He abundance, and applying reddening. Those variables are then fitted to the data in a Bayesian framework, treating distance as an additional free parameter constrained by the \textit{Gaia} parallax and a prior tailored to a white dwarf population.

\begin{figure*}
	\includegraphics[width=2\columnwidth]{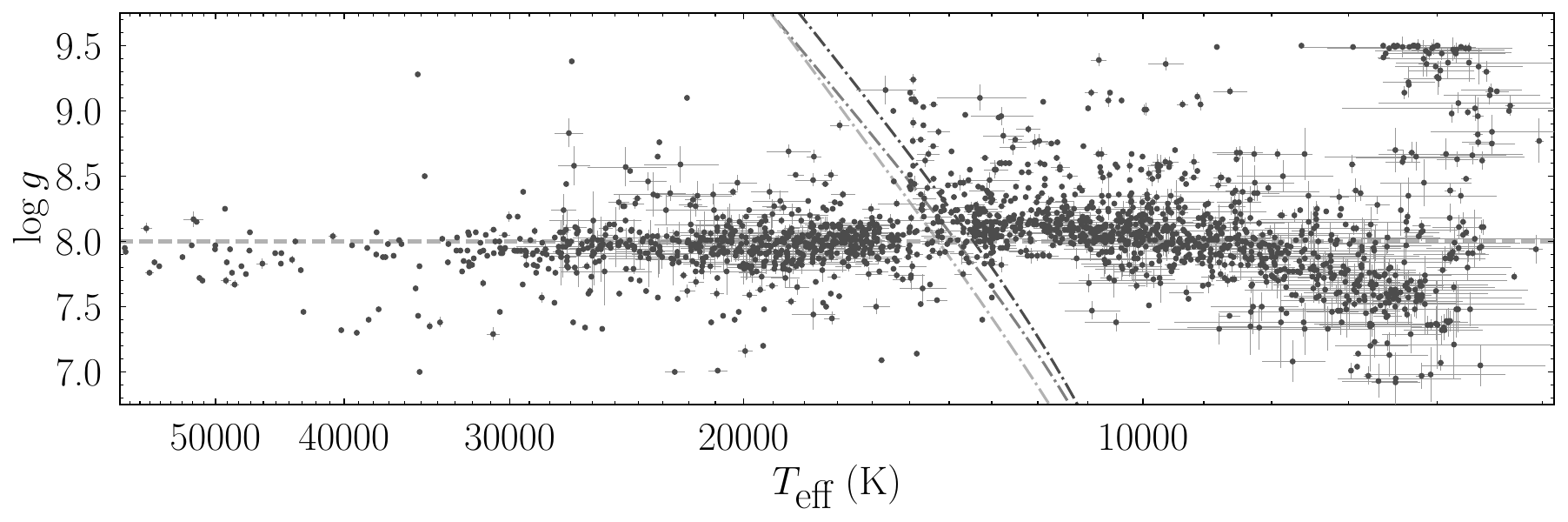}
	\includegraphics[width=2\columnwidth]{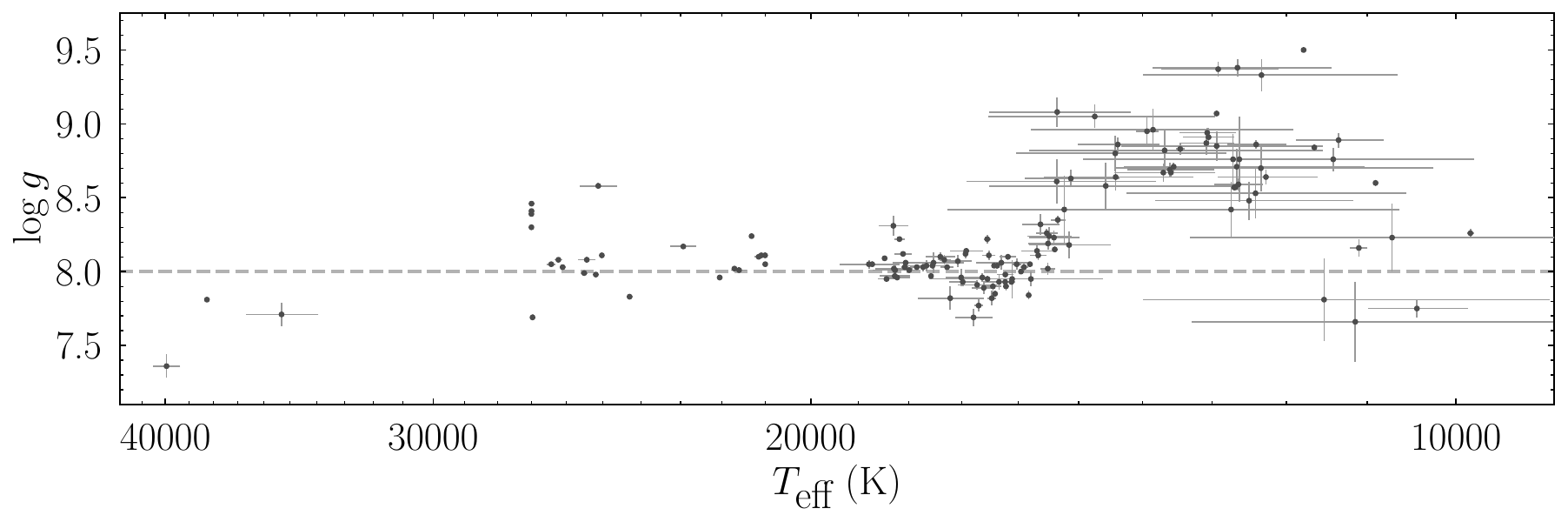}
    \caption{Effective temperatures against surface gravities of DA (top) and DB (bottom) white dwarfs measured from the DESI EDR spectra. The horizontal dashed lines indicate $\log g=8$, the average surface gravity of white dwarfs. For the DA white dwarfs, we applied the 3D corrections of \protect\cite{tremblayetal13-1} applied. The diagonal dot-dashed lines show the maximum equivalent widths of the H$\beta$, H$\gamma$, and H$\delta$ lines, going from dark grey (right) to light grey (left), in effective temperature as a function of surface gravity for the models we use in our fitting procedure.}
    \label{fig:Tvlogg}
\end{figure*}

\subsubsection{The DA sample}

The distribution of \logg\ as a function of \Teff\ drawn from the \numdas\ DAs within the DESI EDR spectra is presented in Fig.\ref{fig:Tvlogg} (top panel). 

This distribution should largely be smooth and closely follow the canonical value of $\log g = 8.0$\,dex for DA white dwarfs. However, there are three clear deviations from this that we explain below: a scarcity of systems between 15\,000 and 16\,000\,K, a small population of cool and massive systems, and a deviation from the canonical value of 8.0\,dex for the cooler objects. These are both artefacts related to the analysis with no physical meaning.

The clear decrease in the number of systems between 15\,000 and 16\,000 K is also present in other population studies \citep[see e.g.,][]{gianninasetal11-1,genest-beaulieu+bergeron19-1} and coincides with the boundary between hot and cold solutions. In fact, we are using the maximum EW of H$\beta$ line to choose between these two sets of solutions, but we have repeated this analysis for H$\gamma$ and H$\delta$. These two Balmer lines define new boundaries that move towards hotter temperatures, tentatively explaining the gap. A group of unusually cool and (apparently) massive white dwarfs exists in a region around \Teff\,$\simeq 6000$\,K and $\log g$\,$\simeq 9.5$\,dex. Inspection of these spectra show unusually broad Balmer features, and these systems are likely helium-dominated atmosphere DAs. Detailed modelling of these stars is beyond the scope of this study, and they do not affect the overall conclusions we draw from the main population of DA white dwarfs.

On the other hand, while the hotter white dwarfs are closely gathered around the canonical value of $\logg = 8.0$, their cooler counterparts ($\Teff \lesssim 15\,000$\,K) suffer from the so-called high-\logg\ problem \citep{tremblayetal10-1}. This deviation is an artefact related to the spectroscopic fitting approach, since white dwarfs are expected to cool at constant radii and the abrupt change in \logg\ is not seen in photometric studies. \cite{tremblayetal13-1} presented a grid of pure-H 3D model atmospheres with a correct treatment of the convection, and they demonstrated that the high-\logg\ problem is related to the use of the 1D mixing-length approximation. 3D model atmospheres are still too computationally expensive but \cite{tremblayetal13-1} derived analytical functions to convert spectroscopic 1D \Teff\ and \logg\ to 3D atmospheric parameters. This correction was applied to the sample but there persists a small deviation from the canonical $\logg = 8.0$.


\subsubsection{The DB and DBA samples}

The spectroscopic parameters derived for the 141 DB white dwarfs within DESI EDR are displayed in Fig.\,\ref{fig:Tvlogg} (bottom panel). The warmer stars ($\Teff \geq 15\,000$\,K) are closely clustered around $\logg\simeq8$, the typical value for white dwarfs. This is not the case for the cooler objects where there is a known issue with the implementations of the broadening mechanisms for the neutral He, which results in fitted white dwarfs having unphysically high surface gravities \citep[see e.g.,][]{bergeronetal11-1, koester+kepler15-1, cukanovaiteetal21-1}. Above $\Teff \simeq 23\,000$\,K the \logg\ values exhibit larger scatter: the spectroscopic parameters are difficult to estimate in the region where the EWs of the He\,\textsc{i} transitions reach their maximum ($23\,000 - 30\,000$\,K) and this translates to less accurate parameters. Additionally, there are four objects with $\Teff \simeq 27\,000$\,K, which corresponds to the top boundary of the cool models, and likely have hotter temperatures. Above 30\,000\,K, which coincides with the top end of the instability strip of DBs, we find three objects that have $\log g<8.0$.

\subsection{DQ white dwarfs}\label{sec:DQ}

\begin{figure}
	\includegraphics[width=1\columnwidth]{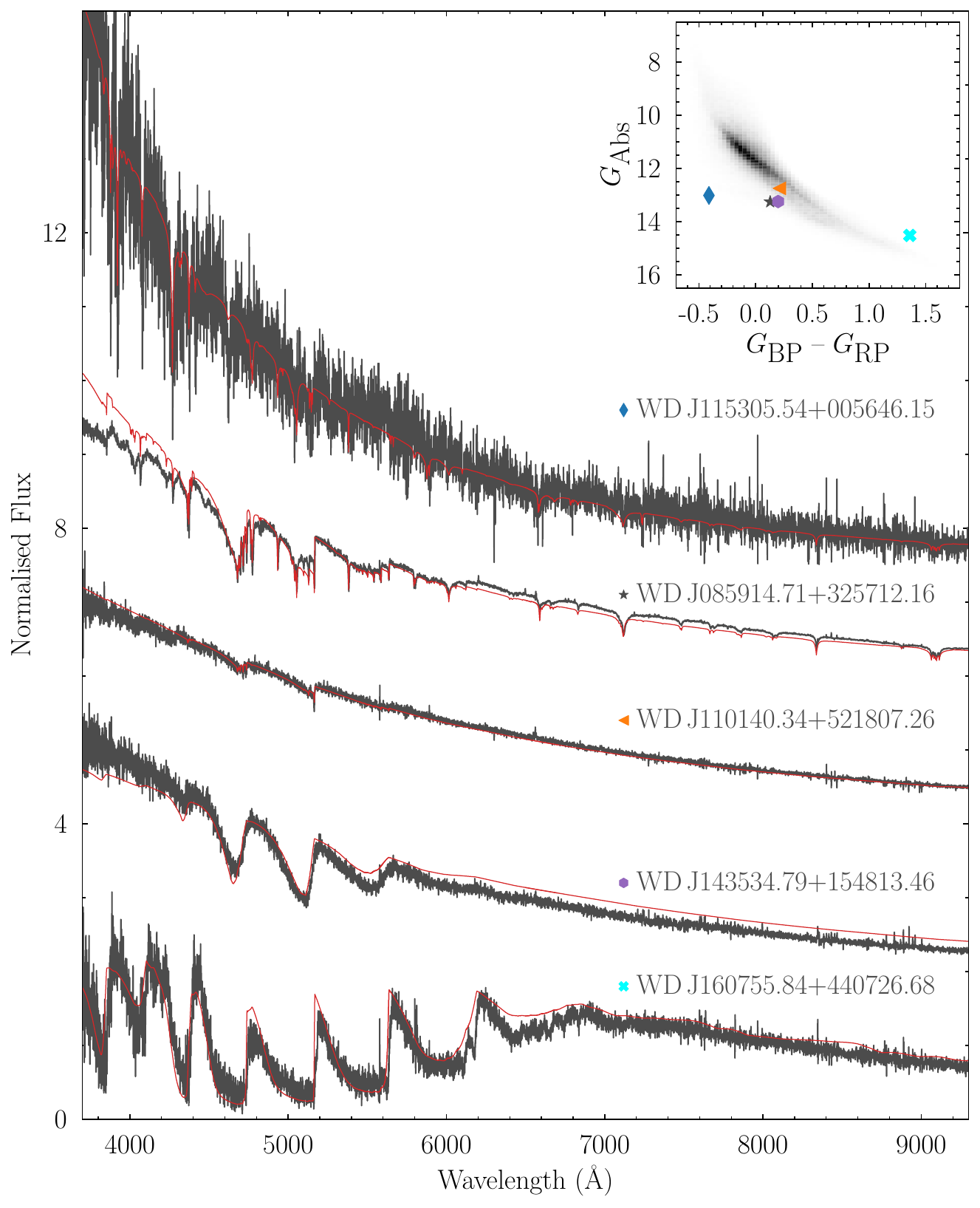}
    \caption{Examples of different types of DQs (gray) with model fits (red) overplotted. Spectra are vertically offset for clarity. The inset in the top right shows the location of these systems on the \textit{Gaia} HRD as coloured markers.}
    \label{fig:DQ_fits}
\end{figure}

The DESI EDR sample contains 55 DQ white dwarfs (Table\,\ref{t-DQx}), of which 43 are new discoveries (78\,per\,cent). We fitted all DQ spectra following the methods of \cite{koester+kepler19-1}, and using the models computed with the code of \cite{koester10-1} with updates described in \cite{koesteretal20-1}. The spectral appearance of DQ white dwarfs strongly depends on their effective temperature, with hotter white dwarfs showing atomic carbon lines and the cooler ones $C_2$ Swan bands (Fig.\,\ref{fig:DQ_fits}). Within the \textit{Gaia} HRD, the hotter DQs are located on the Q-branch, and are correspondingly massive ($M_\mathrm{wd} > 1\,\Msun$, Table\,\ref{t-DQx}). In contrast, the cooler DQs largely lie on the standard cooling sequence, having average masses.

\begin{figure}
	\includegraphics[width=1\columnwidth]{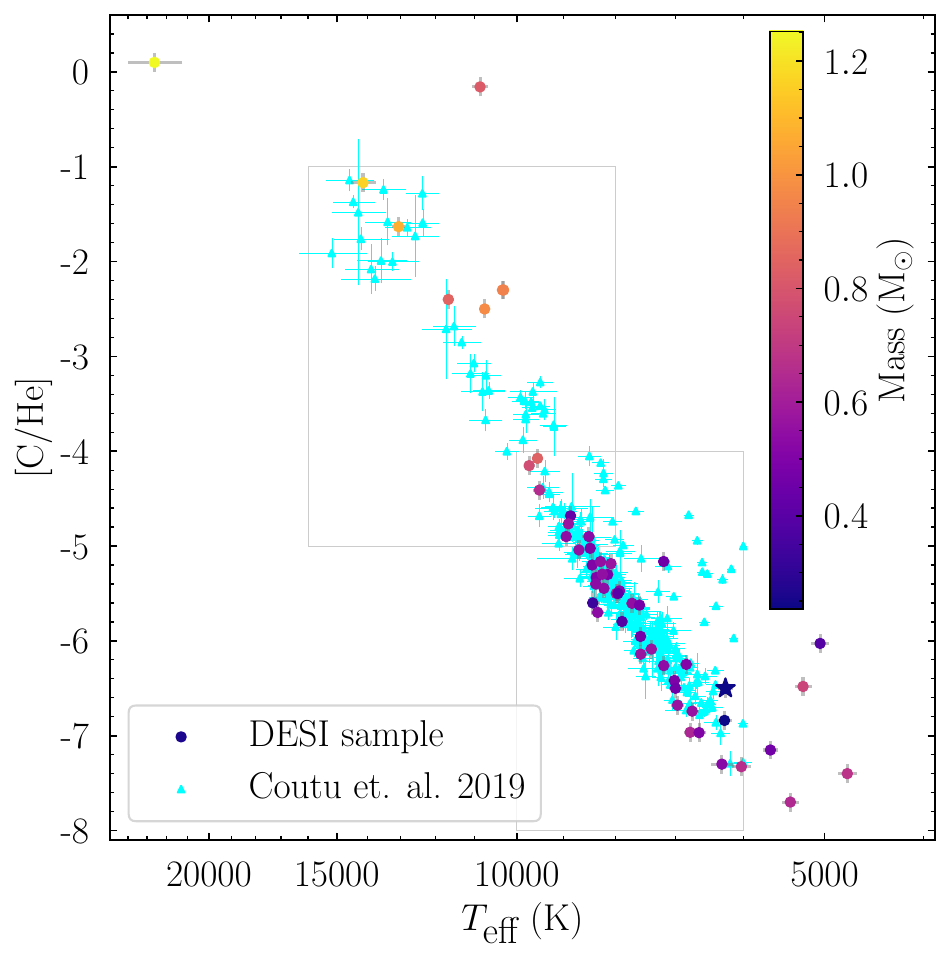}
    \caption{Fitted parameters of carbon-rich white dwarfs (DQs) in our sample (purple, orange, and yellow circles) compared to the sample of 317 DQs studied by \protect\cite{coutuetal19-1} (cyan triangles). Gray boxes denote the parameter space of the model grids used in \protect\cite{coutuetal19-1}, and are presented here to highlight the regions where comparisons between the samples should be made. The DA+DQ binary WD\,J0920+6926 is included (star), although the fit assumes the spectrum is of a single object.}
    \label{fig:DQs}
\end{figure}

The carbon abundances show a strong anti-correlation with effective temperature (Fig.\,\ref{fig:DQs}), following the trend found earlier by \cite{coutuetal19-1}. It is noteworthy that the spectrum of the hottest DQ in the EDR sample, WD\,J085601.74+321354.51, is consistent with a carbon-dominated atmosphere.

\begin{figure}
	\includegraphics[width=1\columnwidth]{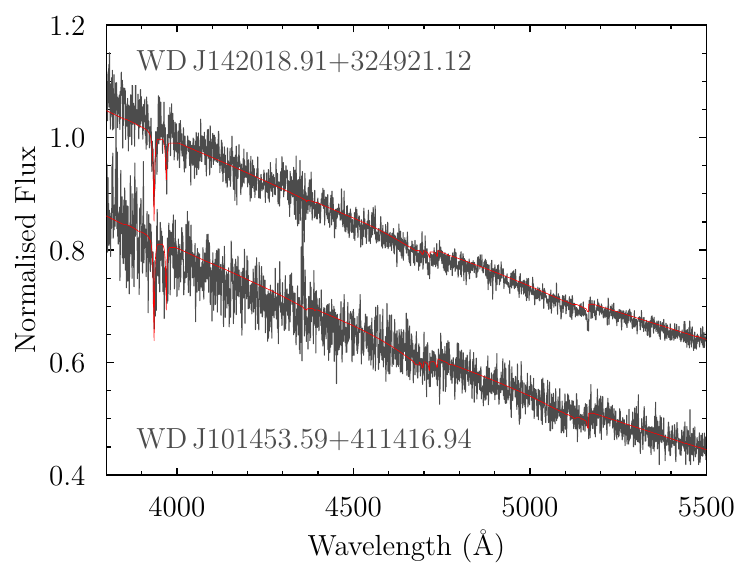}
    \caption{DESI spectra (gray) of DZQ WD\,J1014+4114 and the DQZ WD\,J1420+3249 with model fit including metals overplotted (red).}
    \label{fig:DQZ_plot}
\end{figure}

We also identified two DQs that show traces of photospheric metals in addition to carbon: WD\,J101453.59+411416.94 and  WD\,J142018.91+324921.12 (hereafter WD\,J1014+4114 and WD\,J1420+3249 respectively) which we classify as DZQ and DQZ, respectively (Fig.\,\ref{fig:DQZ_plot}). The fraction of DQ white dwarfs that exhibit metals in their spectra is smaller than that among DA or DB white dwarfs. This is thought to be the result of an interplay between the accreted metals and the atmosphere structure, leading to the rapid transformation of DQ white dwarfs into DZ white dwarfs \citep{hollandsetal22-1, blouinetal22-1}.

Finally, we identified one DA+DQ double degenerate (WD\,J0920+6926 see Fig.\ref{fig:wdj0920} and Sect.\ref{sec:hrd-exotic}) and detect Zeeman splitting in one of the warm DQs (WD\,J112513.32+094029.68, hereafter WD\,J1125+0940, see Fig.\,\ref{fig:DQH} and Sect.\,\ref{sec:DQH}).


\subsection{Metal line DxZ white dwarfs}\label{sec:DxZ}

Among the 152 DxZ (where ``x'' is a placeholder for other spectral identifiers, e.g. A, B, Q) white dwarfs that show metal lines in their DESI EDR spectra (Table\,\ref{t-DxZ1}) 121 are new discoveries (80\,per\,cent) and 31 were previously known. This sample is very likely dominated by white dwarfs that are actively accreting, or have recently accreted planetary material \citep{zuckermanetal03-1,zuckermanetal07-1,barstowetal14-1,koesteretal14-1, wilsonetal19-1}, in particular in nearby white dwarfs with strong metal absorption features. However, narrow absorption lines in hotter white dwarfs, in particular of Ca~H/K and Na~D,  could also be present due to absorption in the interstellar medium (ISM) as these intrinsically bright stars can be observed at greater distances. To disentangle enrichment of the white dwarf photosphere and absorption in the ISM requires detailed modelling of the DESI spectra that takes into account the ionisation and excitation within the white dwarf atmosphere, as well as the line-of-sight velocities of both the white dwarf and the ISM, and is beyond the scope of this work. We briefly discuss two noteworthy metal-enriched white dwarfs among the DESI EDR sample.

\subsubsection{WD\,J085035.17+320804.29}

\begin{figure*}
    \includegraphics[width=1\columnwidth]{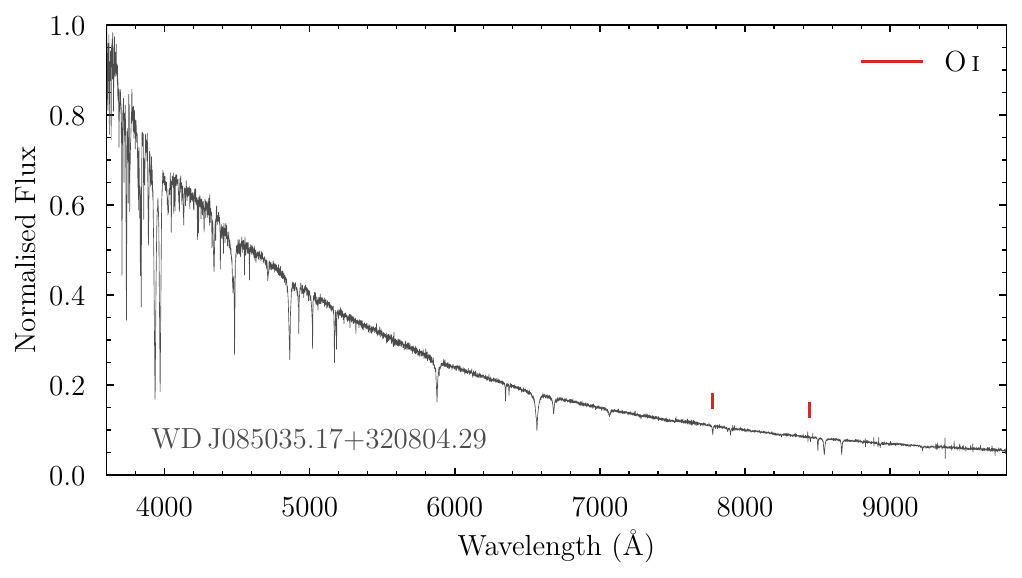}
    \includegraphics[width=1\columnwidth]{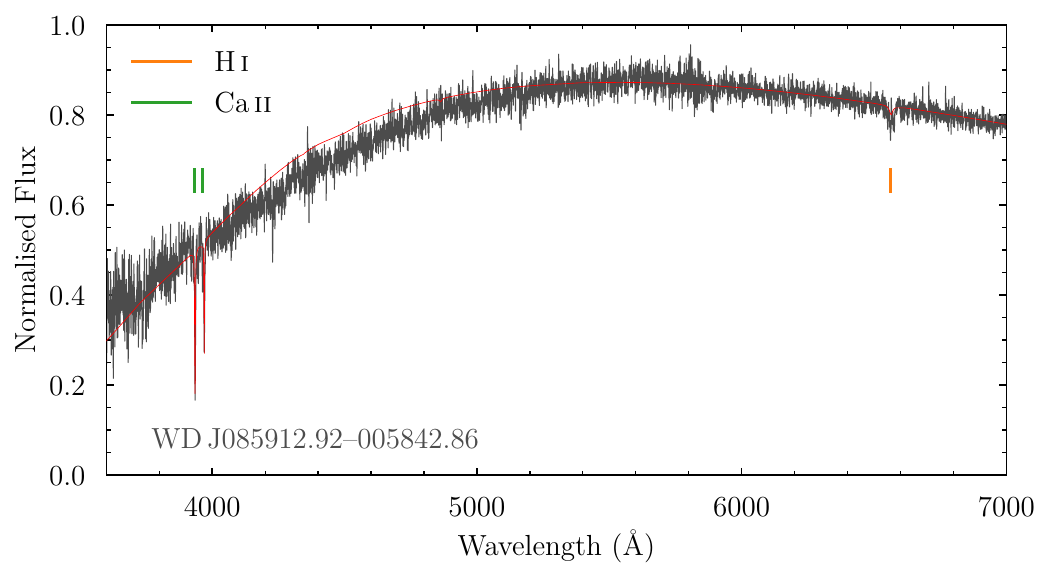}
    \caption{Two examples of white dwarfs within the DESI EDR sample that accrete planetary debris. Left: the spectrum of the extremely metal-enriched DZAB white dwarf WD\,J0850+3208 exhibits absorption lines of many elements, including oxygen (indicated by the red tick marks), which along with the detection of Balmer lines may be indicative of water accretion. Right: the DESI spectrum (gray) of the DZA WD\,J0859--0058 contains calcium and hydrogen absorption features indicated by vertical tick marks. The best-fit model (red) implies $\Teff=4832 \pm 10$\,K,  $\logg = 7.87 \pm 0.10$, and [Ca/H]\,$ = -10.25 \pm 0.10$. WD\,J0859--0058 is a known 20\,pc member with a previous DC spectral classification based on lower-quality spectroscopy \citep{subasavageetal08-1}.}
    \label{fig:dxz}
\end{figure*}

WD\,J085035.17+320804.29 (hereafter WD\,J0850+3208, Fig.\,\ref{fig:dxz}) is a DZAB\footnote{WD\,J0850+3208 is a known metal-enriched white dwarf, classified as a DABZ by \protect\cite{kongetal19-1}.} which shows extremely strong metal lines. O\,\textsc{i} absorption at 7775\,\AA\ and 8446\,\AA, along with the strong H$\alpha$ line indicated significant amounts of H in  what is likely a He-dominated atmosphere suggest that this white dwarf may have accreted a water-rich body \citep{kleinetal10-1,farihietal13-1,fusilloetal17-1}. Further bespoke modelling of WD\,J0850+3208 is required to determine the composition of the planetary body it accreted. 

\subsubsection{WD\,J085912.92--005842.86 }

WD\,J085912.92--005842.86 (hereafter WD\,J0859--0058, Fig.\,\ref{fig:dxz}) is a known 20\,pc member which has been previously reported as a featureless DC \citep{subasavageetal08-1, hollandsetal18-2}, however, the DESI spectrum clearly reveals Ca H/K lines. Inspection of spectrum of WD\,J0859--0058 obtained by \cite{subasavageetal08-1} and available in the MWDD shows no signs of metal absorption features, likely because of the much lower spectral resolution compared to the one obtained by DESI. This system is among the closest sources to the Sun observed by DESI, and this classification shows that new and interesting systems are still being identified in well-studied samples.


\subsection{Magnetic white dwarfs}\label{sec:DxH} 

\begin{figure}
	\includegraphics[width=1\columnwidth]{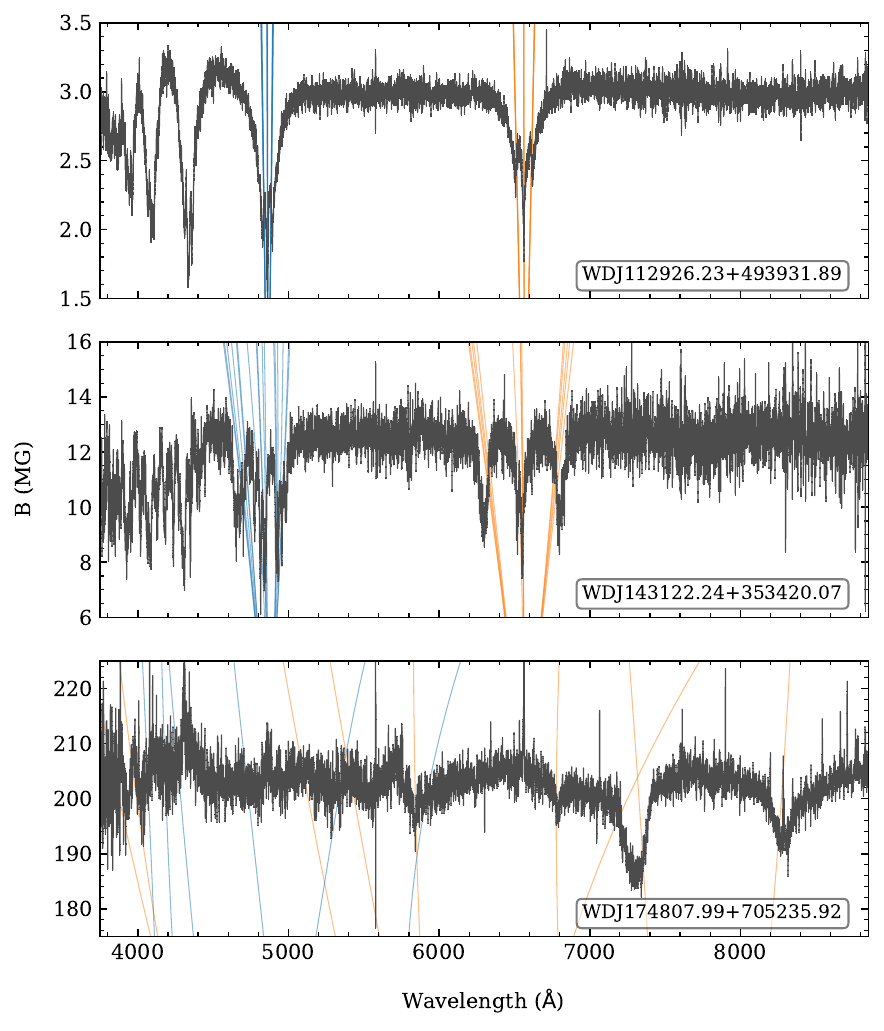}
    \caption{Continuum normalised spectra of three DAH white dwarfs showcasing the range of magnetic fields observed. Each spectrum has been normalised to one, and then multiplied by the field strength, $B$, determined from the spectrum. Transition wavelengths for H$\beta$ and H$\alpha$ as a function of $B$ are provided by \protect\cite{schimeczek+wunner14-1, schimeczek+wunner14-2} and are presented as blue and orange lines respectively. \textbf{Top panel:} The field strengths on a few MG level show clear three-components consistent with the linear Zeeman-splitting regime. \textbf{Middle panel:} As $B$ increases, the energy degeneracy due to orbital angular momentum, $l$, will lift and these three components will split further, resulting in 18 (15) transitions in H$\beta$ (H$\alpha$). \textbf{Bottom panel:} When the field reaches a few hundred MG, the transitions spread across the optical spectral range.}
    \label{fig:DAH_example}
\end{figure}

For the 56 systems presenting clear Zeeman-split spectral features we estimated their surface-averaged field strengths which are provided in Table\,\ref{t-DxH}. 39  (70\,per\,cent) of the magnetic white dwarfs identified in the DESI EDR sample are new magnetic identifications from DESI. All but one appear to show Zeeman-splitting of the Balmer lines evident in the spectrum (Fig.\,\ref{fig:DAH_example}). For these systems, we used the transition wavelengths of the Balmer series as a function of field strength $B$ provided by \cite{schimeczek+wunner14-1, schimeczek+wunner14-2} and fit the Zeeman-split H$\beta$ and H$\alpha$ profiles, as these are usually the strongest features present. The fitting was performed by splitting the magnetic features into resolvable components and determining their location in wavelength-space, and then performing a least-squared fit with the theoretical wavelengths of the Balmer transitions to find a magnetic field strength. Two magnetic white dwarfs show Zeeman-split features in emission, classified as DAHe white dwarfs \citep{manseretal23-1}, and were fit by the authors using the same method.

While we assume in our fitting a single field strength, the field geometry of white dwarfs are known to be more complicated, having dipolar, quadrupole or even higher order poles and non-trivial combinations \citep{martin+wickramasinghe84-1, achilleos+wickramasinghe89-1, achilleosetal92-2,euchneretal02-1,euchneretal05-1,euchneretal06-1}. This is clearly evident in the spectra of high-field white dwarfs above $\sim$\,200\,MG, where only transitions that are close to $\textrm{d}\lambda/\textrm{d}B \simeq 0$ are identifiable, where $\lambda$ is the wavelength of the transition, with others being smeared out across the spectrum (see fig.\,2 of \citealt{schmidtetal03-1}). 

\begin{figure}
	\includegraphics[width=1\columnwidth]{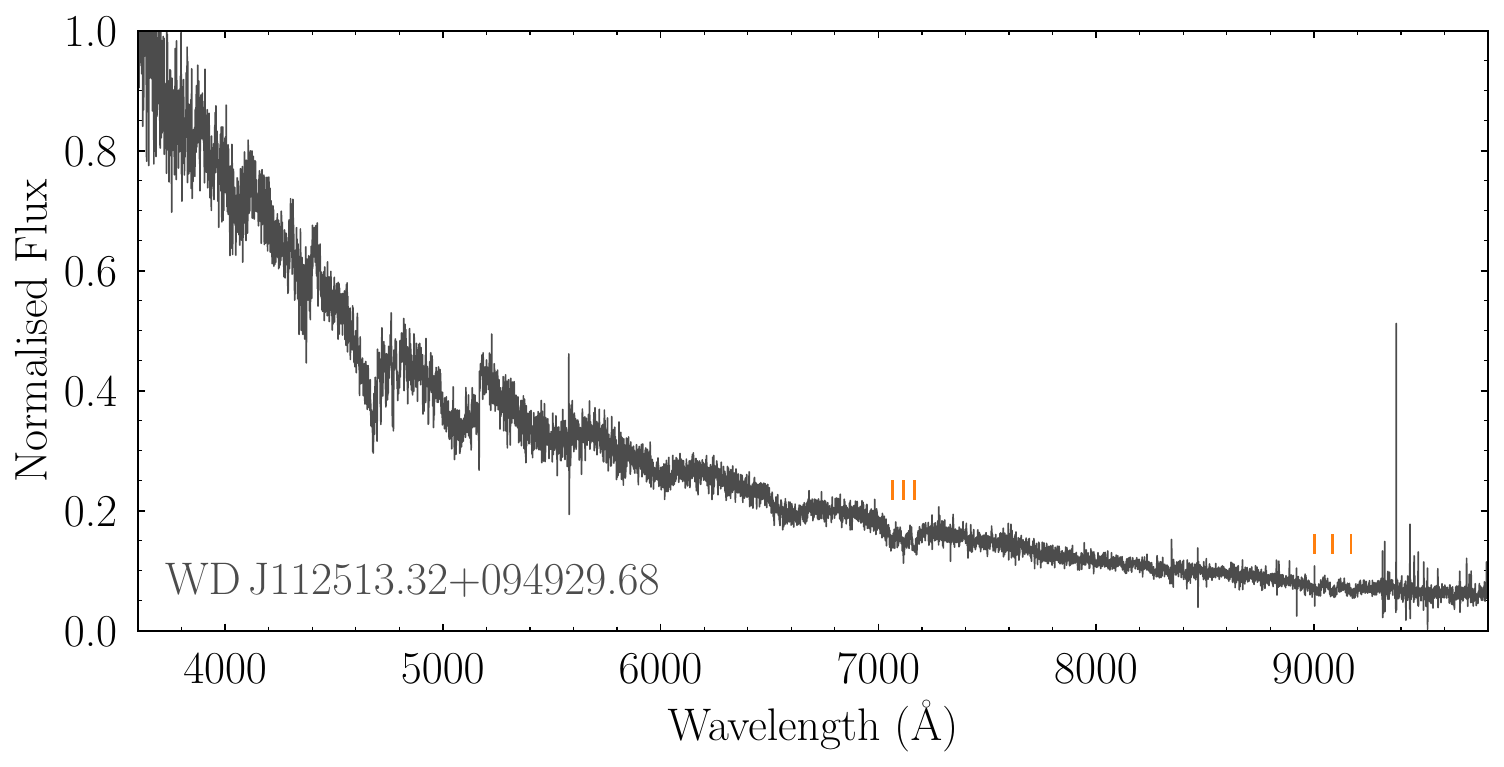}
    \caption{The DQH WD\,J1125+0949. Orange vertical tabs highlight clear Zeeman-split components, which we used to determine the surface averaged field strength of $B=2.19$\,MG.}
    \label{fig:DQH}
\end{figure}

Below we briefly discuss three systems of further interest:

\subsubsection{WD\,J112513.32+094029.68}
\label{sec:DQH}
WD\,J1125+0940 is a DQH white dwarf that shows Zeeman splitting clearly identified in the $\simeq7100$\,\AA\ and $\simeq9100$\,\AA\ C lines. We fit these Zeeman-split features and obtain a field strength of 2.19\,MG (Fig.\,\ref{fig:DQH}).

\subsubsection{WD\,J113357.66+515204.69}

WD\,J113357.66+515204.69 (hereafter WD\,J1133+5152) shows five components in H$\alpha$. Individual exposures obtained by DESI for WD\,J1133+5152 only show three components, as expected for Zeeman-splitting, and the profiles shift from one field strength to another and back again over a time-frame of $\simeq$\,20\,mins. This sharp transition between field strengths in the observed spectrum is similar to that observed at G183--35 \citep{kilicetal19-1}, and it has been suggested that a combination of a complex magnetic field structure in addition to an inhomogenous chemical distribution across the white dwarf surface can explain the profiles. We estimate the two identified field strengths observed for WD\,J1133+5152 from the five components assuming they are a superposition of two sets of Zeeman-split profiles as $3.04 \pm 0.07$MG and $4.4 \pm 0.1$MG, and these are both reported in Table\,\ref{t-DxH}.

WD\,J1133+5152 was previously identified as a DAH from SDSS spectroscopy \citep{schmidtetal03-1, kuelebietal09-1}, however there is no documentation of five Zeeman components. It has been modelled to have a polar field strength of $\simeq$\,8\,MG, slightly varying depending on the complexity of the dipolar model used, and a viewing angle compared to the dipolar field configuration as looking along the equator. For a centred-dipolar field configuration, the field strength near the magnetic equator is a factor two lower than the field strength at the poles \citep{achilleosetal92-2}, and our measured field strengths for WD\,J1133+5152 are in rough agreement with a factor two drop compared with the previously reported polar field strength.

\subsubsection{WD\,J070253.76+553733.64}

WD\,J070253.76+553733.64 (hereafter WD\,J0702+5537) also shows five components in H$\alpha$, similar to WD\,J1133+5152. By fitting the 5 components assuming they come from two sets of Zeeman-split profiles, we obtain field strengths of $6.89 \pm 0.07$\,MG and $15.5 \pm 0.2$\,MG.

\subsection{Cataclysmic Variables (CVs)}\label{sec:CV}

\begin{figure*}
\centerline{\includegraphics[width=\columnwidth]{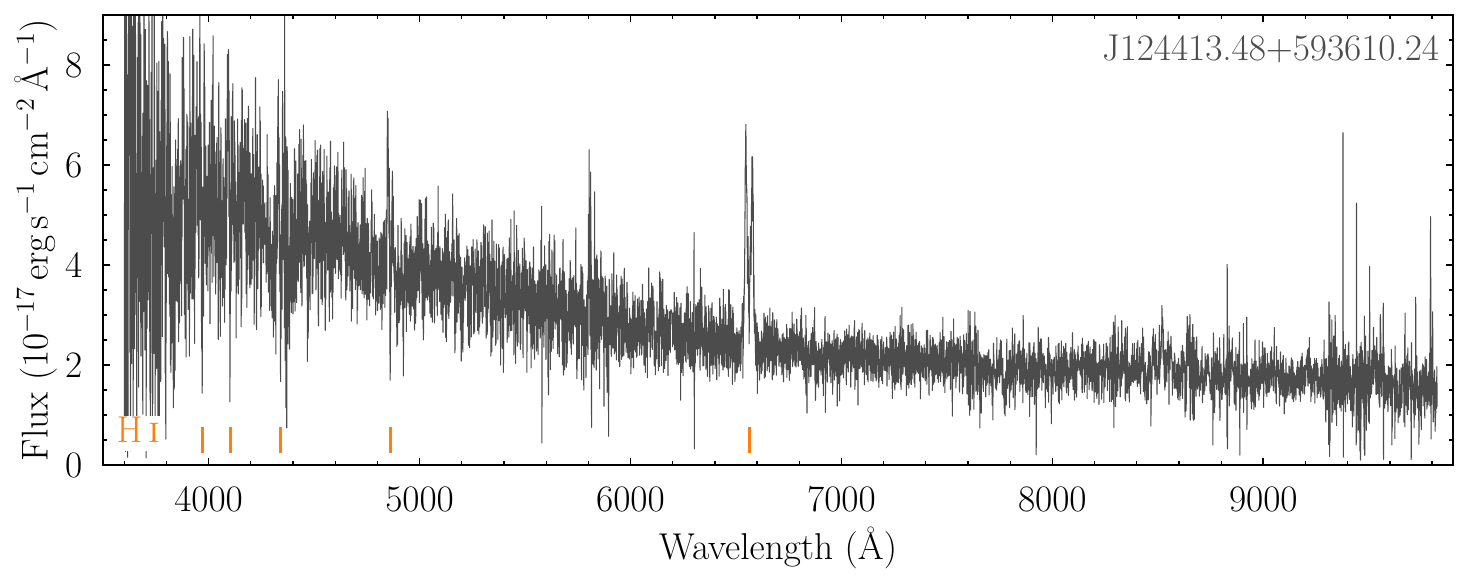}
\includegraphics[width=\columnwidth]{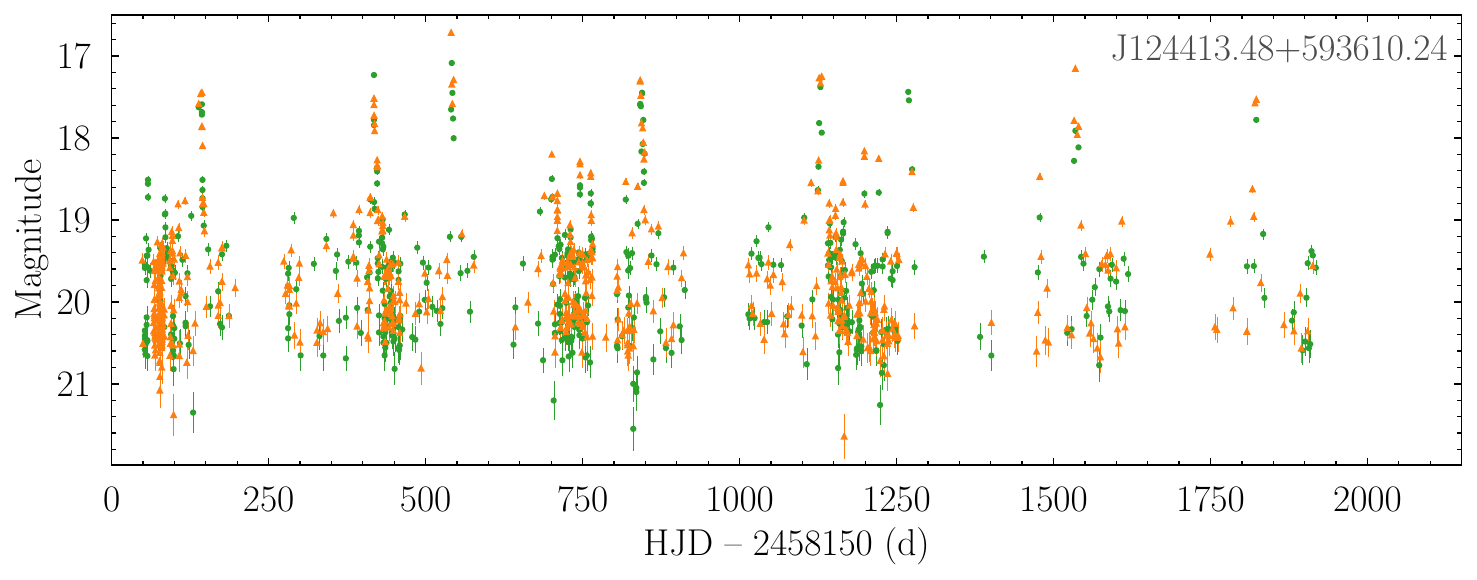}
}
\centerline{\includegraphics[width=\columnwidth]{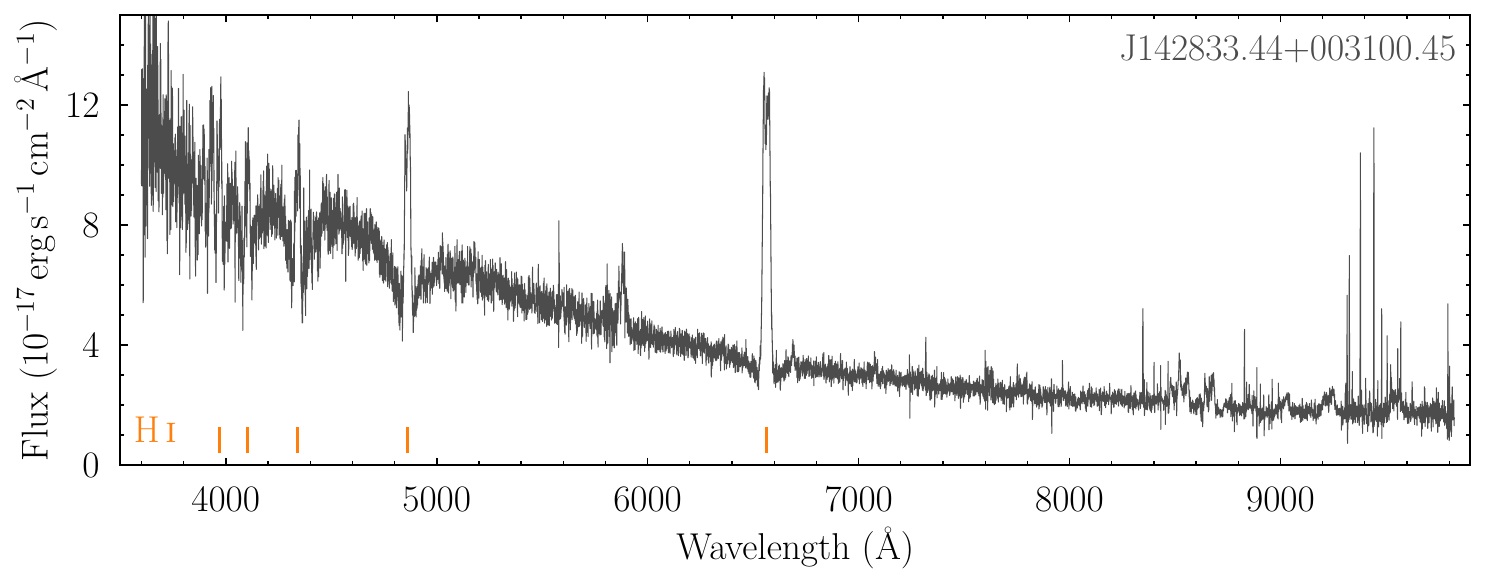}
\includegraphics[width=\columnwidth]{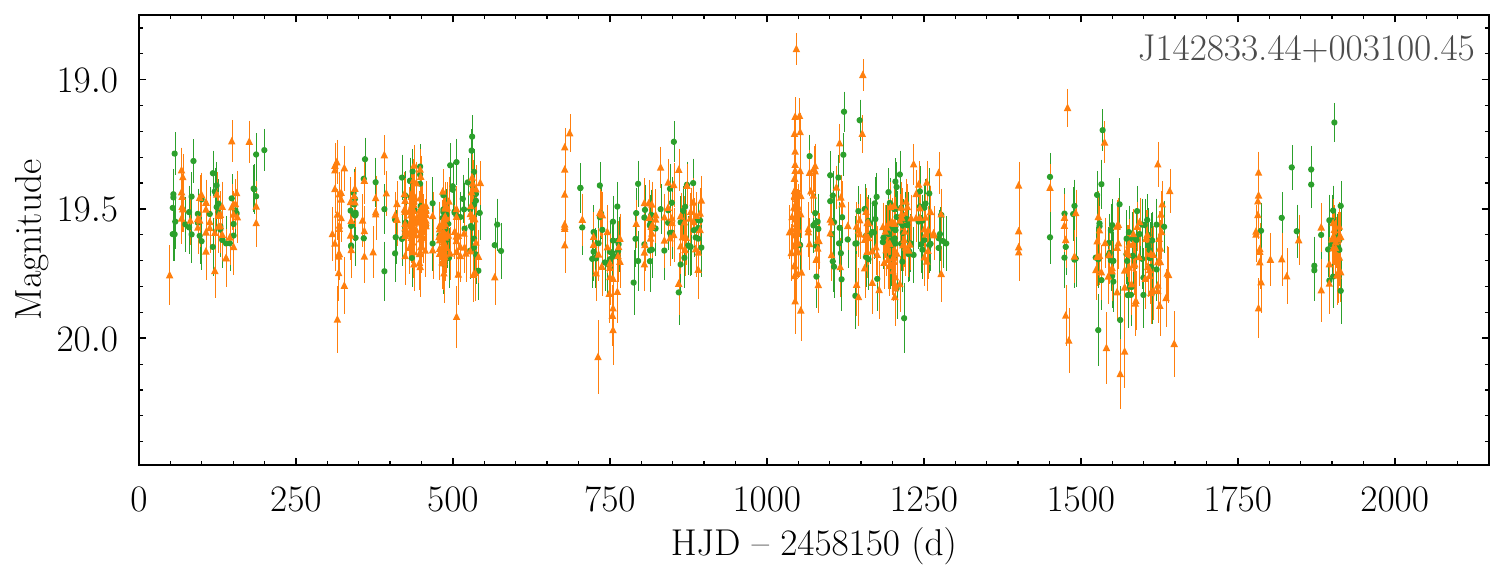}
}
\centerline{\includegraphics[width=\columnwidth]{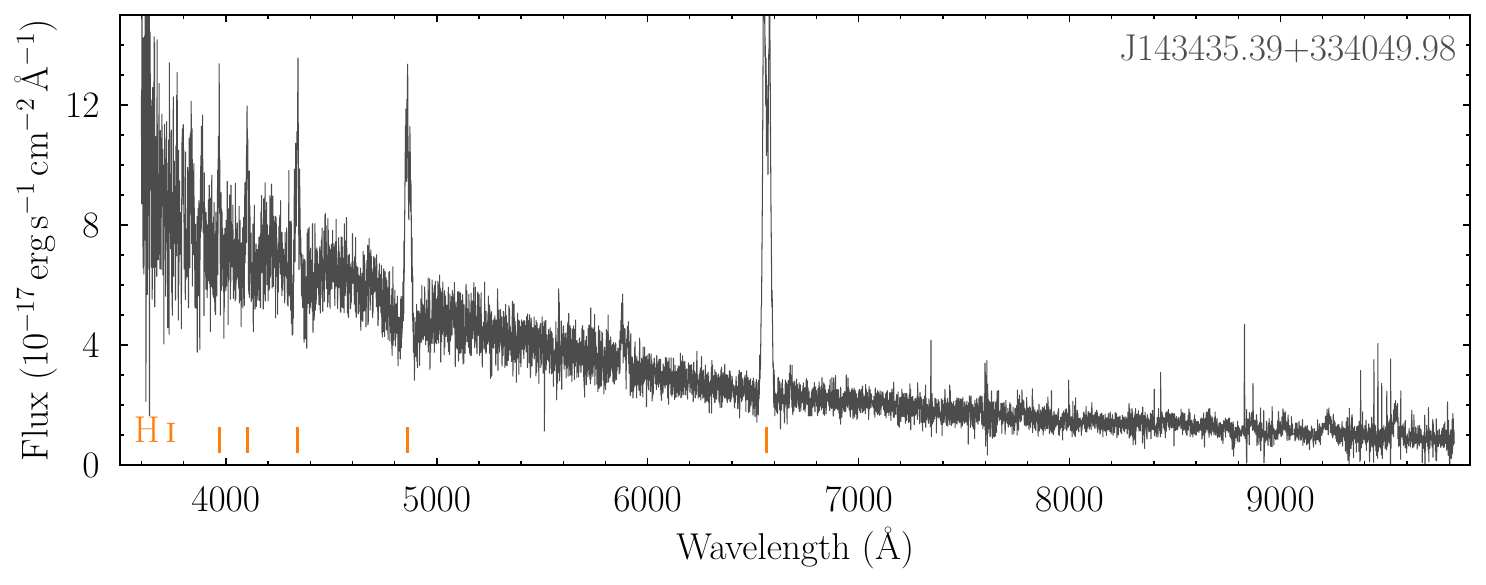}
\includegraphics[width=\columnwidth]{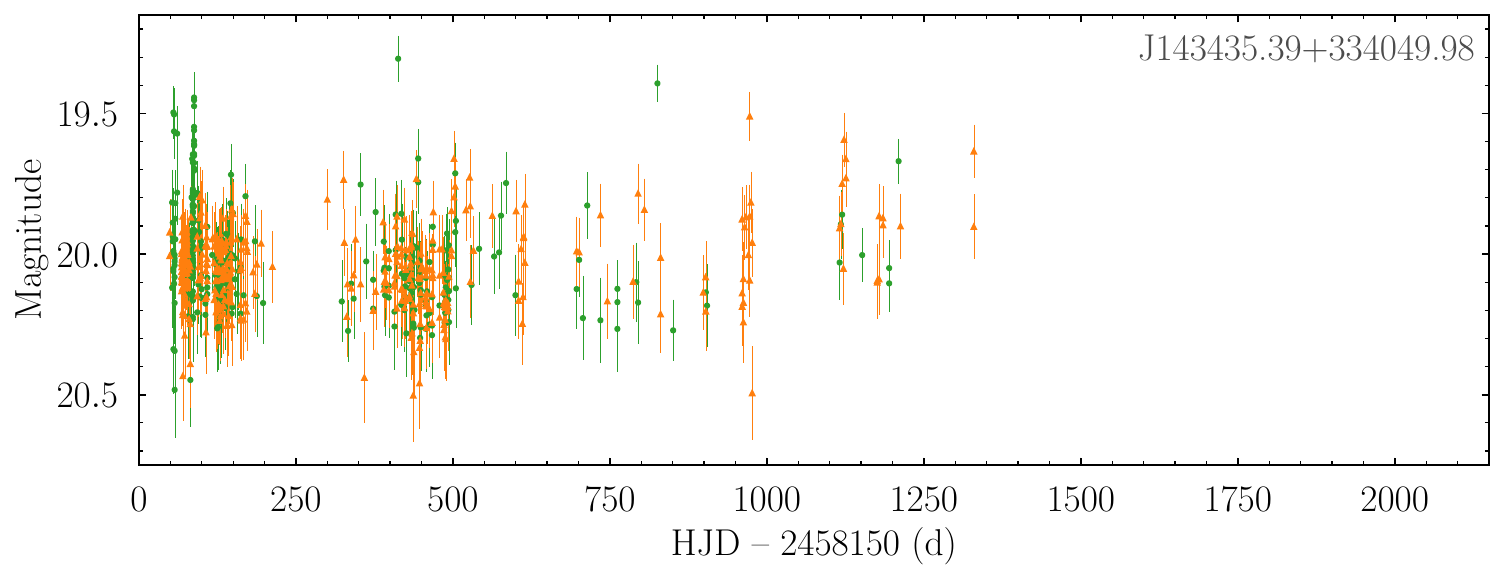}
}
\centerline{\includegraphics[width=\columnwidth]{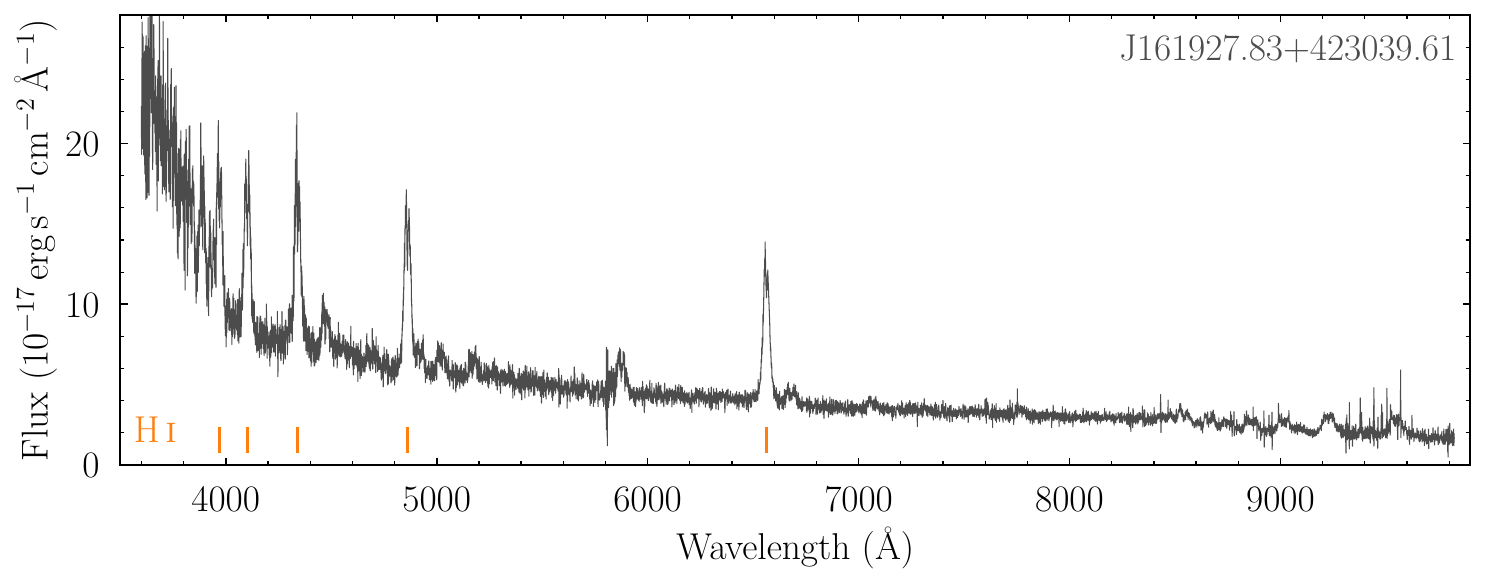}
\includegraphics[width=\columnwidth]{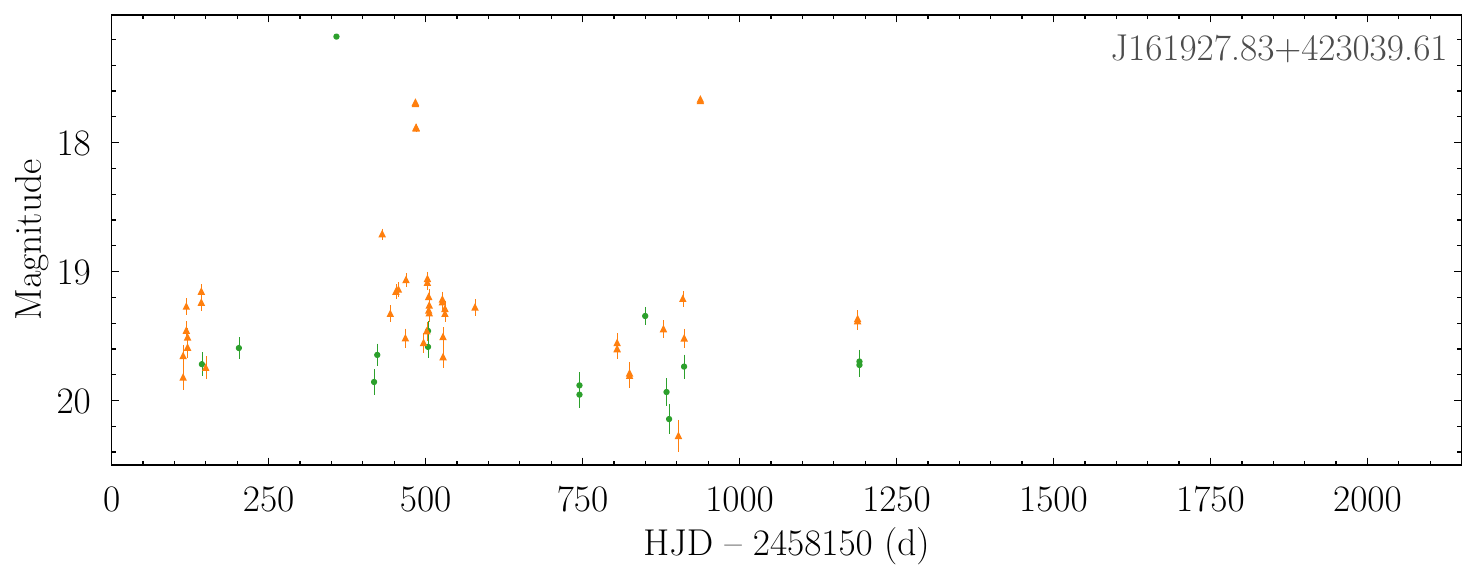}
}
\centerline{\includegraphics[width=\columnwidth]{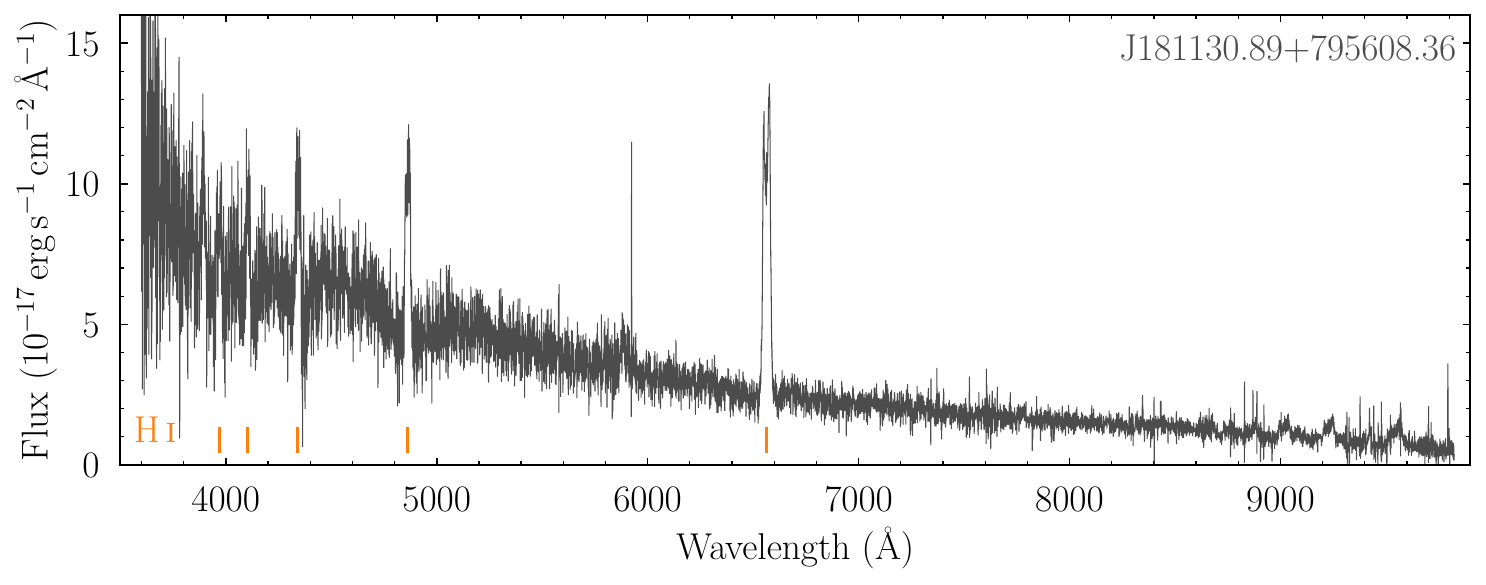}
\includegraphics[width=\columnwidth]{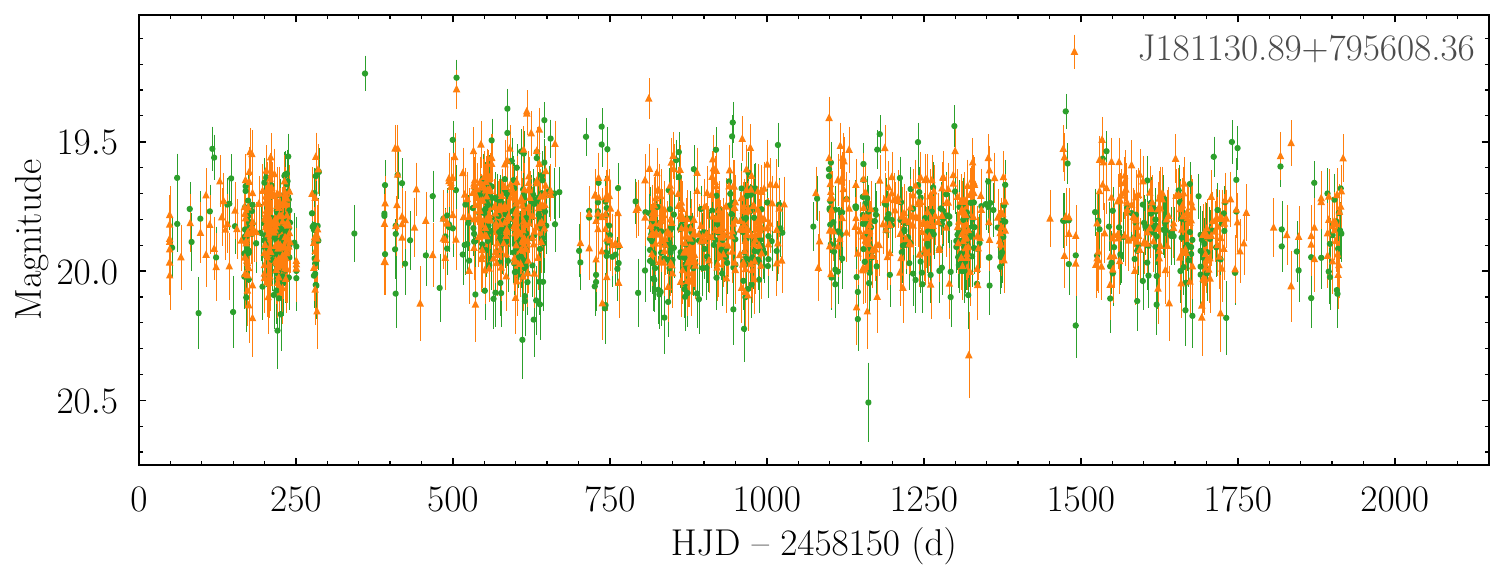}
}
\caption{\label{f-CV1} DESI spectra (left panels) and ZTF light curves (right panels) of five new CVs identified in the DESI EDR. The rest-wavelength of the first five Balmer lines are denoted by vertical orange tabs for the spectra. $g$-band and $r$-band data points are denoted by green circles and orange triangles respectively.}
\end{figure*}

Table\,\ref{t-CVs} lists the twelve CVs identified by DESI, of which five are new discoveries, and seven are previously known systems. DESI spectra along with ZTF light curves are shown in Fig.\,\ref{f-CV1} for the five new systems, and we briefly discuss them below.

\subsubsection{J124413.48+593610.24} The DESI spectrum reveals a blue continuum superimposed with moderately strong H$\alpha$ emission, and weaker H$\beta$ emission. The deep central absorption feature within H$\alpha$ is suggestive of a high inclination \citep{horne+marsh86-1}. The slope of the spectrum displays a break around 7000\,\AA, which may indicate the contribution of the donor star, however, the signal-to-noise ratio is too low to identify any of its spectral signatures. The ZTF light curves of this system reveal multiple outbursts, several of which lasting $\simeq10$\,d, corroborating a classification as SU\,UMa dwarf nova \citep{brun+petit52-1,vogt80-1}. The ZTF data also reveals deep eclipses, and we measure an orbital period of $P=1.74953778(98)$\,h.

\subsubsection{J142833.44+003100.45} The DESI spectrum shows a blue slope with broad Balmer absorption lines from the white dwarf, which are partially filled in by double-peaked emission lines indicating the presence of an accretion disc. The spectrum reveals no signature from the donor star. The ZTF light curves do not exhibit any outbursts, and a period analysis fails to detect a coherent signal in the data. This CV is most likely a WZ\,Sge dwarf nova with a long outburst recurrence time \citep{bailey79-2}. 

\subsubsection{J143435.39+334049.98} The DESI and ZTF data of this CV are overall similar to J142833.44+003100.45, though with weaker Balmer absorption lines from the white dwarf, and stronger Balmer emission lines, suggestive of a larger contribution of the accretion disc. It is also most likely a WZ\,Sge dwarf nova.

\subsubsection{J161927.83+423039.61} The DESI spectrum shows a blue continuum with the Balmer jump in emission, and double-peaked emission lines from an accretion disc, with no signature of either the white dwarf or the donor star. The spectral appearance is typical of an SU\,UMa dwarf nova. The ZTF light curve is sparse, but captured one outburst, confirming the SU\,UMa classification. 

\subsubsection{J181130.89+795608.36} The DESI and ZTF data of this system closely resemble those of J142833.44+003100.45 and J143435.39+334049.98, and this CV is also most likely a WZ\,Sge dwarf nova.

These CVs have been serendipitously included in the selection cuts made by \cite{fusilloetal19-1} and later by \cite{cooperetal23-1} to target isolated white dwarfs. CVs in this area of the \textit{Gaia} HRD are dominated by WZ\,Sge and SU\,UMa subtypes (compare our Fig.\,\ref{fig:HRD_WDsystems2} top left, and fig.\,15 of \citealt{inightetal23-1}), consistent with our identifications.

%
%
%
%
%
%
%

\section{Conclusions}\label{sec:conc}

We present here the DESI EDR sample of 2706 white dwarfs and 66 binaries including white dwarfs selected from cuts made on \textit{Gaia} astrometry and photometry. The resulting \textit{Gaia} selection function is relatively simple, making the DESI sample significantly more unbiased than previous surveys such as the SDSS and facilitating its use in statistical studies of white dwarf samples. Many of the white dwarfs in the DESI EDR have new spectroscopic classifications, including 43 DQs, 121 metal-line DxZ white dwarfs, 39 magnetic white dwarfs, and 5 CVs. The forthcoming DESI DR1 sample contains over 47\,000 white dwarf candidates, and will revolutionise statistical studies of white dwarf samples.

\section*{Acknowledgements}

The authors acknowledge financial support from Imperial College London through an Imperial College Research Fellowship grant awarded to CJM. This project has received funding from the European Research Council (ERC) under the European Union’s Horizon 2020 research and innovation programme (Grant agreement No. 101020057). 
PI was supported by a Leverhulme Trust Research Project Grant. 
SX is supported by NOIRLab, which is managed by the Association of Universities for Research in Astronomy (AURA) under a cooperative agreement with the National Science Foundation.
SK acknowledges support from the Science \& Technology Facilities Council (STFC)
grant ST/Y001001/1.
For the purpose of open access, the author has applied a Creative Commons Attribution (CC BY) licence to any Author Accepted Manuscript version arising from this submission.
In identifying new white dwarfs and classifications, we have made use of the Montreal White Dwarf Database \citep{dufouretal17-1} and the SIMBAD database, operated at CDS, Strasbourg, France \citep{wengeretal00-1}.
This research is supported by the Director, Office of Science, Office of High Energy Physics of the U.S. Department of Energy under Contract No. DE–AC02–05CH11231, and by the National Energy Research Scientific Computing Center, a DOE Office of Science User Facility under the same contract; additional support for DESI is provided by the U.S. National Science Foundation, Division of Astronomical Sciences under Contract No. AST-0950945 to the NSF’s National Optical-Infrared Astronomy Research Laboratory; the Science and Technologies Facilities Council of the United Kingdom; the Gordon and Betty Moore Foundation; the Heising-Simons Foundation; the French Alternative Energies and Atomic Energy Commission (CEA); the National Council of Science and Technology of Mexico (CONACYT); the Ministry of Science and Innovation of Spain (MICINN), and by the DESI Member Institutions: \url{https://www.desi.lbl.gov/collaborating-institutions}.
The authors are honoured to be permitted to conduct scientific research on Iolkam Du’ag (Kitt Peak), a mountain with particular significance to the Tohono O’odham Nation.
This work has made use of data from the European Space Agency (ESA) mission {\it Gaia} (\url{https://www.cosmos.esa.int/gaia}), processed by the \textit{Gaia} Data Processing and Analysis Consortium (DPAC, \url{https://www.cosmos.esa.int/web/gaia/dpac/consortium}). Funding for the DPAC
has been provided by national institutions, in particular the institutions participating in the {\it Gaia} Multilateral Agreement.
Funding for the Sloan Digital Sky Survey IV has been provided by the Alfred P. Sloan Foundation, the U.S. Department of Energy Office of Science, and the Participating Institutions. 
SDSS-IV acknowledges support and resources from the Center for High Performance Computing  at the University of Utah. The SDSS website is \url{www.sdss4.org}.
SDSS-IV is managed by the Astrophysical Research Consortium for the Participating Institutions of the SDSS Collaboration including the Brazilian Participation Group, the Carnegie Institution for Science, Carnegie Mellon University, Center for Astrophysics | Harvard \& Smithsonian, the Chilean Participation Group, the French Participation Group, Instituto de Astrof\'isica de Canarias, The Johns Hopkins University, Kavli Institute for the Physics and Mathematics of the Universe (IPMU) / University of Tokyo, the Korean Participation Group, Lawrence Berkeley National Laboratory, Leibniz Institut f\"ur Astrophysik Potsdam (AIP),  Max-Planck-Institut f\"ur Astronomie (MPIA Heidelberg), Max-Planck-Institut f\"ur Astrophysik (MPA Garching), Max-Planck-Institut f\"ur Extraterrestrische Physik (MPE), National Astronomical Observatories of China, New Mexico State University, New York University, University of Notre Dame, Observat\'ario Nacional / MCTI, The Ohio State University, Pennsylvania State University, Shanghai Astronomical Observatory, United Kingdom Participation Group, Universidad Nacional Aut\'onoma de M\'exico, University of Arizona, University of Colorado Boulder, University of Oxford, University of Portsmouth, University of Utah, University of Virginia, University of Washington, University of Wisconsin, Vanderbilt University, and Yale University.
Based on observations obtained with the Samuel Oschin Telescope 48-inch and the 60-inch Telescope at the Palomar Observatory as part of the Zwicky Transient Facility project. ZTF is supported by the National Science Foundation under Grants No. AST-1440341 and AST-2034437 and a collaboration including current partners Caltech, IPAC, the Weizmann Institute for Science, the Oskar Klein Center at Stockholm University, the University of Maryland, Deutsches Elektronen-Synchrotron and Humboldt University, the TANGO Consortium of Taiwan, the University of Wisconsin at Milwaukee, Trinity College Dublin, Lawrence Livermore National Laboratories, IN2P3, University of Warwick, Ruhr University Bochum, Northwestern University and former partners the University of Washington, Los Alamos National Laboratories, and Lawrence Berkeley National Laboratories. Operations are conducted by COO, IPAC, and UW.
This research has used data, tools or materials developed as part of the EXPLORE project that has received funding from the European Union’s Horizon 2020 research and innovation programme under grant agreement No 101004214.

\section*{Data availability}
The data presented here are all available from the public archives of DESI, \textit{Gaia}, ZTF, SDSS, and PanSTARRS. Data and python scripts used to produce the figures presented in this paper, along with the FITS file containing the DESI EDR white dwarf catalogue, are available here \url{https://zenodo.org/records/10620344}.



\bibliographystyle{mnras}
\bibliography{aamnem99,aabib,Swan}




\newpage

\appendix

%
%
%

\section{Description of the DESI EDR white dwarf catalogue}\label{sec:catalogue}

\begin{table*} 
\centering
\caption{Catalogue FITS extension one: the DESI EDR white dwarf catalogue. Parameters with a prefix of `dr2' and `edr3' are reproduced from the \textit{Gaia} white dwarf catalogues of \protect\cite{fusilloetal19-1} and \protect\cite{fusilloetal21-1} respectively.\label{t-WD_EDR_catalogue}}
\begin{tabular}{lll} 
\hline
Column & Heading                                      & Description \\
\hline
1      & wdj\_name                                    & WD J + J2000 RA (hh mm ss.ss) + dec (dd mm ss.s), equinox and epoch 2000 \\
2      & desi\_sp\_class                              & Spectral class determined from DESI EDR spectra \\
3      & desi\_id                                     & DESI target ID (unique within DESI targeting) \\
4      & desi\_nexp                                   & Number of exposures included in the DESI EDR \\
5      & desi\_stn\_aiq                               & Signal-to-noise ratios of the DESI EDR blue-arm individual exposures added in quadrature \\
6      & dr2\_source\_id                              & \textit{Gaia} DR2 unique source identifier (unique within a particular Data Release) \\
7      & dr2\_ra                                      & \textit{Gaia} DR2 Right Ascension (J2015.5) [deg] \\
\hline
\end{tabular}
\end{table*}

\clearpage

\setcounter{table}{0}

\begin{table*} 
\centering
\caption{Continued.\label{t-WD_EDR_catalogue-2}}
\begin{tabular}{lll} 
\hline
Column & Heading                                      & Description \\
\hline
58    & edr3\_pmra\_error                             & \textit{Gaia} EDR3 Standard error of proper motion in Right Ascension [mas yr$^{-1}$] \\
59    & edr3\_pmdec                                   & \textit{Gaia} EDR3 Proper motion in Declination [mas yr$^{-1}$] \\
60    & edr3\_pmdec\_error                            & \textit{Gaia} EDR3 Standard error of proper motion in Declination [mas yr$^{-1}$] \\
61    & edr3\_astrometric\_excess\_noise              & \textit{Gaia} EDR3 Measure of the residuals in the astrometric solution for the source [mas] \\
62    & edr3\_astrometric\_excess\_noise\_sig         & \textit{Gaia} EDR3 significance of the measure of the residuals in the astrometric solution \\
63    & edr3\_astrometric\_params\_solved             & \textit{Gaia} EDR3 Number of parameters solved for the astrometric solution \\
64    & edr3\_astrometric\_sigma5d\_max               & \textit{Gaia} EDR3 Five-dimensional equivalent to the semimajor axis of the Gaia position error ellipse [mas] \\
\hline
\end{tabular}
\end{table*}

\clearpage

\setcounter{table}{0}

\begin{table*} 
\centering
\caption{Continued.\label{t-WD_EDR_catalogue-3}}
\begin{tabular}{lll} 
\hline
Column & Heading                                      & Description \\
\hline
114    & edr3\_r\_med\_geo                             & Median of the geometric distance posterior (pc) \protect\citep{bailer-jonesetal21-1} \\
115    & edr3\_r\_lo\_geo                              & 16th percentile of the geometric distance posterior (pc) \protect\citep{bailer-jonesetal21-1} \\
116    & edr3\_r\_hi\_geo                              & 84th percentile of the geometric distance posterior (pc) \protect\citep{bailer-jonesetal21-1} \\
117    & edr3\_r\_med\_photogeo                        & Median of the photogeometric distance posterior (pc) \protect\citep{bailer-jonesetal21-1} \\
118    & edr3\_r\_lo\_photogeo                         & 16th percentile of the photomgeometric distance posterior (pc) \protect\citep{bailer-jonesetal21-1} \\
119    & edr3\_r\_hi\_photogeo                         & 84th percentile of the photomgeometric distance posterior (pc) \protect\citep{bailer-jonesetal21-1} \\
120    & edr3\_fidelity\_v1                            & `Astrometric fidelity' metric from \protect\cite{rybizkietal21-1} \\
\hline
\end{tabular}
\end{table*}

\begin{table*} 
\centering
\caption{Catalogue FITS extension two: crossmatched SDSS spectra.\label{t-WD_EDR_catalogue_e2}}
\begin{tabular}{lll} 
\hline
Column & Heading              & Description \\
\hline
1      & wdj\_name            & WD J + J2000 ra (hh mm ss.ss) + dec (dd mm ss.s), equinox and epoch 2000 \\
2      & plate                & Plate identifier for a unique SDSS spectrum \\
3      & mjd                  & MJD identifier for a unique SDSS spectrum \\
4      & fiberid              & Fiber identifier for a unique SDSS spectrum \\
5      & desi\_sdss\_separation & Separation between DESI source and cross-matched source in the SDSS sample \\
6      & n\_sdss\_spec          & Number of unique SDSS spectra cross-matched to a given white dwarf in the DESI EDR sample \\
\hline
\end{tabular}
\end{table*}

\begin{table*} 
\centering
\caption{Catalogue FITS extension three: photometric fits to DESI EDR DA white dwarfs.\label{t-WD_EDR_catalogue_e3}}
\begin{tabular}{lll} 
\hline
Column & Heading              & Description \\
\hline
1      & wdj\_name      & WD J + J2000 ra (hh mm ss.ss) + dec (dd mm ss.s), equinox and epoch 2000 \\
2      & teff           & Effective temperature [K] obtained from fit \\
3      & teff\_err      & Error on effective temperature [K] obtained from fit \\
4      & logg           & Surface gravity [cm\,s$^{-2}$] obtained from fit\\
5      & logg\_err      & Error on surface gravity [cm\,s$^{-2}$] obtained from fit \\
6      & d\_fitted      & Distance [pc] obtained from the fit \\
7      & d\_fitted\_err & Error on the distance [pc] obtained from the fit \\

\hline
\end{tabular}
\end{table*}

\begin{table*} 
\centering
\caption{Catalogue FITS extension four: spectroscopic fits to DESI EDR DA white dwarfs.\label{t-WD_EDR_catalogue_e4}}
\begin{tabular}{lll} 
\hline
Column & Heading              & Description \\
\hline
1      & wdj\_name     & WD J + J2000 ra (hh mm ss.ss) + dec (dd mm ss.s), equinox and epoch 2000 \\
2      & teff          & Effective temperature [K] obtained from fit \\
3      & teff\_err     & Error on effective temperature [K] obtained from fit \\
4      & logg          & Surface gravity [cm\,s$^{-2}$] obtained from fit\\
5      & logg\_err     & Error on surface gravity [cm\,s$^{-2}$] obtained from fit \\
6      & teff\_3D\_0.8 & Effective temperature [K] obtained from fit after applying 3D corrections of \protect\cite{tremblayetal11-1} \\ 
7      & logg\_3D\_0.8 & Surface gravity [cm\,s$^{-2}$] obtained from fit after applying 3D corrections of \protect\cite{tremblayetal11-1} \\

\hline
\end{tabular}
\end{table*}

\begin{table*} 
\centering
\caption{Catalogue FITS extension five: spectroscopic fits to SDSS DA white dwarfs.\label{t-WD_EDR_catalogue_e5}}
\begin{tabular}{lll} 
\hline
Column & Heading              & Description \\
\hline
1      & wdj\_name       & WD J + J2000 ra (hh mm ss.ss) + dec (dd mm ss.s), equinox and epoch 2000 \\
2      & mjd             & MJD identifier for a unique SDSS spectrum \\
3      & plate           & Plate identifier for a unique SDSS spectrum \\
4      & fiberid         & Fiber identifier for a unique SDSS spectrum \\
5      & teff            & Effective temperature [K] obtained from fit \\
6      & teff\_err       & Error on effective temperature [K] obtained from fit \\
7      & logg            & Surface gravity [cm\,s$^{-2}$] obtained from fit\\
8      & logg\_err       & Error on surface gravity [cm\,s$^{-2}$] obtained from fit \\
9      & teff\_3D\_0.8   & Effective temperature [K] obtained from fit after applying 3D corrections of \protect\cite{tremblayetal11-1} \\ 
10     & logg\_3D\_0.8   & Surface gravity [cm\,s$^{-2}$] obtained from fit after applying 3D corrections of \protect\cite{tremblayetal11-1} \\

\hline
\end{tabular}
\end{table*}

\begin{table*} 
\centering
\caption{Catalogue FITS extension six: photometric fits to DESI EDR DB white dwarfs.\label{t-WD_EDR_catalogue_e6}}
\begin{tabular}{lll} 
\hline
Column & Heading              & Description \\
\hline
1      & wdj\_name      & WD J + J2000 ra (hh mm ss.ss) + dec (dd mm ss.s), equinox and epoch 2000 \\
2      & teff           & Effective temperature [K] obtained from fit \\
3      & teff\_err      & Error on effective temperature [K] obtained from fit \\
4      & logg           & Surface gravity [cm\,s$^{-2}$] obtained from fit\\
5      & logg\_err      & Error on surface gravity [cm\,s$^{-2}$] obtained from fit \\
6      & d\_fitted      & Distance [pc] obtained from the fit \\
7      & d\_fitted\_err & Error on the distance [pc] obtained from the fit \\

\hline
\end{tabular}
\end{table*}

\begin{table*} 
\centering
\caption{Catalogue FITS extension seven: spectroscopic fits to DESI EDR DB white dwarfs.\label{t-WD_EDR_catalogue_e7}}
\begin{tabular}{lll} 
\hline
Column & Heading              & Description \\
\hline
1      & wdj\_name     & WD J + J2000 ra (hh mm ss.ss) + dec (dd mm ss.s), equinox and epoch 2000 \\
2      & teff          & Effective temperature [K] obtained from fit \\
3      & teff\_err     & Error on effective temperature [K] obtained from fit \\
4      & logg          & Surface gravity [cm\,s$^{-2}$] obtained from fit\\
5      & logg\_err     & Error on surface gravity [cm\,s$^{-2}$] obtained from fit \\
6      & teff\_3D\_0.8 & Effective temperature [K] obtained from fit after applying 3D corrections of \protect\cite{cukanovaiteetal21-1} \\ 
7      & logg\_3D\_0.8 & Surface gravity [cm\,s$^{-2}$] obtained from fit after applying 3D corrections of \protect\cite{cukanovaiteetal21-1} \\

\hline
\end{tabular}
\end{table*}

\begin{table*} 
\centering
\caption{Catalogue FITS extension eight: spectroscopic fits to SDSS DB white dwarfs.\label{t-WD_EDR_catalogue_e8}}
\begin{tabular}{lll} 
\hline
Column & Heading              & Description \\
\hline
1      & wdj\_name       & WD J + J2000 ra (hh mm ss.ss) + dec (dd mm ss.s), equinox and epoch 2000 \\
2      & mjd             & MJD identifier for a unique SDSS spectrum \\
3      & plate           & Plate identifier for a unique SDSS spectrum \\
4      & fiber           & Fiber identifier for a unique SDSS spectrum \\
5      & teff            & Effective temperature [K] obtained from fit \\
6      & teff\_err       & Error on effective temperature [K] obtained from fit \\
7      & logg            & Surface gravity [cm\,s$^{-2}$] obtained from fit\\
8      & logg\_err       & Error on surface gravity [cm\,s$^{-2}$] obtained from fit \\
9      & teff\_3D\_0.8   & Effective temperature [K] obtained from fit after applying 3D corrections of \protect\cite{cukanovaiteetal21-1} \\ 
10     & logg\_3D\_0.8   & Surface gravity [cm\,s$^{-2}$] obtained from fit after applying 3D corrections of \protect\cite{cukanovaiteetal21-1} \\

\hline
\end{tabular}
\end{table*}

\begin{table*} 
\centering
\caption{Catalogue FITS extension nine: photometric fits to DESI EDR DBA/DAB white dwarfs.\label{t-WD_EDR_catalogue_e9}}
\begin{tabular}{lll} 
\hline
Column & Heading              & Description \\
\hline
1      & wdj\_name      & WD J + J2000 ra (hh mm ss.ss) + dec (dd mm ss.s), equinox and epoch 2000 \\
2      & teff           & Effective temperature [K] obtained from fit \\
3      & teff\_err      & Error on effective temperature [K] obtained from fit \\
4      & logg           & Surface gravity [cm\,s$^{-2}$] obtained from fit\\
5      & logg\_err      & Error on surface gravity [cm\,s$^{-2}$] obtained from fit \\
6      & d\_fitted      & Distance [pc] obtained from the fit \\
7      & d\_fitted\_err & Error on the distance [pc] obtained from the fit \\

\hline
\end{tabular}
\end{table*}

\begin{table*} 
\centering
\caption{Catalogue FITS extension ten: spectroscopic fits to DESI EDR DBA/DAB white dwarfs.\label{t-WD_EDR_catalogue_e10}}
\begin{tabular}{lll} 
\hline
Column & Heading              & Description \\
\hline
1      & wdj\_name     & WD J + J2000 ra (hh mm ss.ss) + dec (dd mm ss.s), equinox and epoch 2000 \\
2      & teff          & Effective temperature [K] obtained from fit \\
3      & teff\_err     & Error on effective temperature [K] obtained from fit \\
4      & logg          & Surface gravity [cm\,s$^{-2}$] obtained from fit\\
5      & logg\_err     & Error on surface gravity [cm\,s$^{-2}$] obtained from fit \\
6      & teff\_3D\_0.8 & Effective temperature [K] obtained from fit after applying 3D corrections of \protect\cite{cukanovaiteetal21-1} \\ 
7      & logg\_3D\_0.8 & Surface gravity [cm\,s$^{-2}$] obtained from fit after applying 3D corrections of \protect\cite{cukanovaiteetal21-1} \\

\hline
\end{tabular}
\end{table*}

\begin{table*} 
\centering
\caption{Catalogue FITS extension eleven: spectroscopic fits to DESI EDR DQx white dwarfs.\label{t-WD_EDR_catalogue_e11}}
\begin{tabular}{lll} 
\hline
Column & Heading              & Description \\
\hline
1      & wdj\_name  & WD J + J2000 ra (hh mm ss.ss) + dec (dd mm ss.s), equinox and epoch 2000 \\
2      & teff       & Effective temperature [K] obtained from fit \\
3      & teff\_err  & Error on effective temperature [K] obtained from fit \\
4      & logg       & Surface gravity [cm\,s$^{-2}$] obtained from fit\\
5      & logg\_err  & Error on surface gravity [cm\,s$^{-2}$] obtained from fit \\
6      & mass       & Mass (\Msun) obtained from fit \\ 
7      & c\_he      & [C/He] abundance obtained from fit \\
8      & c\_he\_err & Error on the [C/He] abundance obtained from fit \\

\hline
\end{tabular}
\end{table*}

\clearpage

\section{Details and validation of the DA, DB, and DBA fitting routines}\label{sec:fitting_methods}

The synthetic spectra used in the fitting procedure are generated by the latest version of the \cite{koester10-1} code. The convection zones are modelled employing the 1D mixing-length (ML) approximation, with the ML2 parametrization and a fixed convective efficiency $\alpha$ set to 0.8 and 1.25 for the DAs and DBs/D(AB)s, respectively \citep[the canonical values in the treatment of H- and He-dominated photospheres,][]{tremblayetal10-1,bergeronetal11-1}.

The ranges and steps in effective temperature $\Teff$ that cover each of the employed grids of synthetic spectra vary with chemical composition. Pure H models are used to fit the DA spectra, spanning $5000 \leq \Teff \leq 80\,000$\,K in steps of 250, 1000, 2000 and 5000\,K up to 20\,000, 30\,000, 40\,000 and 80\,000\,K, respectively. We employed pure He models for DBs, which covered $9000 \leq \Teff \leq 40\,000$\,K in steps of 250, 1000 and 2000\,K up to 20\,000, 30\,000 and 40\,000\,K, since the He\,\textsc{i} transitions disappear from the spectrum below $9000$\,K and above 40\,000\,K He\,\textsc{ii} transitions are visible. The D(AB)s were fit with a H+He grid, equivalent to the pure He one in terms of $\Teff$ since the same reasoning applies. The three grids spanned surface gravities $7.0 \leq \logg \leq 9.5$\,dex in steps of 0.25\,dex. To obtain estimates on the H content, the H+He grid covered
$-7.0 \leq \htohe \leq -3.0$\,dex in steps of 0.25\,dex. 

For both the spectroscopic and photometric analyses, we used the Markov-Chain Monte Carlo (MCMC) \textsc{emcee} package within \textsc{python} \protect\citep{Foreman-Mackey2013}, following procedures established in previous work \protect\citep{izquierdo23} with some modifications that are explained in the following subsections. In all cases, chains were initialised with 30 walkers distributed randomly across the parameter space. Autocorrelation times were calculated and used to test for convergence (following the guidelines in the \textsc{emcee} documentation), to identify and discard a burn-in section, and to select 10000 independent samples from the ends of the chains.

\subsection{Spectroscopic fitting}

We performed spectroscopic analyses to the DESI coadded spectra and, when available, the SDSS archival spectroscopy, using up to three free parameters, namely \Teff, \logg\ and \htohe. The fitting technique is analogous to that presented in \cite{izquierdo23} except for the normalisation procedure. In this work we do not normalise the spectra by the continuum. Instead, the useful spectral regions (Balmer and/or He\,\textsc{i} transitions) were split in smaller windows. Each of these windows sample, at least, an absorption line and its surroundings. The total flux of each window is integrated and each data point contained in the window is divided by this value. The same normalisation is applied to both observed and synthetic spectra. The windows span $220$ and $200$\,\AA\ for the DAs and DB/D(AB)s, respectively. These optimal widths were identified through various internal tests with independent studies.

A purely spectroscopic fit usually has two degenerate solutions that can differ significantly in $T_{\textrm{eff}}$, and are called the hot and cold solutions. This is due to the equivalent widths (EW) of the Balmer absorption lines reaching a maximum around 13\,000\,K \protect\citep{rebassa-mansergasetal07-1}; and because the EW of the He\,\textsc{i} transitions reach a plateau between 22\,000 to 33\,000\,K \protect\citep[dependent on the \logg,][]{bergeronetal01-1}. In order to overcome this degeneracy, we performed two fits to each white dwarf: one using only cool synthetic spectra, and a second fit employing hot synthetic spectra, with some overlap between the two subsets. For the DAs these translated into cool models up to 15\,000\,K and hot ones down to 13\,000\,K. For the DBs and D(AB)s the cool models reached to 27\,000\,K and the hot ones went down to 22\,000\,K.

Due to this degeneracy, the default \textsc{emcee} setup is unsuitable as it is tuned to problems with unimodal parameters. To ensure proper exploration of the posteriors, we used differential evolution proposals, so that walkers are able to move freely between widely separated modes \protect\citep{terBraak2008, Nelson2014}. Specifically, we used \texttt{DEMove} with stretch factor $\gamma=1.17$ for 81~per cent of the moves, \texttt{DESnookerMove} ($\gamma=2.38/\sqrt{2}$) for 9~per cent, \texttt{DEMove} ($\gamma=1$) for 9~per cent,  and \texttt{DESnookerMove} ($\gamma=1$) for 1~per cent. 

To decide between the cold and hot solution for each H-dominated white dwarf, we followed the procedures presented in \protect\cite{rebassa-mansergasetal07-1} to evaluate the true value (the photometric result, since it is not subject to the EW degeneracy) against the theoretical value. We measured the EW of the Balmer lines in our models, and fit a polynomial in $\Teff-\logg$ space to the peak values. That way, if the independently-fit photometric solution falls on the hot side of that polynomial, we select the hot spectroscopic fit, similarly for the cold side/fit.

For the DB/D(AB)s there is no such study performed, so we chose between the hot and cold solutions by proximity to the photometric result of each star. 

We applied the 3D corrections to convert 1D spectroscopic values to 3D atmospheric parameters, following the prescriptions of \cite{tremblayetal11-1, cukanovaiteetal21-1}. These results are available in the online catalogue and used throughout the paper when specified. 

\subsection{Photometric fitting}

We modelled photometric archival data from SDSS, Pan-STARRS, \textit{Gaia} DR2 and EDR3 to test our spectroscopic results. These photometric analyses fitted for the \Teff, \logg, and distance, $D$, as free parameters and followed the same prescription as in \cite{izquierdo23} with a few variations. The model fluxes were scaled by the solid angle of the star $\pi(R_{\textrm{WD}}/D)$, with $R_{\textrm{WD}}$ the radius of the white dwarf derived from evolutionary models from \cite{Althaus2013}. The photometric data were supplemented by \textit{Gaia} parallaxes, but some of the white dwarfs have imprecise measurements. We therefore used a distance prior constructed following the popular prescription of \cite{bailer-jonesetal21-1} but generated only from sources with $\log{g} > 7$ from the mock \textit{Gaia}~EDR3 catalogue \citep{Rybizki2020}. The synthetic spectra were reddened by the values from the G-Tomo 3D dust map \citep{Fitzpatrick1999, Lallement2022, Vergely2022}, queried at the inverse-parallax distance, and then integrated over the SDSS bandpasses.

\subsection{Validation of DA fitting method}

\begin{figure*}
	\includegraphics[width=2\columnwidth]{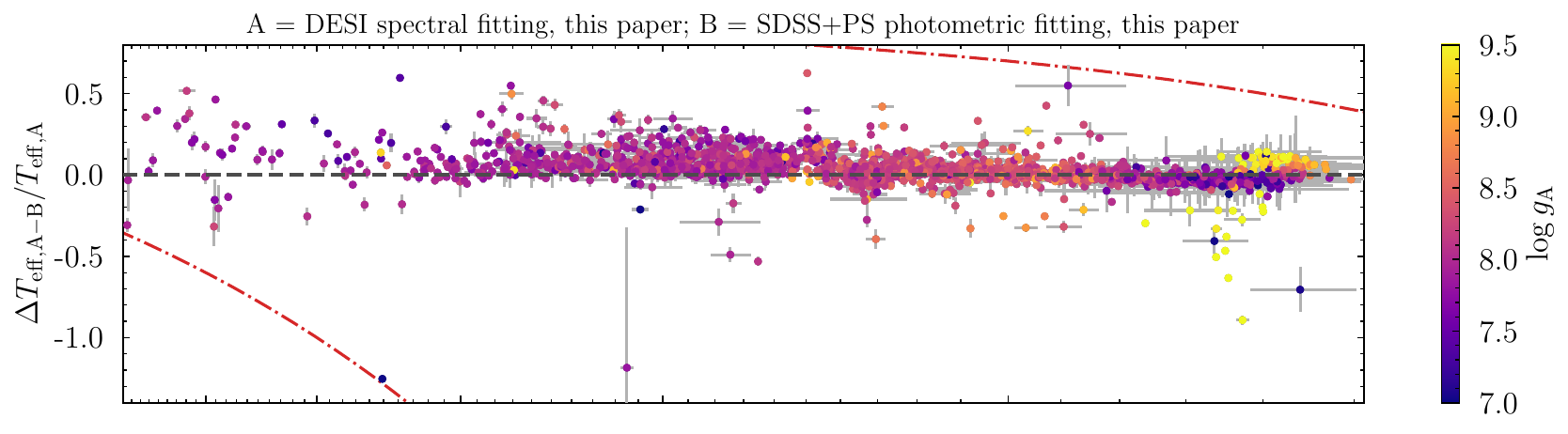}
 	\includegraphics[width=2\columnwidth]{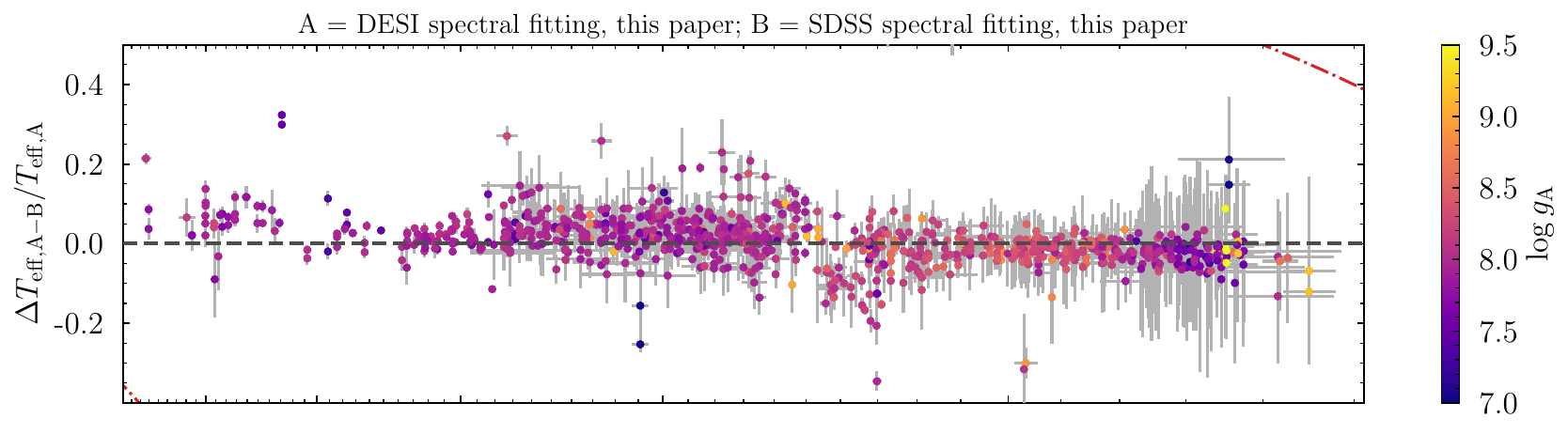}
  	\includegraphics[width=2\columnwidth]{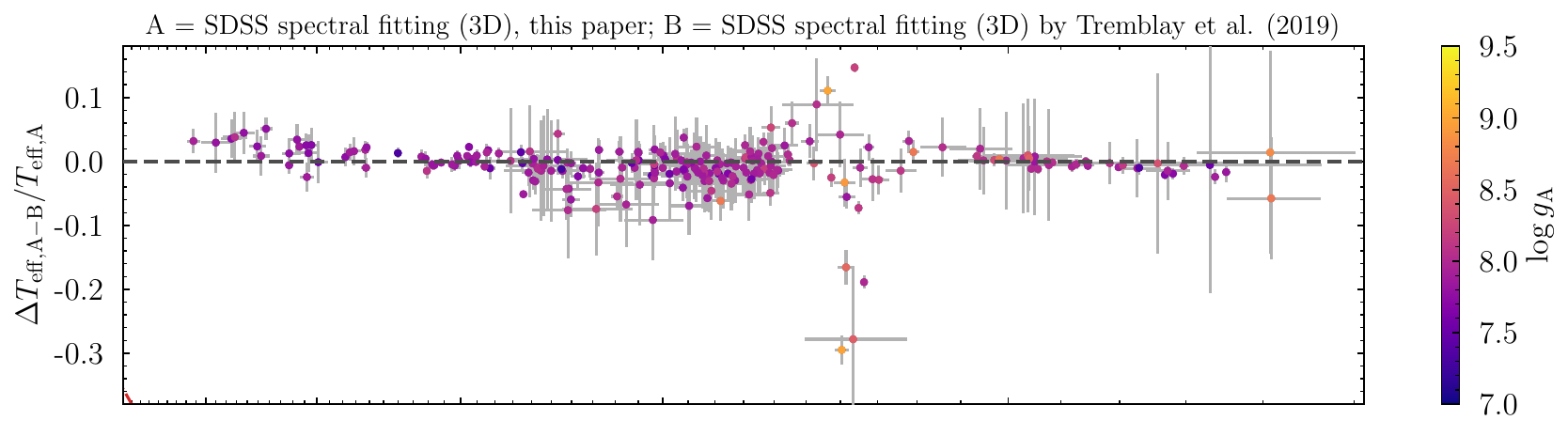}
   	\includegraphics[width=2\columnwidth]{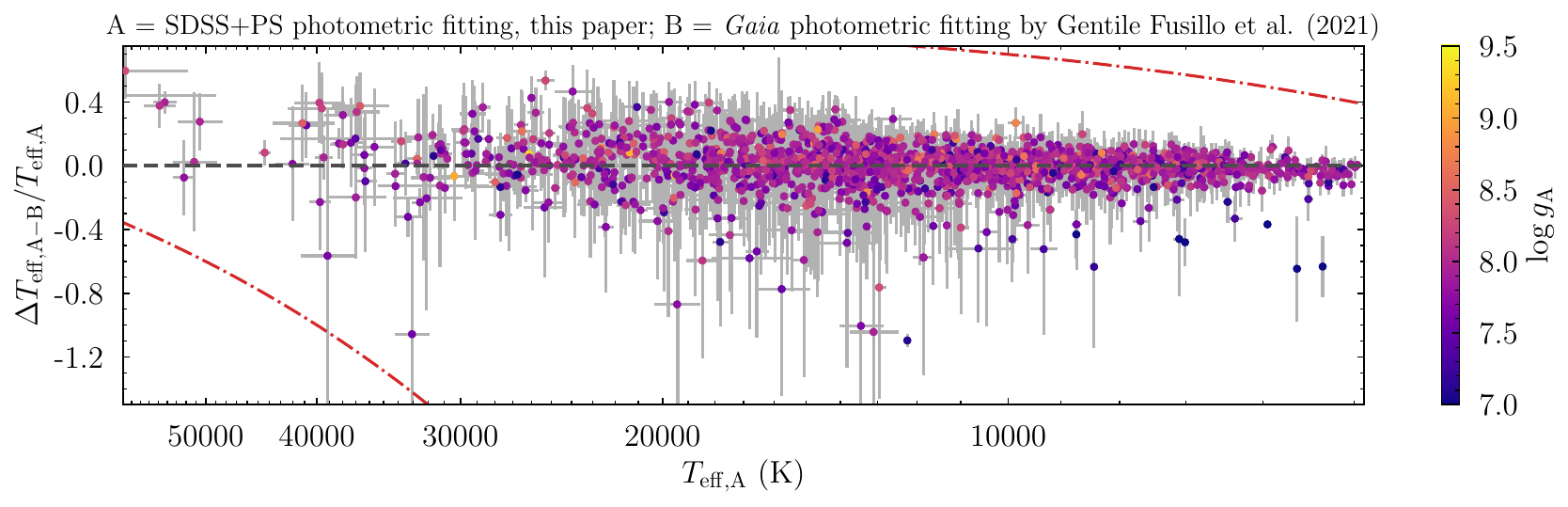}
    \caption{Comparisons of our measured spectroscopic and photometric effective temperatures for DA white dwarfs observed by DESI with several other samples. Data sets are labelled `A' and `B', with definitions given in the heading of each panel. Data points colour-coded by their determined surface gravities. Red dot-dashed lines show the curves $y = (T_\mathrm{eff,A} - 3000)/T_\mathrm{eff,A}$, and $y = (T_\mathrm{eff,A} - 80000)/T_\mathrm{eff,A}$, indicating the bounds of our model grid. From top to bottom, we compare: (1) our spectroscopic fitting of DESI spectra with our photometric fitting using SDSS and PanSTARRS photometry; (2) our spectroscopic fitting of DESI spectra with our spectroscopic fitting of SDSS spectra where a given DA has been observed by both DESI and SDSS; (3) our spectroscopic fitting of SDSS spectra with spectroscopic 3D-fitting by \protect\cite{tremblayetal19-1} with DAs observed by DESI within the sample of stars modelled by \protect\cite{tremblayetal19-1}; (4) our photometric fitting using SDSS and PanSTARRS photometry with \textit{Gaia} photometric fits of GF21 for DA white dwarfs observed by DESI.}
    \label{fig:DAs_temptest}
\end{figure*}

\begin{figure*}
	\includegraphics[width=2\columnwidth]{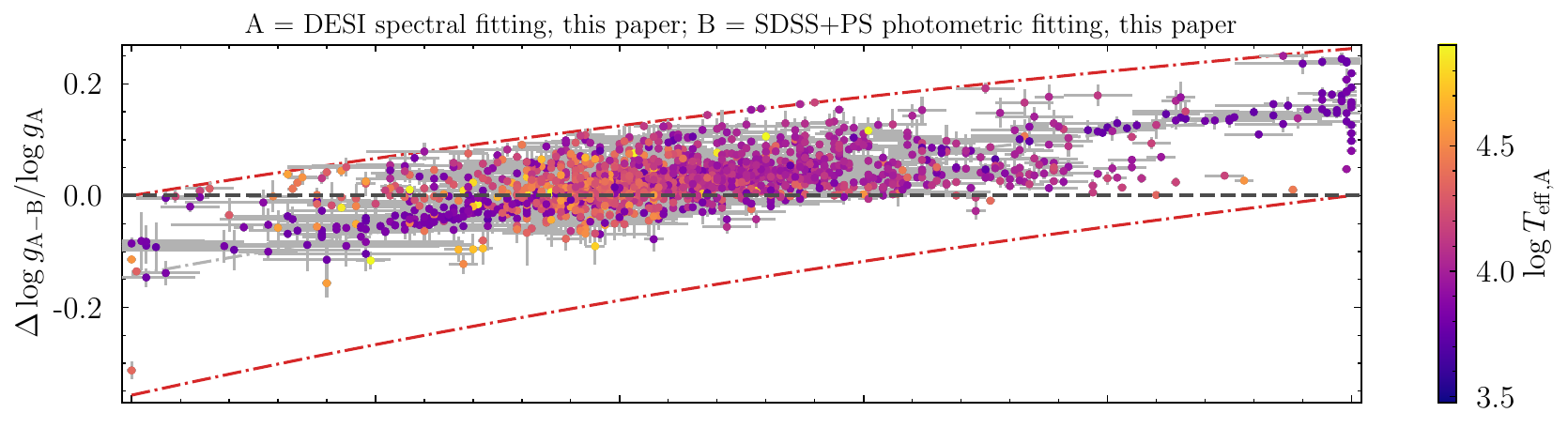}
 	\includegraphics[width=2\columnwidth]{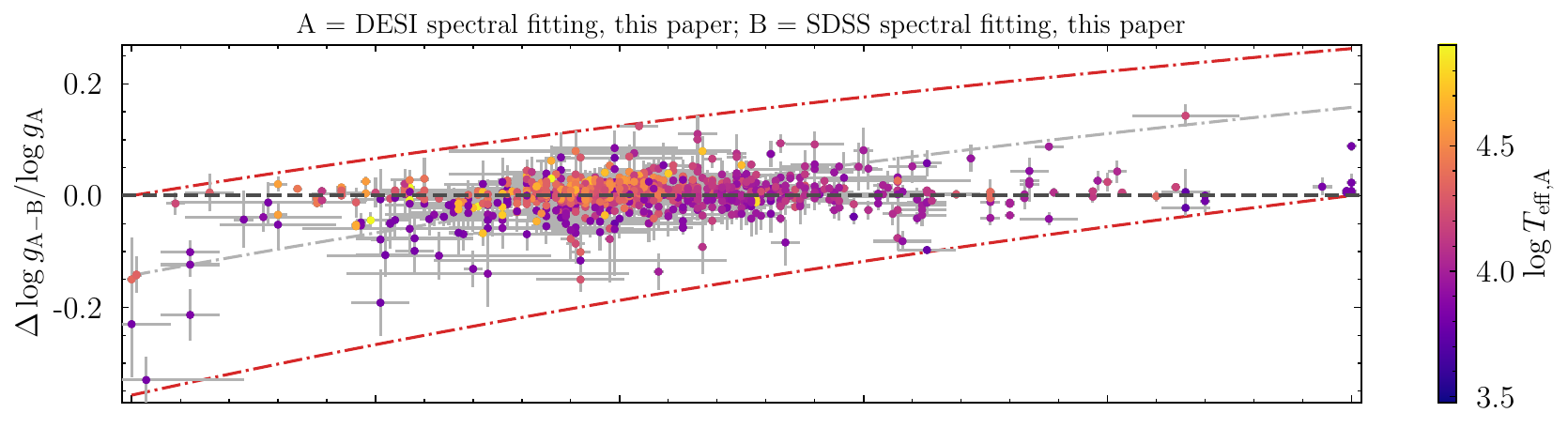}
  	\includegraphics[width=2\columnwidth]{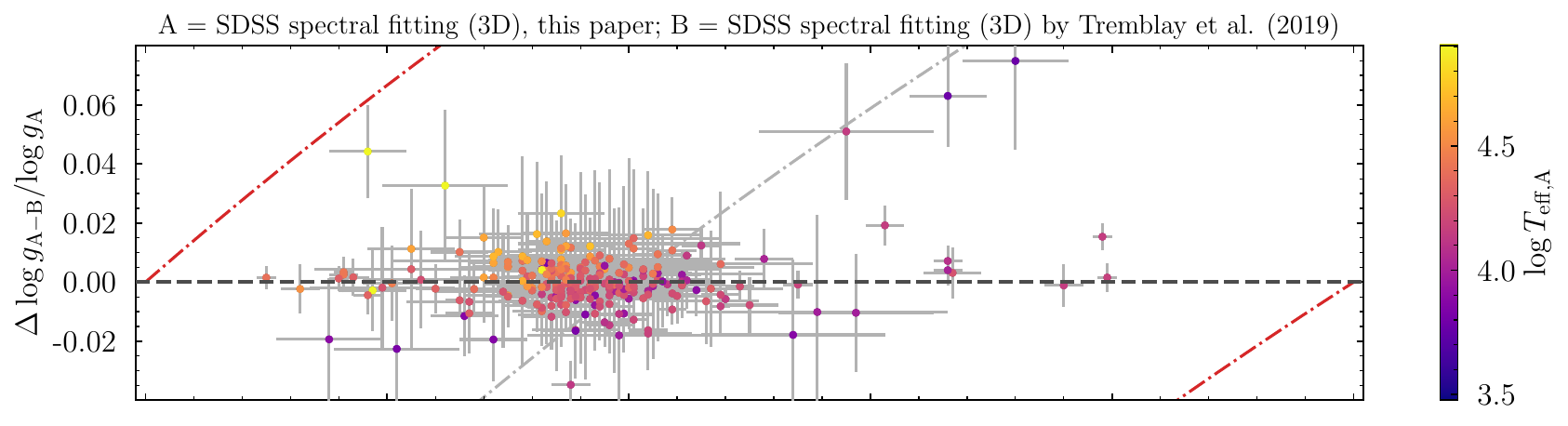}
   	\includegraphics[width=2\columnwidth]{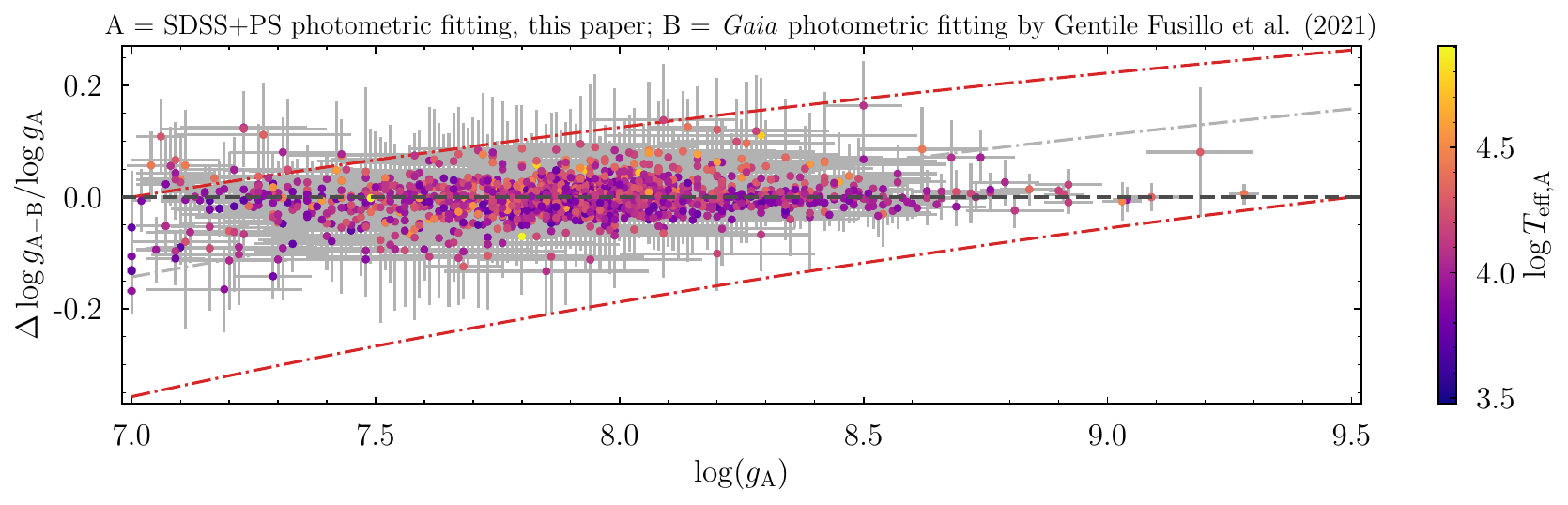}
    \caption{As in Fig.\,\ref{fig:DAs_temptest} but comparisons are shown for surface gravities with points colour-coded by their determined effective temperatures. Red dot-dashed lines show the curves $y = (\log{g}_\mathrm{A} - 7.0)/\log{g}_\mathrm{A}$, and $y = (\log{g}_\mathrm{A} - 9.5)/\log{g}_\mathrm{A}$, indicating the bounds of our model grid. The grey dot-dashed line marks the canonical $y = (\log{g}_\mathrm{A} - 8.0)/\log{g}_\mathrm{A}$.}
    \label{fig:DAs_loggtest}
\end{figure*}

In order to test the accuracy of our spectroscopic \Teff\ and \logg\ drawn from the DESI EDR spectra we compared them to different sets of data in Figs.\,\ref{fig:DAs_temptest} and \ref{fig:DAs_loggtest}. We use the same notation as in Figs.\,\ref{fig:DAs_temptest} and \ref{fig:DAs_loggtest}, where A and B refer to the two samples that are being compared in a respective panel.

For the comparison of our spectroscopic DESI results (A) with an internal photometric test we chose the SDSS and Pan-STARRS catalogues (B). In general, there is good agreement between the spectroscopic and photometric results, with a weighted mean $\langle\Delta T_\mathrm{eff,A-B}/T_\mathrm{eff,A}\rangle = 0.15 \pm 0.22$ and $\langle\Delta \log g_\mathrm{A-B}/\log g_\mathrm{A}\rangle= 0.04 \pm 0.05$. In particular, we are obtaining higher \Teff\ from the DESI spectroscopic analysis than that derived from photometry, which is most significant above $\Teff \geq $15\,000\,K, which is also observed in other DA population studies that compare spectroscopic and photometric solutions \citep{tremblayetal19-1,genest-beaulieu+bergeron19-1,obrienetal23-1}. This is also illustrated in the individual mean and standard deviations for the hotter objects ($\langle\Delta T_\mathrm{eff,A-B}/T_\mathrm{eff,A}\rangle = 0.34 \pm 0.24$), compared to the objects with $\Teff \leq 15\,000$\,K of $0.056 \pm 0.13$. The large discrepancies obtained for white dwarfs with $\Teff \geq 30\,000$\,K are likely related to the loss of sensitivity to \Teff\ that affects optical photometry as \Teff\ increases, becoming unreliable as it reaches the Rayleigh-Jeans regime.


We fit the archival SDSS spectra of the 963 DA white dwarfs (B) in the same way as the DESI EDR spectra (A) to validate our results. If we remove from the sample two low signal-to-noise ratio SDSS spectra (those corresponding to WDJ140151.43+032946.46 and WDJ140547.76-013454.94), which lead to meaningless values both compared to our DESI spectroscopic results and the literature values obtained from photometry, we find $\langle\Delta T_\mathrm{eff,A-B}/T_\mathrm{eff,A}\rangle = 0.02 \pm 0.05$ and $\langle\Delta \log g_\mathrm{A-B}/\log g_\mathrm{A}\rangle= 0.01 \pm 0.02$. This is an excellent agreement given the low signal-to-noise ratio of several of the SDSS spectra compared to DESI. In fact, these figures are in line with the ones seen in literature for independent analyses performed on different spectroscopic data sets \citep{genest-beaulieu+bergeron14-1}.

Some of the SDSS stars that also feature in the DESI EDR sample were fit by \cite{tremblayetal19-1} and the comparison between their analysis (B) and ours (A), both with 3D corrections applied, yields $\langle\Delta T_\mathrm{eff,A-B}/T_\mathrm{eff,A}\rangle = - 0.002 \pm 0.031$ and $\langle\Delta \log g_\mathrm{A-B}/\log g_\mathrm{A}\rangle= -0.0003 \pm 0.0072$. The biggest discrepancies are found in the region where the hot and cold solutions are split and hence, systematic uncertainties due to the different approaches may arise. 

The spectroscopic and photometric comparison performed in this work relies on the validity of both types of analyses. We have compared our DESI EDR spectroscopic results with two independent samples and thus we need to contrast too the goodness of our photometric fit. This comparison between our photometric fitting (A), and the photometric catalogue of \cite{fusilloetal21-1} (B), is displayed in the bottom panels of Figs.\,\ref{fig:DAs_temptest} and \ref{fig:DAs_loggtest}. We obtained $\langle\Delta T_\mathrm{eff,A-B}/T_\mathrm{eff,A}\rangle = 0.007 \pm 0.087$ and $\langle\Delta \log g_\mathrm{A-B}/\log g_\mathrm{A}\rangle= -0.002 \pm 0.016$, which are both on the order of the uncertainties obtained when fitting independent photometric catalogues (see figs.\,7 to 10 of GF19). 

\begin{figure*}
	\includegraphics[width=2\columnwidth]{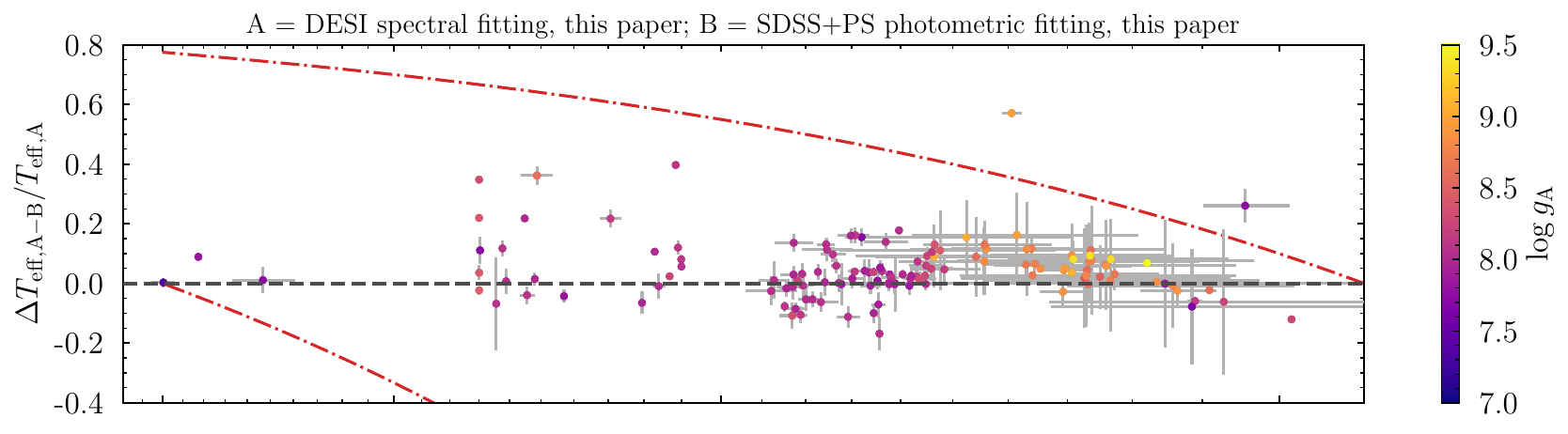}
 	\includegraphics[width=2\columnwidth]{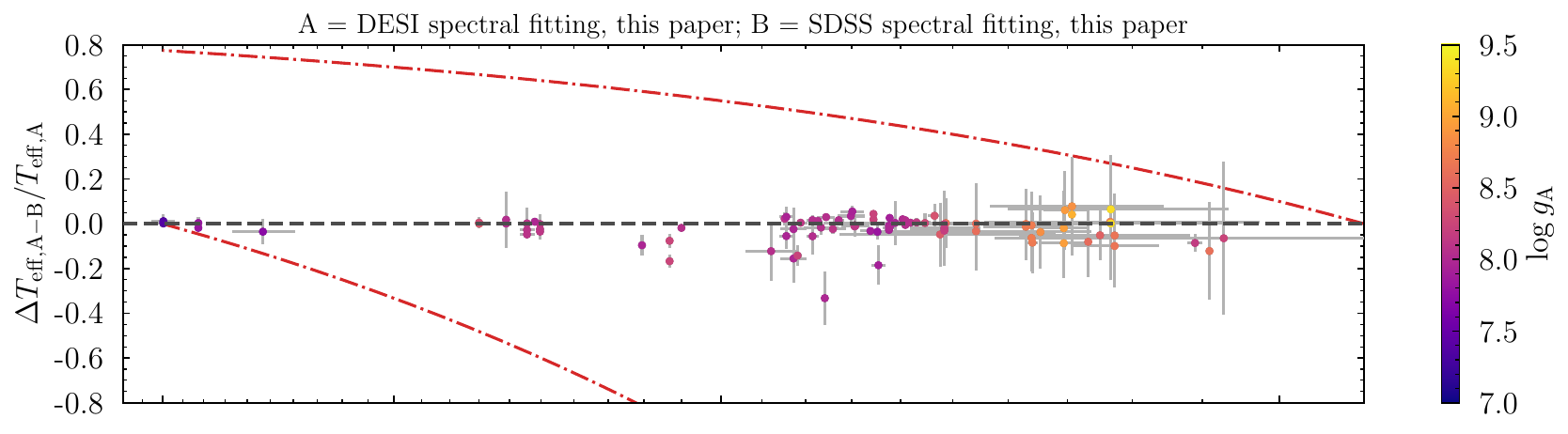}
   	\includegraphics[width=2\columnwidth]{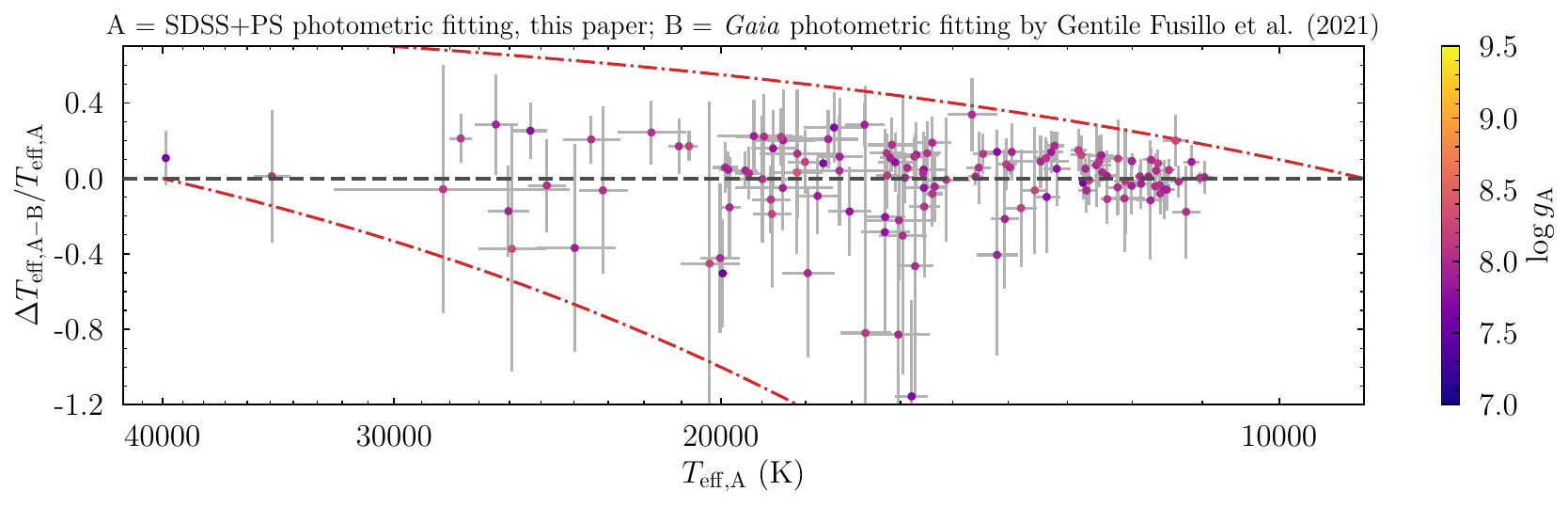}
    \caption{Comparisons of our measured spectroscopic and photometric effective temperatures for DB white dwarfs observed by DESI with several other samples. Data sets are labelled `A' and `B', with definitions given in the heading of each panel. Data points colour-coded by their determined surface gravities. Red dot-dashed lines show the curves $y = (T_\mathrm{eff,A} - 40\,000)/T_\mathrm{eff,A}$, and $y = (T_\mathrm{eff,A} - 9000)/T_\mathrm{eff,A}$, indicating the bounds of our model grid. From top to bottom, we compare: (1) our spectroscopic fitting of DESI spectra with our photometric fitting using SDSS and PanSTARRS photometry; (2) our spectroscopic fitting of DESI spectra with our spectroscopic fitting of SDSS spectra where a given DB has been observed by both DESI and SDSS; (3) our photometric fitting using SDSS and PanSTARRS photometry with \textit{Gaia} photometric fits of GF21 for DB white dwarfs observed by DESI.}
    \label{fig:DBs_temptest}
\end{figure*}

\begin{figure*}
	\includegraphics[width=2\columnwidth]{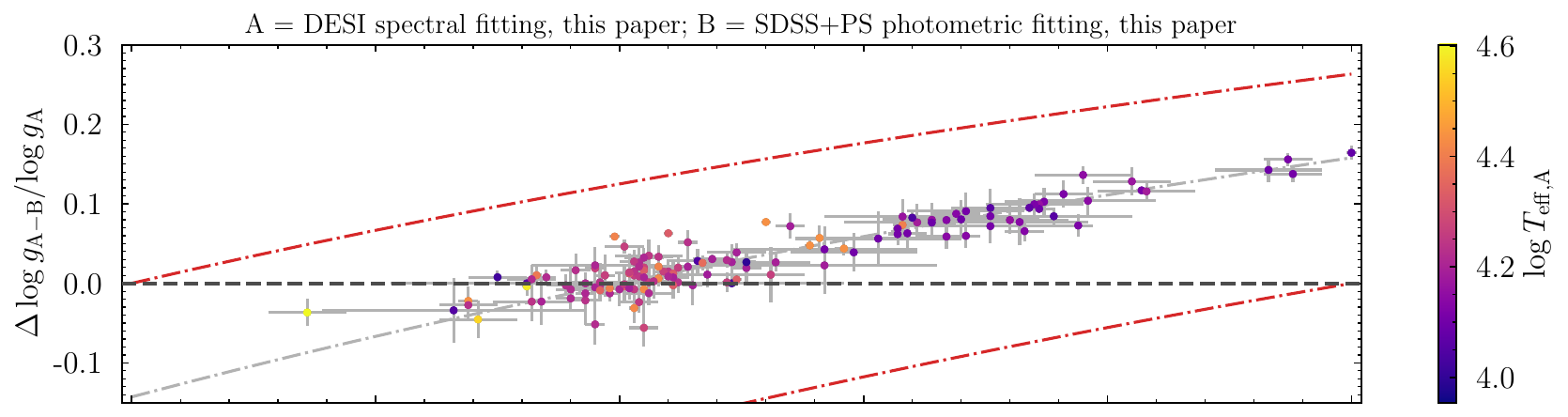}
 	\includegraphics[width=2\columnwidth]{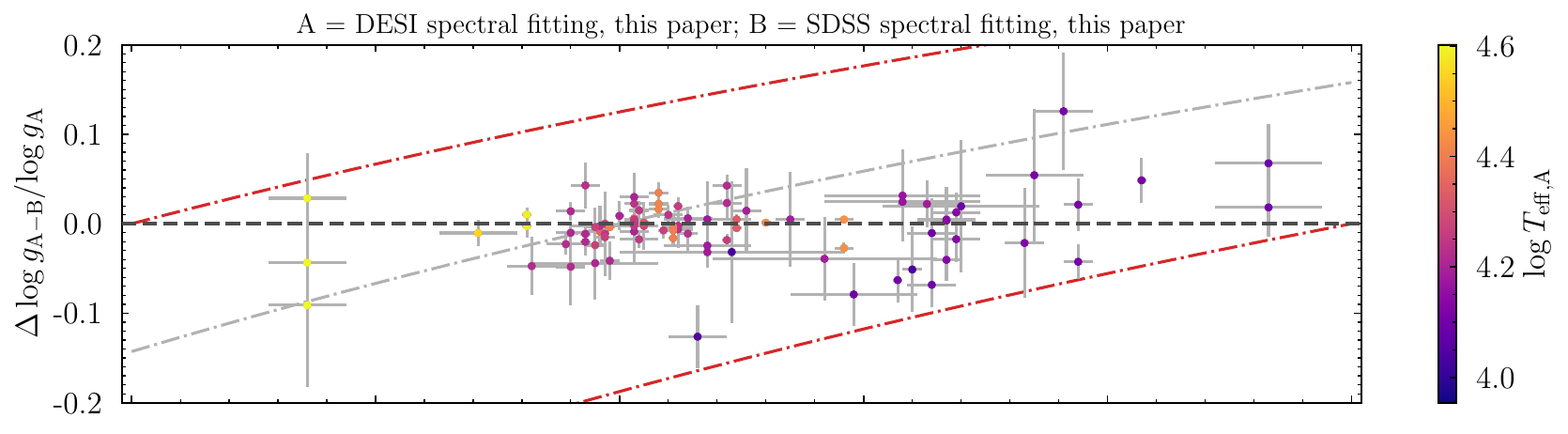}
   	\includegraphics[width=2\columnwidth]{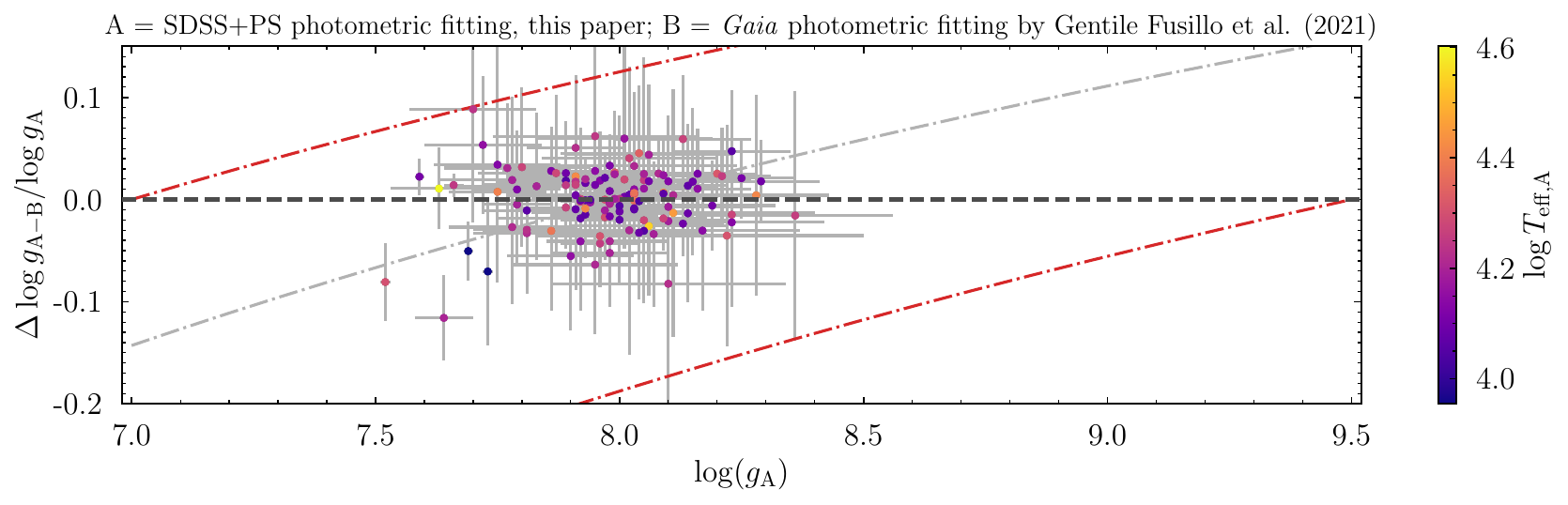}
    \caption{As in Fig.\,\ref{fig:DBs_temptest} but comparisons are shown for surface gravities with points colour-coded by their determined effective temperatures. Red dot-dashed lines show the curves $y = (\log{g}_\mathrm{A} - 7.0)/\log{g}_\mathrm{A}$, and $y = (\log{g}_\mathrm{A} - 9.5)/\log{g}_\mathrm{A}$, indicating the bounds of our model grid. The grey dot-dashed line marks the canonical $y = (\log{g}_\mathrm{A} - 8.0)/\log{g}_\mathrm{A}$.}
    \label{fig:DBs_loggtest}
\end{figure*}

Our spectroscopic results drawn from the DESI EDR spectra for the DB white dwarfs are contrasted to independent analyses in Figs.\,\ref{fig:DBs_temptest} and \ref{fig:DBs_loggtest}. As for the DA white dwarfs, we use the notation of Figs.\,\ref{fig:DBs_temptest} and \ref{fig:DBs_loggtest} when describing the comparison between two given samples.

The comparison between the spectroscopic (A) and photometric (B) techniques we developed (top panels in Figs.\,\ref{fig:DBs_temptest} and \ref{fig:DBs_loggtest}) highlights the non-physical increase in spectroscopic \logg\ below $\Teff \simeq 15\,000$\,K. Within the same regime, we have identified two objects whose photometrically derived \Teff\ fall outside our spectroscopic grid boundaries, as indicated by the red dot-dashed lines. This is because the He\,\textsc{i} transitions are only visible above $\Teff \simeq 9000$\,K but the photometric solutions can cover cooler temperatures as well, including DC spectral types. Inspection of these two systems in more details revealed that the observed photometry did not appear consistent, and that the spectroscopic solutions are preferred. The \Teff\ differences between 20\,000 and 30\,000\,K originate from the difficulty to obtain accurate photometric parameters in the instability strip of DBs (where they pulsate) and its coincidence with the maximum of the EW for He\,\textsc{i} lines, where the hot and cold technique is necessary but forces sharp boundaries in \Teff. In general, we find a weighted mean of $\langle\Delta T_\mathrm{eff,A-B}/T_\mathrm{eff,A}\rangle = 0.15 \pm 0.19$ and $\langle\Delta \log g_\mathrm{A-B}/\log g_\mathrm{A}\rangle= 0.05 \pm 0.04$, similar to the values seen in the literature \citep{tremblayetal19-1}. If we calculate those values for the objects with photometric \Teff\ smaller than 15\,000\,K, we get $\langle\Delta T_\mathrm{eff,A-B}/T_\mathrm{eff,A}\rangle = 0.11 \pm 0.23$ and $\langle\Delta \log g_\mathrm{A-B}/\log g_\mathrm{A}\rangle= 0.09 \pm 0.04$; and for the hotter ones: $\langle\Delta T_\mathrm{eff,A-B}/T_\mathrm{eff,A}\rangle = 0.17 \pm 0.17$ and $\langle\Delta \log g_\mathrm{A-B}/\log g_\mathrm{A}\rangle= 0.04 \pm 0.03$, which illustrates the bigger discrepancy in \Teff\ for hotter objects and the significant difference in \logg\ for the cooler ones. 


The refitted archival SDSS spectroscopy (B) of crossmatched DESI EDR (A) DB white dwarfs shows a good agreement, both in \Teff\ and \logg, with $\langle\Delta T_\mathrm{eff,A-B}/T_\mathrm{eff,A}\rangle = 0.000 \pm 0.018$ and $\langle\Delta \log g_\mathrm{A-B}/\log g_\mathrm{A}\rangle= -0.003 \pm 0.014$. The comparison between our SDSS+Pan-STARRS photometric (A) parameters and the \textit{Gaia}-obtained values published by GF21 (B) results in $\langle\Delta T_\mathrm{eff,A-B}/T_\mathrm{eff,A}\rangle = 0.05 \pm 0.09$ and $\langle\Delta \log g_\mathrm{A-B}/\log g_\mathrm{A}\rangle= 0.003 \pm 0.018$. Given the small number of D(AB) white dwarfs no population study was performed, but the derived spectroscopic and photometric parameters are available in the online catalog.

\clearpage

\section{Tables of DQ classifications}

\begin{table*}
\centering
\caption{Carbon-rich atmosphere white dwarfs. Statistical errors from our fitting routines are given in brackets, which in some cases are underestimates. Realistic errors are 100\,K in $T_\mathrm{eff}$ and 0.1 dex in $\log g$.}\label{t-DQx}
\begin{tabular}{lllllll} 
\hline
Name                   & Class & $T_\mathrm{eff}$ (K) & $\log g$      & Mass (M$_\odot$ ) & [C/He]               & References/Notes                             \\
\hline                                                                                                               
\multicolumn{2}{l}{Newly confirmed carbon-rich systems} &               &               &                   &                      &                                              \\
\hline                                                                                                                                     
WDJ072922.34+451425.96 & DQ    & 7182 (62)            & 7.943 (0.155) & 0.541             & $-$6.261 (0.004)       &                                              \\
WDJ074019.13+314234.94 & DQ    & 7385 (30)            & 7.988 (0.078) & 0.568             & $-$6.087 (0.006)       &                                              \\
WDJ084131.55+332915.47 & DQ    & 6736 (23)            & 7.978 (0.087) & 0.561             & $-$6.741 (0.002)       &                                              \\
WDJ085049.40+315656.41 & DQ    & 8859 (70)            & 7.724 (0.222) & 0.427             & $-$4.682 (0.007)       &                                              \\
WDJ085601.74+321354.51 & DQ    & 22611 (1379)         & 9.286 (0.102) & 1.252             & $\sim$0.100 (0.100)      &                                              \\
WDJ091350.42+833906.33 & DQ    & 8334 (119)           & 7.926 (0.094) & 0.533             & $-$5.700 (0.015)       &                                              \\
\end{tabular}
\end{table*}

\clearpage

\section{Tables of metal-line classifications}

\begin{table*} 
\centering
\caption{White dwarfs exhibiting metal lines in the DESI sample. \label{t-DxZ1}}
\begin{tabular}{lll} 
\hline
Name                   & Class & Reference for DxZ candidacy \\
\hline                               
\multicolumn{2}{l}{Newly confirmed metal-line systems} &  \\
\hline 
WD\,J013105.68+314755.63 & DAZ  & \\ 
WD\,J014145.83+311049.96 & DAZ  & \\ 
WD\,J014211.57+314603.62 & DBZ  & \\ 
WD\,J020559.33--042230.18 & DZB  & \\ 
WD\,J054545.32--190245.54 & DAZ  & \\ 
WD\,J071959.42+402122.13 & DBZ  & \cite{sionetal88-1} \\
\hline
\end{tabular}
\end{table*}

\clearpage

\setcounter{table}{0}

\begin{table*}
\centering
\caption{Continued.\label{t-DxZ2}}
\begin{tabular}{lll} 
\hline
Name                   & Class & Reference for DxZ candidacy \\
\hline                                                                                                               
\multicolumn{2}{l}{Newly confirmed metal-line systems} &  \\
\hline 
WD\,J114201.76+562949.00 & DAZ  & \\ 
WD\,J115040.63+552956.66 & DZ   & \\ 
WD\,J115802.16+010537.12 & DZ   & \\ 
WD\,J115803.66--011257.60 & DZA  & \\ 
WD\,J115908.60+525653.55 & DZ   & \\ 
WD\,J120538.84+524626.80 & DAZ  & \\
\end{tabular}
\end{table*}

\clearpage

\setcounter{table}{0}

\begin{table*}
\centering
\caption{Continued.\label{t-DxZ3}}
\begin{tabular}{lll} 
\hline
Name                     & Class & Reference for DxZ candidacy               \\
\hline                                                                                                               
\multicolumn{2}{l}{Newly confirmed metal-line systems} &                      \\
\hline 
WD\,J155300.64+692620.03 & DZA  &                                             \\ 
WD\,J161316.63+552125.97 & DAZ  &                                             \\ 
WD\,J161502.83+544603.38 & DBZ  &                                             \\ 
WD\,J162344.53+744313.12 & DAZ  &                                             \\ 
WD\,J163054.20+434009.43 & DBZA &                                             \\ 
WD\,J163405.62+314336.23 & DAZ  &                                             \\
\hline
\end{tabular}
\end{table*}

\clearpage

\section{Tables of magnetic classifications}

\begin{table*}
\centering
\caption{Magnetic field strengths, $B$, determined for magnetic white dwarfs in our sample. Statistical errors from our fitting routines are given in brackets which do not take into account systematics or more complex field structures, and a more reasonable lower bound is $\simeq$\,5\,per\,cent of the derived $B$.\label{t-DxH}}
\begin{tabular}{llll} 
\hline
Name                                       & Class & $B$ (MG)                 & References                               \\
\hline
Newly identified magnetic systems          &       &                          &                                          \\
\hline
WD\,J022723.13--055305.72                  & DAH   & 20.5 (0.4)               &                                          \\
WD\,J042613.14--034701.99$^{\textrm{a}}$   & DAH   & 250\,-\,1000             &                                          \\
WD\,J070253.76+553733.64$^{\textrm{b}}$    & DAH   & 6.89 (0.07) / 15.5 (0.2) &                                          \\
WD\,J082302.41+334534.27                   & DAH   & 1.344 (0.008)            &                                          \\
WD\,J084546.99+194542.45                   & DAH   & 7.01 (0.05)              &                                          \\
WD\,J085045.22+110553.50                   & DAH   & 542 (1)                  &                                          \\
\end{tabular}
\end{table*}

\clearpage

\section{Tables of CVs classifications}

\begin{table*}
\centering
\caption{Table of twelve cataclysmic variables (CVs) identified in the DESI EDR which includes three new systems, two confirmed candidates, and seven previously known systems. All systems except J161927.83+423039.61 are present in the \protect\cite{fusilloetal21-1}  \textit{Gaia} EDR3 white dwarf catalogue. \label{t-CVs}}
\begin{tabular}{llllll}
\hline
Name                  & Alternative name         & Simbad Type   & References                              \\
\hline
Newly confirmed CVs  & & & & & \\
\hline
J124413.48+593610.24  & Gaia 16adi               & CV candidate  & \cite{hodgkinetal21-1}                  \\
J142833.44+003100.45  &                          & CV candidate  & \cite{croometal04-1}                    \\
J143435.39+334049.98  &                          & WD candidate  &                                         \\
J161927.83+423039.61  &                          & Star          &                                         \\
J181130.89+795608.36  &                          & WD candidate  &                                         \\
\hline
Known CVs & & & & & \\
\hline
J022319.73--052002.02 & 2XLSSd J022319.8--052002 & CV            & \cite{kepleretal16-1, fusilloetal19-1}  \\
J090350.73+330036.11  & CRTS J090350.7+330036    & CV            & \cite{szkodyetal05-1}                   \\
J125510.56+264226.59  & V* MT Com                & CV            & \cite{watsonetal96-1}                   \\
J131223.48+173659.21  & 2XMMp J131223.4+173659   & CV            & \cite{vogeletal08-1}                    \\
J150241.00+333424.00  & V* NZ Boo                & CV            & \cite{szkodyetal06-1}                   \\
J151333.10+703721.98  & Gaia 17bcx               & CV            & \cite{hodgkinetal21-1,kelleretal22-1}   \\
J162608.15+332827.83  &                          & CV            & \cite{szkodyetal04-1}                   \\
\hline
\end{tabular}
\end{table*}

\clearpage

\section{DESI EDR white dwarf systems not in the \textit{Gaia} EDR3 white dwarf catalogue}

\begin{figure*}
	\includegraphics[width=1.8\columnwidth]{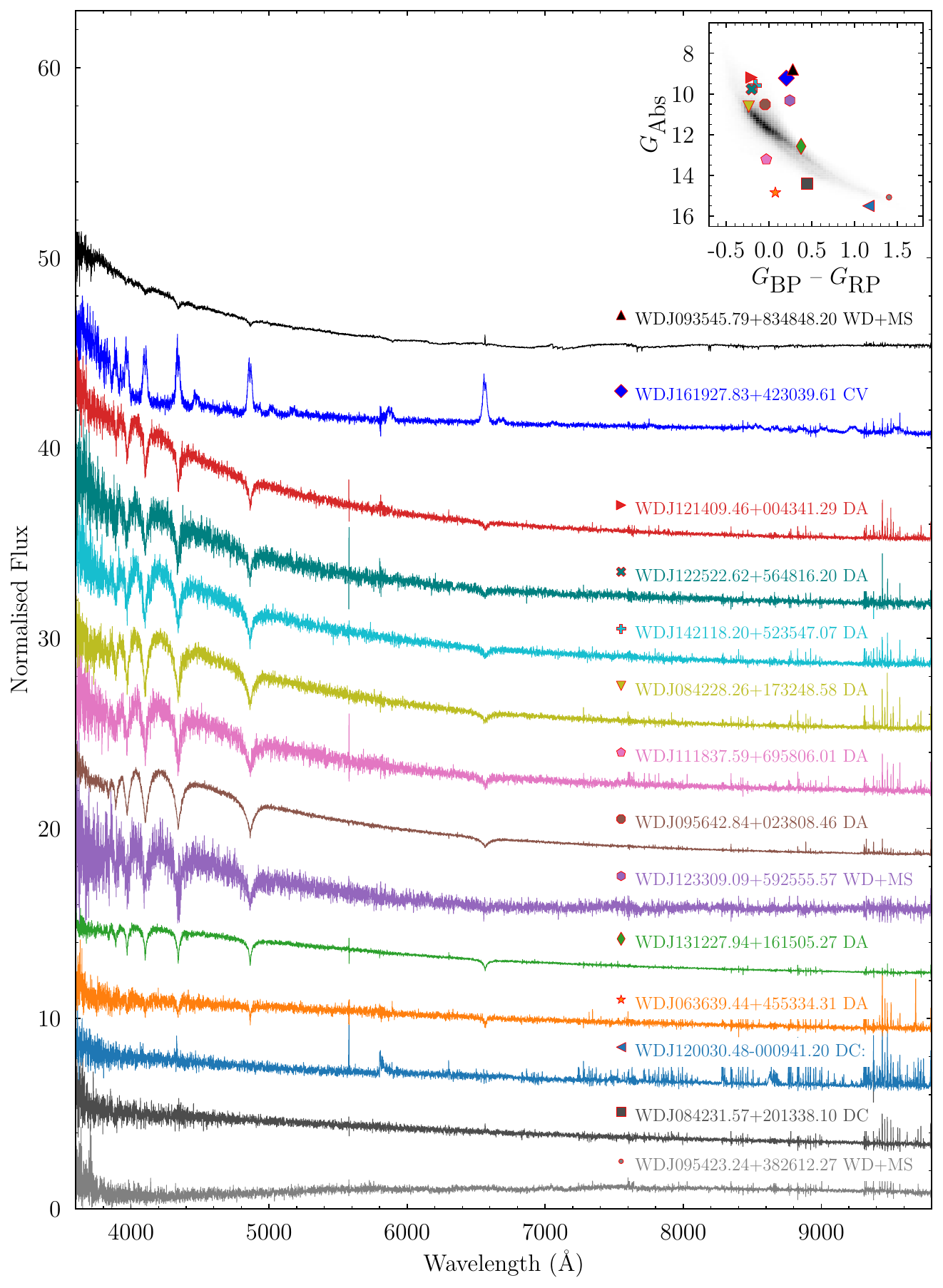}
    \caption{Normalised DESI spectra of white dwarf systems in the DESI EDR sample that do not have an entry in the \textit{Gaia} EDR3 white dwarf catalogue of \protect\cite{fusilloetal21-1}. The spectra are offset by arbitrary amounts for clarity. A few systems have emission features that appear due to imperfect flux calibration and/or sky subtraction, which are particularly strong in the spectrum of WD\,J120030.48--000941.20. The inset shows the \textit{Gaia} HRD, where each system is displayed with a corresponding colored symbol outlined in red. For reference, the high-confidence ($P_{\textrm{WD}} > 0.95$) \textit{Gaia} white dwarfs brighter than $G=20$ identified by \protect\cite{fusilloetal21-1} are shown as a 2D gray-scale histogram.}
    \label{fig:sysindr2}
\end{figure*}



\bsp	
\label{lastpage}
\end{document}